\documentclass[aps, prb,superscriptaddress, twocolumn,showpacs,preprintnumbers,amsmath,amssymb]{revtex4}
\usepackage{graphicx}
\usepackage{bm}
\usepackage{amssymb, latexsym, amsmath, textcomp}
\usepackage{hyperref}

\def\Tr{{\text{Tr}}\,}

\renewcommand{\v}[1]{{\bf #1}}
\renewcommand{\c}[1]{{\cal #1}}

\newcommand{\s}{{\sigma}}
\newcommand{\gr}{{\nabla}}
\def\be{\begin{eqnarray}}
\def\ee{\end{eqnarray}}
\newcommand{\nn}{\nonumber\\}

\newcommand{\Eq}[1]{Eq.~(\ref{#1})}
\newcommand{\e}{\epsilon}
\newcommand{\p}{\partial}
\newcommand{\ua}{\uparrow}
\newcommand{\da}{\downarrow}
\newcommand{\ra}{\rightarrow}

\begin{document}
\title{A direct transition between
a Neel ordered Mott insulator and a $d_{x^2-y^2}$ superconductor on
the square lattice}
\author{Ying Ran}
\author{Ashvin Vishwanath}
\author{Dung-Hai Lee}
\affiliation{Department of Physics, University of California at
Berkeley, Berkeley, CA 94720, USA} \affiliation{Material Science
Division, Lawrence Berkeley National Laboratory, Berkeley, CA 94720,
USA}
\date{\today}

\begin{abstract}
In this paper we study a bandwidth-controlled direct, continuous,
phase transition from a Mott insulator, with easy plane Neel order,
to a fully gapped $d_{x^2-y^2}$ superconductor with a doubled unit cell on the square lattice, a transition that is
forbidden according to the Landau paradigm. This transition is made
possible because the vortices of the antiferromagnet are charged and
the vortices of the superconductor carry spins. These nontrivial
vortex quantum numbers arise because the ordered phases are
intimately related to a topological band insulator. We describe the
lattice model as well as the effective field theory.

\end{abstract}
\maketitle

One of the central questions in the theory of the high T$_c$ cuprates
is how the antiferromagnetic (AF) insulator is converted into a
d-wave superconductor (SC) upon doping. If a continuous transition between
these, in some sense, diametrically opposite phases could be realized, it would be possible to
construct an unified description of the SC and AF in terms of a single critical theory\cite{senthil:174515}.

An immediate indication that such a theory would be rather
unconventional, is that it violates the `Landau' rules which
dictates that continuous phase transitions are only possible when
the symmetry groups of the two phases have a subgroup relationship.
It has been recently proposed that Landau forbidden transitions
could in principle occur in quantum systems\cite{senthil-2004-303},
via `deconfined' quantum critical points. In this area, early work
on the AF to valence bond solid transition on the square lattice
\cite{senthil-2004-303,PhysRevB.42.4568} has been generalized to a
number of other transitions
\cite{balents:144508,grover:247202,burkov:134502,grover:156804}.
However, except an early numerical claim\cite{PhysRevLett.77.4592},
the physically most interesting case of a direct transition between
an AF insulator and a SC has not been discussed before. Here, we
provide a theory for such a transition.

The essential physics here is that vortices of the
AF (in our case the easy-plane variety) are charged and
the vortices of the SC carry spin. The condensation of
either type of vortices drives the system between the
two phases. The reason that the topological defects of these
two rather conventional phases carry unconventional
quantum numbers is because both phases are intricately related to a
topological band insulator. The later is a special type of band
insulator which can be differentiated from conventional ones by a topological quantum number.

The field theory that emerges at criticality is the easy plane version of the non-compact
CP$_1$ (NCCP$_1$) model\cite{senthil-2004-303,
PhysRevB.70.075104}. An important theoretical feature is that in
contrast to most earlier theories
\cite{balents:144508,grover:247202} there is no emergent
$U(1)$ symmetry at criticality. An interesting critical point
between an magnetic insulator and a d-wave SC with
nodal quasiparticles was proposed in Ref.\cite{senthil:174515}. However,
in contrast to the present theory, it is hard to
establish that the insulating state is truly an antiferromagnet.

In contrast to the cuprates, where the transition is triggered by
doping, here we consider a bandwidth controlled Mott transition,
where the system remains at half filling throughout. Moreover, in
addition to the usual d-wave symmetry, our SC breaks lattice
translation and time reversal symmetry due to a staggered spin
dependent hopping term. This leads to a gap appearing at the nodes.
Doping triggers a gap closing transition and gap nodes are restored
at sufficiently high doping. Finally, our AF phase has a spin
anisotropy favoring easy plane magnetic order. Topological defects with nontrivial quantum numbers were also discussed in Ref.\cite{PhysRevLett.80.5401,kou:235102,weng-2007-21}, in the context of cuprate superconductors.

While in mean-field theory the easy-plane NCCP$_1$ model clearly
predicts a continuous transition, the effect of fluctuations has
been actively
debated\cite{PhysRevB.70.075104,Prokofiev,sandvik:227202,melko:017203,motrunich-2008,kuklov-2008}.
If we consider the general family NCCP$_n$ models, these are know to
have continuous transitions for $n=0$ and $n\rightarrow \infty$. For
the case of $n=1$, with full $SU(2)$ symmetry, the case for a
continuous transition has been made in some microscopic realizations
\cite{sandvik:227202,melko:017203,motrunich-2008}, although other
authors have concluded that weak first order transitions are present
in the entire parameter range \cite{kuklov-2008}. Since the nature
of the transition depends on the microscopic realization, it is
clear that more numerical studies on a variety of models, including
those with easy plane symmetry, are required to settle this question
conclusively.
In this paper, we do not further address this issue, rather we aim
at establishing whether a continuous phase transition between the d-wave SC and AF insulator that is possible even in principle.

\begin{figure}
{\includegraphics[scale=0.45]{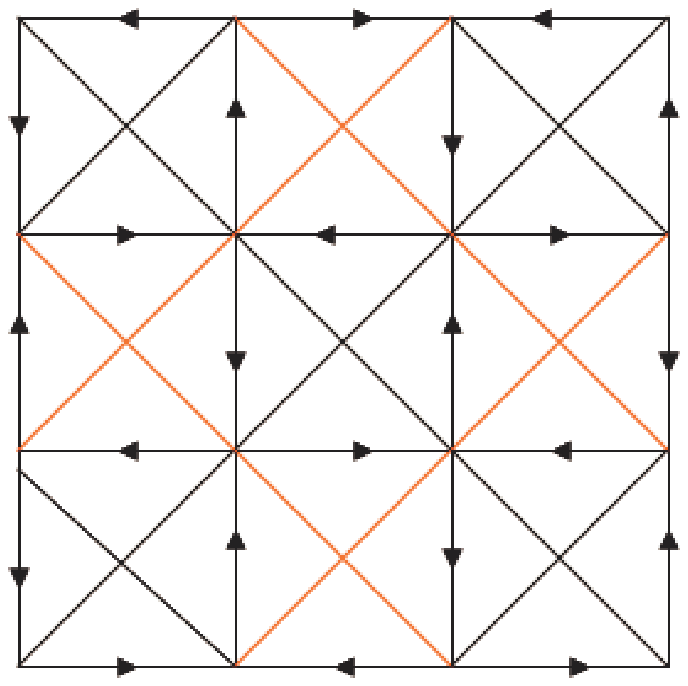}{{\bf
a}}\includegraphics[scale=0.45]{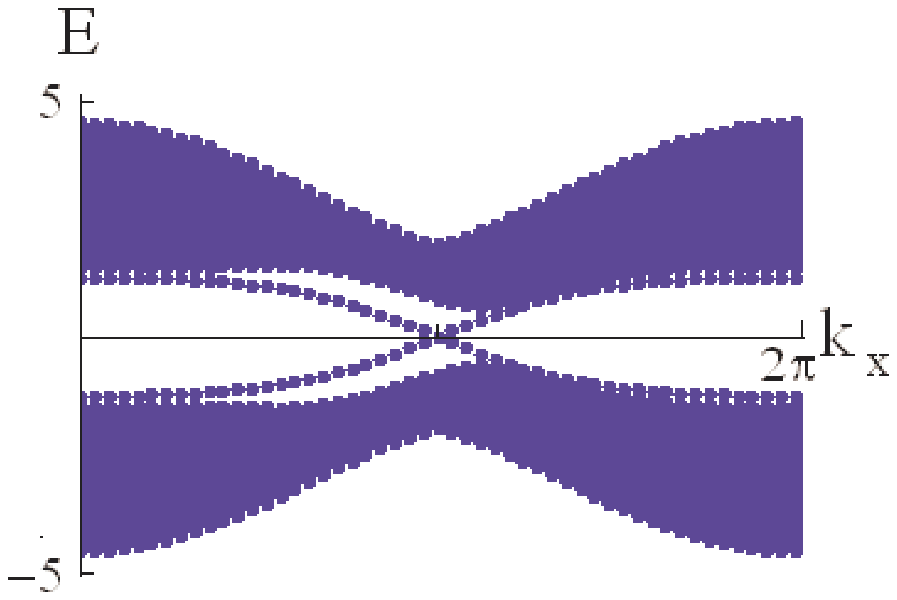}}{{\bf b}} \caption{(a)The
arrows indicate the direction in which the nearest neighbor hopping
amplitude is $e^{i\phi}$ (hopping in the reverse direction has
amplitude $e^{-i\phi}$. The color indicate the sign of the next
nearest neighbor hopping, Red means negative. (b)Bulk and edge
spectra of \Eq{sc} and \Eq{TBI}. We used parameters $\phi=\pi/5$ and
$t=0.2$ (see the discussions below \Eq{sc}).} \label{hop}
\end{figure}

The plan of the paper is as follows. We start with the Bogoliubov-de
Gennes description of a d$_{x^2-y^2}$  superconductor with a
particular unit cell doubling shown in Fig. \ref{sc} (see later) which leads to
gapped nodes. It turns out that this is a topological superconductor, which in
the presence of particle-hole symmetry, has stable edge modes. We
show that vortices of this superconductor carry spin and are bosons.
We ask what happen if these vortices condense. This is answered by
studying the field theory of the vortex excitations, which predicts
that the vortex condensed phase is an easy-plane antiferromagnet and
the critical theory is an easy plane NCCP$_1$ model.
A duality transformation allows us to study the transition in
reverse, where condensation of charged vortices of the AF order leads to superconductivity.
The self duality
of the NCCP$_1$ model \cite{PhysRevB.70.075104} leads to an
interesting spin-charge duality, whose consequences are described in
Section \ref{spincharge}. As a consistency check, we study the same problem
from the insulating side. Quite interestingly the most natural slave boson formulation leads to the
same phase diagram and critical theory. This alternative approach also
clarifies the role of symmetries required to realize this
transition, and allows us to generalize this to other lattices. In
the conclusion we discuss the similarity of the phenomenon described in this work and that of the cuprate  superconductors. Eight
appendices contain the important technical details of various part
of the paper, including Appendix \ref{sec:monopole} where the
nature of the magnetic order are studied in details and Appendix
\ref{sec:honeycomb} describes an equivalent theory on the honeycomb
lattice, where a transition between an easy plane AF and a {\it triplet}
superconductor occurs.

\section{The topological d-wave superconductor}\label{sec:SC}
Let us begin with a mean-field description of one side of the phase
transition: the d-wave superconductor. The usual d-wave
superconductor has gapless nodal quasi-particles;
here we introduce a modification which leads to a unit cell
doubling, and a full gap around the Fermi surface.
The mean-field Hamiltonian describing this superconductor is given by 
\be
 H^{SC}&&=\sum_{\langle
 ij\rangle \s}{\rm Re}(\chi_{ij})c_{i\s}^\dagger c_{j\s}-\sum_{\langle
 ij\rangle \s\s'}\eta_i{\rm
 Im}(\chi_{ij})\e_{\s\s'}c_{i\s}c_{j\s'}
 \nn&&+\sum_{\langle\langle ij\rangle\rangle\s}
 \s t_{ij}c_{i\s}^\dagger c_{j\s}+h.c.\label{sc} \ee

Here $i,j$ run through the sites of a square lattice, $\eta_i$ is
the AF staggered sign, and $\langle ij\rangle$,$\langle\langle
ij\rangle\rangle$ denote the nearest and second neighbors,
respectively; $\chi_{ij}=e^{\pm i\phi}$ and $t_{ij}=\pm t$ ($t=$
real).
We note, the nearest neighbor hopping, $Re(\chi_{ij})$, is the same as the
hopping parameter associated with the ``staggered flux phase'' of
the Heisenberg antiferromagnet\cite{PhysRevB.37.3774} (for a review,
see Ref.\cite{9015906}). We illustrate the $\chi_{ij}$ and $t_{ij}$
patterns in Fig(\ref{hop}(a)). It is important to note that the
pairing term has $d_{x^2-y^2}$ symmetry; due to the presence of $\eta_i$ in the second term of \Eq{sc}, the
pairing is translation-invariant.

However, because of the
spin-dependent second neighbor hopping, both lattice translation $T$ and
time reversal $\mathcal{TR}$ are broken. However, $T\circ \mathcal{TR}$ and reflection
along the vertical(horizontal) line passing the center of each plaquette ($\mathcal{P}$)  are
symmetries of the Hamiltonian. ((By looking at Fig(\ref{hop}(a)) one
might think that reflection is not a symmetry. However this figure
merely illustrates the pattern of $\chi_{ij}$ and $t_{ij}$ not the
symmetry of the Hamiltonian.) In addition, \Eq{sc} has spin $S_z$ rotation symmetry, and the charge U(1)
symmetry is, of course, spontaneously broken.

We will see in later discussions that $T\circ \mathcal{TR}$ and $\mathcal{P}$ together
with the $U(1)-S_z$ conservation and the $U(1)$-charge conservation
(which is spontaneously broken in the current superconductor phase),
ensures that the direct SC$\leftrightarrow$AF transition can be
continuous at half-filling. There are two more symmetries of the Hamiltonian \Eq{sc} which simplify
our analysis but are not necessary for the direct transition
to occur: the particle hole conjugation $PH$ ($
c_{i\uparrow}\rightarrow \eta_i
c_{i\downarrow}^{\dagger},c_{i\downarrow}\rightarrow -\eta_i
c_{i\uparrow}^{\dagger}$) and and translation together with
$c_{i\s}\leftrightarrow c_{i-\s}$ ($T\circ\ua\leftrightarrow\da$).

\Eq{sc} is closely related to a special type of topological
insulator (TBI) Hamiltonian \be
 H^{TBI}=\sum_{\langle i,j\rangle,\s}\chi_{ij}f_{i\sigma}^{\dagger}f_{j\sigma}
 +\sum_{\langle\langle i,j\rangle\rangle,\s}
 \s t_{ij} f_{i\s}^{\dagger}f_{j\s}
&+h.c..\label{TBI} \ee Indeed, one can transform \Eq{sc} into
\Eq{TBI} by making the following transformation \be
c_{i\ua}=e^{i{\pi\over 4}}\left({f_{i\ua}-\eta_i
f^\dagger_{i\da}\over\sqrt{2}}\right),~~c_{i\da}=e^{i{\pi\over
4}}\left({f_{i\da}+\eta_i f^\dagger_{i\ua}\over\sqrt{2}}\right)
\label{trans}.\ee If we drop the $\s$ in front of the $t_{ij}$,
\Eq{TBI} turns into the Hamiltonian for the ``chiral spin
liquid''\cite{PhysRevB.39.11413}. With the $\s$, the chirality of the spin up and
spin down fermions are opposite. In that case \Eq{TBI} describes a
band insulator with opposite ``TKNN index''\cite{PhysRevLett.49.405} ($\pm 1$) for
the spin up/down fermions. Hence this type of topological insulator
will exhibit the spin hall effect. However, unlike the spin hall
insulator introduced in Ref.\cite{kane:226801}, our spin hall insulator
breaks time reversal symmetry. (To appreciate the full relationship
between \Eq{sc} and \Eq{TBI}, see Sec. \ref{slave rotor})

Since \Eq{trans} is an
unitary transformation the excitation spectra of \Eq{TBI} and
\Eq{sc} are the same. One only needs to reinterpret the
electron/hole excitations in the TBI as the Bogoliubov quasiparticle
excitations in the superconductor. Interestingly, the conserved
$S_z$ of \Eq{TBI} translates into conserved $S_z$ of \Eq{sc} as
well.
Among other things the above correspondence implies that the
superconductor described by \Eq{sc} has gapless edge excitations
(Fig.\ref{hop}(b)). Thus it is justified to regard this type of
superconductor as ``topological''. In Appendix
\ref{subsec:pseudospin} and near the end of
Sec.\ref{subsec:charge-fluctuation} we show that the edge modes
requires the $PH$ symmetry which is difficult to realize
in a material. Therefore in general there is no gapless edge modes
and the SC phase is not topological. On the other hand the physics
of the direct SC$\leftrightarrow$AF transition, which is the main
result of the current paper, does nog require the existence of
gapless edge modes.

\begin{figure}
\includegraphics[scale=0.45]{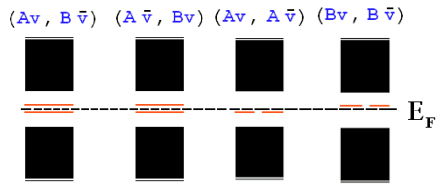}
\caption{The Bogoliubov quasiparticle (q.p.) spectra of a pair
vortex and anti vortex; The red lines mark the core states, and the
black solid line represents the Fermi energy. In the limit of infinite vortex-antivortex separation each
core level is doubly degenerate. We used parameters
$\phi=\pi/5$ and $t=0.2$ (see the discussions below \Eq{sc}).
\label{vc}}
\end{figure}

\begin{center}
\begin{figure}
\includegraphics[scale=0.5]{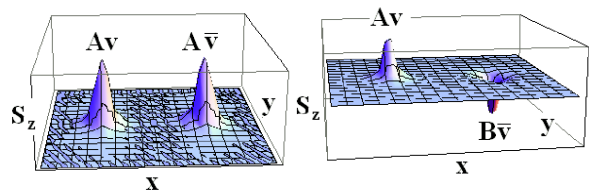}
\caption{The $S_z$ distribution of an (AV,A$\bar{{\rm V}}$) and
(AV,B$\bar{{\rm V}}$) pair. The total integrated values of $S_z$
around each vortex is $S_z=\pm 1/2$.} \label{prof}
\end{figure}
\end{center}

Now let us study the vortices of the superconductor. Direct numerical calculations for a vortex and antivortex pair gives
the spectra illustrated in Fig.(\ref{vc}), where each core level is doubly degenerate (in the limit
of infinite separation between the vortices)/
 There AV and B$\bar{{\rm V}}$ stand for a vortex located on A dual sublattice and an
antivortex on B dual sublattice, and etc. Here A/B dual sublattice
locate at the center of plaquette in Fig.(\ref{hop}(a)) marked by
red/black cross lines. The core levels of these two types of vortices lie on opposite side of zero. In the ground state all levels below/above zero are filled/empty. For each vortex/antivortex, there are three low lying excited states arise from different way of occupying the two core levels associated with it.
Explicit calculation of the ground state value of  $S_z$ associated with the four panels of
Fig.(\ref{vc}) gives\be S^z_{{\rm AV}}=S^z_{{\rm A\bar{\rm
V}}}=1/2,~{\rm and~}S^z_{{\rm BV}}=S^z_{{\rm B\bar{\rm
V}}}=-1/2.\label{vsp}\ee The spin distribution associated with
(AV,A$\bar{{\rm V}}$) and (AV,B$\bar{{\rm V}}$) is shown in
Fig.(\ref{prof}).
 The fact that vortices carry spins is the most
interesting aspect of the superconductor under consideration.

In the above numerical study, the size of the vortex core (the region in which the pairing amplitude is
suppressed to zero) is one lattice constant.
As the core size of the vortices increase, their core levels move toward zero. As the core size diverge, a continuum BdG equation applies. In that limit all core levels are situated at zero energy (Appendix \ref{sec:SC-BdG}) and the previous three (low lying) excited states become degenerate with the ground state. When that happen it is no longer possible to differentiate the spectrum of A and B dual sublattice vortices. In this limit each vortex/antivortex has four degenerate states: two with $S_z=\pm 1/2$ and the other two with spin zero.

Next we determine the statistics of the vortices. For this purpose
it is important to note that under
\begin{align} \mbox{$PH$: }& \mbox{AV}_{\uparrow}\leftrightarrow\mbox{A}\bar{\mbox{V}}_{\uparrow},\mbox{BV}_{\downarrow}\leftrightarrow\mbox{B}\bar{\mbox{V}}_{\downarrow}\notag\\
 \mbox{$T\circ\mathcal{TR}$: }&\mbox{AV}_{\uparrow}\leftrightarrow\mbox{B}\bar{\mbox{V}}_{\downarrow},\mbox{BV}_{\downarrow}\leftrightarrow\mbox{A}\bar{\mbox{V}}_{\uparrow}
\label{transv}\end{align}
 Now let $\phi_{{\rm BV}_\da,{\rm
BV}_\da}$, and etc., be the phase factor due to the
(counterclockwise) exchange of two BV$_\da$ vortices. $T\circ\mathcal{TR}$ plus
PH imply $\phi_{{\rm BV}_\da,{\rm BV}_\da}=\phi^*_{{\rm
AV}_\ua,{\rm AV}_\ua}$. On the other hand, $T\circ\uparrow\leftrightarrow\downarrow$ requires
$\phi_{{\rm BV}_\da,{\rm BV}_\da}=\phi_{{\rm AV}_\ua,{\rm
AV}_\ua}$. Put together the above implies $\phi^*_{{\rm
AV}_\ua,{\rm AV}_\ua}=\phi_{{\rm AV}_\ua,{\rm AV}_\ua}$.
Consequently \be \phi_{{\rm AV}_\ua,{\rm AV}_\ua}=\pm
1,\label{bf}\ee i.e., AV$_\ua$ is either boson or fermion (the
same for BV$_\da$). Although we used the symmetry $T\circ\ua\leftrightarrow\da$ to derive this result, the result Eq.(\ref{bf}) does not depend on that symmetry because statistics cannot change so long as the
core levels do not merge into the continuum. % state if we turn off the symmetry continuously.

\begin{figure}
 \includegraphics[scale=0.5]{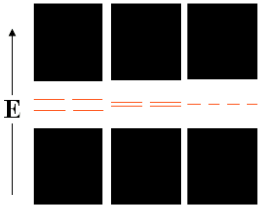}
\caption{The spectra of a AV$_\ua$ and a B$\bar{{\rm V}}_\da$
vortex as a function of $\lambda$. From left to right
$\lambda=1,{1\over 2}$ and 0. } \label{evolve}
\end{figure}

To further distinguish the two we consider the following gauge
transformation $c_{i\s}\ra e^{i\theta_i/2}c_{i\s}$. This
transformation remove the vorticity in the paring order parameter
up to a $\pi$ cut. The hopping parameters acquires a $\pi$ cut as
well as a smooth hopping phase $e^{i \alpha_{ij}}$ where
$\alpha_{ij}=Mod({\theta_i-\theta_j\over 2},\pi).$ This hopping
phase is often referred to as the ``Doppler shift term'' in the theory
of superconductivity. If one gets rid of the hopping phase, and
leaves the $\pi$ cut in both the pairing and hopping parameters, both core levels
of a given vortex/antivortex move to zero energy and there are four degenerate vortex states
, like the case of the diverging core size discussed earlier.  In Fig.(\ref{evolve}) we
illustrate the evolution of the spectra for a AV$_\ua$ and a B$\bar{{\rm V}}_\da$
vortex after the following modification of the hopping phase: \be e^{i \alpha_{ij}}\ra e^{i\lambda \alpha_{ij}}~~0\le
\lambda\le 1.\label{mod}\ee In
the $\lambda=0$ limit \Eq{trans} transforms \Eq{sc} to a spin-Hall
insulator with $Z_2$ flux in the plaquette previously occupied by the
vortices. This is a problem we have previously solved and the $S_z=\pm 1/2$
vortices (which are referred to as spin fluxons in Ref.\cite{ran-2008}) are bosons\cite{ran-2008}. It is simple to show that \Eq{transv} holds true for all $\lambda$ hence \Eq{bf}
remains valid. This, plus the fact that the spin fluxons adiabatically evolve into the spin-carrying vortices, implies \be \phi_{{\rm AV}_\ua,{\rm
AV}_\ua}\Big|_{\lambda=1}=1,\ee , i.e., the AV$\ua$ vortices are
bosons (same argument applies to BV$\da$, A$\bar{{\rm V}}\ua$ and
B$\bar{{\rm V}}\da$.)

\section{The superconductor-vortex field theory}\label{sec:sc-eff}
Since the vortices are bosons, they can condense. The condensation
will trigger the destruction of superconductivity and lead to an
insulator. However, since the vortices carry spin, we expect the resulting
insulator to have magnetic signature. To understand
what happens after vortices have condensed, we examine the effective
field theory describing the phase fluctuations of the
superconductor. Phase fluctuations appears in two different types:
gaussian (spin wave ) type and vortex-antivortex fluctuations. The
latter is responsible for destroying the superconducting state when
it is strong. For example, it is the vortex condensation that
triggers the ordinary superconductor to insulator transition. A
convenient way of describing the vortex physics is to use the dual
form of the phase fluctuation theory\cite{PhysRevLett.47.1556,PhysRevB.39.2756}. In this form, the
vortices are particles and the original boson density  and current
become magnetic and electric fields respectively.
Since this dual form is widely used in literature, we will refer the
readers to references\cite{PhysRevLett.47.1556,PhysRevB.39.2756} for details. The relevant field
theory for the superconducting vortices in three dimensional
Euclidean space takes the form:
\begin{widetext}
\be
  {\cal
L}&=\sum_{\s}\left\{\sum_\mu|(\partial_{\mu}+i\alpha_{\mu}-i\frac{\hbar}{2}A^S_{\mu}\sigma^3)
\Psi_{\s}|^2+m^2|\Psi_{\s}|^2\right\}
+v(\sum_\s|\Psi_{\s}|^2)^2+u|\Psi_{\ua}|^2|\Psi_{\da}|^2+\frac{\kappa}{2}(\nabla
\times \alpha)^2+{ie\over\pi}A\cdot(\nabla\times \alpha).\label{eff}
 \ee
\end{widetext}
Where $\Psi_\ua,\Psi_\da$ are relativistic vortex fields, and for
brevity, we have omitted the space-time  dependence of all fields. In addition, we have used a three vector notation
$\nabla=(\partial_0,\Delta_x,\Delta_y)$ to denote space time derivatives.
The connection between these relativistic fields and the microscopic
vortex fields is discussed in Appendix \ref{duality}. As usual, a
gauge field $\alpha$ is introduced such that
 ${2e\over 2\pi}\gr\times \alpha=$ is the Cooper pair 3-current
caused by the gaussian fluctuations of the pair-condensate phase. Hence, it is coupled to the external electromagnetic
field $A$, contained in the last term of \Eq{eff}.  The minimum
coupling
$\sum_\mu|(\partial_{\mu}+i\alpha_{\mu}-i\frac{\hbar}{2}A^S_{\mu}\sigma^3)\Psi_{\s}|^2$
between $\Psi_\s$ and $\alpha_\mu$ reflects the fact that vortices see
the 3-current of the condensate as electric and magnetic fields. To probe the
spin response we have we coupled a ``spin gauge field'' $A^S_\mu$ to the conserved $S_z$ current. Note that both
$A_\mu$ and $A^S_\mu$ are non-dynamical. The fact that the vortices minimally couple to the spin gauge field reflects the fact that they carry spin. The ${\hbar\over 2}\s^3$ in the coupling between $\Psi_\s$
and $A^S_\mu$ reflects the fact that the $\Psi_{\ua,\da}$ boson
carries $S_z=\pm 1/2$. Finally, we have included standard Ginzburg-Landau type quartic $\Psi_\s$ terms. They describe the interaction
between the order parameters of the two type of vortices.

To ensure the correctness of the low energy effective theory Eq.(\ref{eff}), the following conditions must be met: (1)
$\Psi_{\uparrow}\rightarrow\Psi_{\downarrow}$ symmetry (the two
boson fields are interchangeable), (2)
$\Psi\rightarrow\Psi^{\dagger}$ symmetry (particle and anti-particle
are on the same footing which requires  the effective theory to be relativistic), (3)
$\Psi$ particles do not experience a background magnetic field; i.e.
there is no linear term $h(\partial_x\alpha_y-\partial_y\alpha_x)$ in the
action above. We show in the following that the three requirements
are ensured by the symmetry operations $PH$, $\mathcal{P}$ and
$T\circ \mathcal{TR}$. (Eventually, in Sec.\ref{sec:conclusion}, we will show that even $PH$ is not required: as long as the density is fixed at half-filling, the low energy effective theory Eq.(\ref{eff}) is correct.)

Note that according to Eq.(\ref{transv}) both $PH$ and $\mathcal{P}$
transforms a vortex into an anti-vortex, while preserving spin while
$T\circ \mathcal{TR}$ transforms a vortex into an anti-vortex and
flips spin. Therefore,
\begin{align}
 \mbox{$PH$: }
&\Rightarrow& \Psi_{\uparrow}\rightarrow
\Psi_{\downarrow}^{\dagger},\,&\Psi_{\downarrow}\rightarrow
\Psi_{\uparrow}^{\dagger}\label{eq:sc-ph}\\
 \mbox{$\mathcal{P}$: }
&\Rightarrow& \Psi_{i \uparrow}\rightarrow
\Psi_{\mathcal{P}(i)\downarrow}^{\dagger},\,&\Psi_{\downarrow}\rightarrow
\Psi_{\mathcal{P}(i)\uparrow}^{\dagger}\label{eq:sc-p}\\
\mbox{$T\circ \mathcal{TR}$: }&\Rightarrow&
\Psi_{\uparrow}\rightarrow
\Psi_{\uparrow}^{\dagger},\,&\Psi_{\downarrow}\rightarrow
\Psi_{\downarrow}^{\dagger}\mbox{
(\textbf{anti-unitary})}\label{eq:sc-tr}
\end{align}

Although $\mathcal{P}$ and $PH$ perform similar
transformations on the vortex field, they act very differently on the background flux
$(\partial_x\alpha_y-\partial_y\alpha_x)$. $\mathcal{P}$ changes the sign of this term while $PH$ leaves it invariant.
Therefore the combination of $\mathcal{P}$ and $PH$ ensures
condition (3). Similarly, \Eq{eq:sc-ph} and \Eq{eq:sc-tr}
validate the conditions (1) and (2). Therefore in the presence of $PH$,
$\mathcal{P}$ and $T\circ \mathcal{TR}$ conditions (1),(2) and (3) are satisfied. This dictates the effective
field theory above. In later discussions we will derive the same theory
from a completely different approach Eq.(\ref{Leff}). This will clarify the role of particle hole symmetry,
which will not be necessary as a microscopic symmetry to achieve
this transition. Note, in the absence of the external fields $A,\,
A^S$ and in the presence of easy plane anisotropy ($u<0$), the above theory is
identical to the easy plane NCCP$_1$ of Ref.
\cite{PhysRevB.70.075104}.

\section{The AF order in the vortex condensed phase}\label{sec:psi-cond-af}
Let us now focus on \Eq{eff}; for $m^2$ large and positive, the
$\Psi_\s$ vortices are absent at long wavelength/time. In that limit
\Eq{eff} becomes a gaussian theory in $\alpha_\mu$. Integrating out
$\alpha_\mu$ generates a Higgs effective action for $A_\mu$, in other
words, the system is a superconductor. When $m^2$ changes sign the
spin-carrying vortices will condense. If $u>0$ the $\Psi_\ua$ and
$\Psi_\da$ vortices suppress each other, hence energetically it is
favorable for $\langle\Psi_\ua\rangle\ne 0, \langle\Psi_\da\rangle =
0$ or $\langle\Psi_\ua\rangle= 0, \langle\Psi_\da\rangle\ne 0$ in
the condensed phase. When $-4 v<u<0$ it is favorable for
(1) $\langle\Psi_\ua\rangle=\langle\Psi_\da\rangle \ne 0$ or (2) $\langle \Psi^\dagger_\ua\Psi_\da\rangle \ne 0$ while
 $\langle\Psi_\ua\rangle=\langle\Psi_\da\rangle = 0$.
In case (2) the order parameter has no vorticity (hence is an local operator) while flips the spin.
When it develops expectation value, superconductivity needs not be destroyed. Since $\langle \Psi^\dagger_\ua\Psi_\da\rangle\sim\langle S^+\rangle\ne 0$, the system exhibits SC and magnetic order simultaneously. While this scenario can certainly be realized in certain
parameter regime, in the rest of the paper we focus on the more interesting case where the system
goes from $\langle\Psi_\ua\rangle=\langle\Psi_\da\rangle = 0$ to $\langle\Psi_\ua\rangle=\langle\Psi_\da\rangle \ne 0$
in a single transition.

When single vortices condense, a SC turns into an insulator. When both $\langle\Psi_\ua\rangle$ and $\langle\Psi_\da\rangle$ are none zero, $\langle\Psi_\ua^{\dagger}\Psi_\da\rangle$ hence $\langle S^+\rangle$ are also non vanishing. As a result, the insulating phase has easy plane magnetic order. However, what is the order pattern? Is it AF or ferromagnetic?

\begin{figure}
 \includegraphics[width=0.3\textwidth]{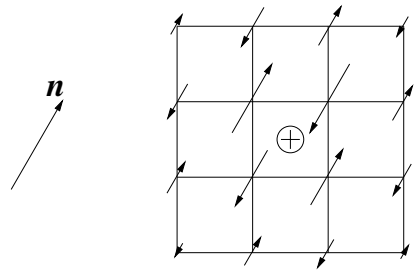}
\caption{For $\mbox{Re}(\chi)=1$, $t=0.8$, and $\mbox{Im}(\chi)=0.3$ (see Eq.(\ref{sc})) and a vortex located in the plaquette labeled by $\oplus$, we plot the magnetization $\langle S^{\hat n}_i\rangle$ on site $i$ in a small in-plane field $\vec B=0.1 \hat n$. The sign of the magnetization is labeled by the arrows, and the length of the arrows is proportional to $|\langle S_{\hat n}\rangle|$. The calculation is performed on a 1152-site system with one vortex and one anti-vortex separated by half system size.}
\label{fig:af-order-in-field}
\end{figure}

To minimize technical complexity, here we present a simple, but less rigorous, way to clarify the nature of the magnetic order. We put the more rigorous calculation, based on computing vortex tunneling amplitude, in Appendix \ref{sec:vortex-tunneling}.  To mimic the coherent tunneling ($\langle\Psi^\dagger_\ua\Psi_\da\rangle\ne 0$) between the spin up and spin down vortex states in the magnetic phase, we turn on a small in plane magnetic field (the field strength needs to be larger than the spacing between the core levels in the first panel of Fig.\ref{vc}) to mix the spin-up and spin-down vortex states in a single static vortex. In Fig.\ref{fig:af-order-in-field} we show the expectation value of the in-plane spin component in the vicinity of the vortex core when there is an in plane magnetic field in $\hat{n}$ direction. The AF order is clearly seen.

\section{Transition from the AF side}\label{sec:transition-from-af}

In the previous section we  have analyzed the SC $\leftrightarrow$
AF transition from SC side using the effective action given in
Eq.(\ref{eff}). In this section we approach the transition from the
AF side, where the AF order is destroyed by the condensation of AF
vortices. Since superconductivity emerges after this condensation,
it suggests that the AF vortices carry electric charge. In the
following we show that this is indeed the case.

Let us assume in vortex condensed phase 
the U(1) order parameter $\langle\Psi_\ua\rangle$ and $\langle\Psi_\da\rangle$ have equal magnitude. In the following
we derive an effective theory describing the vortices of these order parameter fields. The AF$\rightarrow$SC transition is then triggered by the condensation of these new vortices. 
The derivation is accomplished by the performing duality transformation\cite{PhysRevLett.47.1556,PhysRevB.39.2756} on Eq.(\ref{eff}) (for details see
appendix \ref{duality}). In fact, in the absence of the external gauge potentials $A,\,A^S$,
this has been carried out in Ref.\cite{PhysRevB.70.075104} where it
was discovered that the theory at criticality is self-dual. In appendix \ref{duality} we extend it to the case with these potentials present. The end result is the effective theory given in \Eq{eft}. Essentially in the AF phase described by $\langle\Psi_\ua\rangle=\sqrt{\rho}e^{i\phi_{\ua}},\,\Psi_\da=\sqrt{\rho}e^{i\phi_{\da}}$ , there are two types of vortices. One causes the winding of $\phi_ua$ by $2\pi$ and the other the winding of
$\phi_\da$ by $-2\pi$ (here is the minus sign is chosen for convenience).
In the presence of such vortices, $\alpha$ nucleates $\pm\pi$ fluxes to screen the vorticity (so that the kinetic energy is minimized). From the term $i\frac{e}{\pi}A\cdot\nabla\times \alpha$ in \Eq{eff} we find that these two types of screened vortices carry charge $\pm e$. We label these vortices by $z_1$ and $z_2$ respectively. This explains the opposite sign in the minimal coupling to $\v A$ in the first term of \Eq{eft}. The curl of the $a_\mu$, namely $\nabla\times\v a$, describes the spin 3-current caused by the gaussian fluctuation in the phase of $\langle S^-\rangle\sim e^{i(\phi_\ua-\phi_\da)}$. This is reflected in the last term of \Eq{eft}. Since both $z_1$ and $z_2$ creates $2\pi$ winding in  $\phi_\ua-\phi_\da$ it couples to $a_\mu$ minimally as shown by the first term of \Eq{eft}.
Finally $z_1^{\dagger}z_2$ inserts charge $2e$ while has no spin vorticity (hence is an local operator) is the Cooper pair operator.
\begin{widetext}
\be
 {\cal L}_{\rm dual}&&=\sum_{\alpha=1}^2\left[\sum_\mu|(\partial_{\mu}+ia_{\mu}-i e
A_{\mu}\s^3)z_{\alpha}|^2+m^2|z_{\alpha}|^2\right]
+u|z_{1}|^2|z_{2}|^2
+v\left(\sum_\alpha|z_{\alpha}|^2\right)^2+\frac{\kappa}{2}(\nabla
\times a)^2+c_1(\gr\times A)^2\nn&&-\frac{i\hbar}{2\pi}(\gr\times
a)\cdot A^S\label{eft} \ee
\end{widetext}
Note, the form of \Eq{eft} is almost identical to \Eq{eff} except for the exchange $\Psi_\s\leftrightarrow z_\alpha$ and $2eA \leftrightarrow \hbar
A^S$. Since Eq.(\ref{eft}) is the dual theory of Eq.(\ref{eff}), it must be consistent with microscopic symmetries of  Eq.(\ref{eff}). In Appendix \ref{subsec:sym-eft} we further directly checked the consistency between microscopic symmetries and Eq.(\ref{eft}).

Next let us how to describe the magnetically ordered phase using \Eq{eft}. % we want to recover the fact that this phase is XY ordered.
In the $\langle\psi_\s\rangle\ne 0$ phase, $m^2>0$ and the
$z_\alpha$ bosons are absent at long wavelength/time. In that limit
we again have a gaussian theory \be {\cal L}_{\rm gauge}\ra
\frac{\kappa}{2}(\nabla \times
a)^2-\frac{i\hbar}{2\pi}A^S\cdot\nabla\times a.\label{eft2} \ee
describing a ``spin superfluid''. The above action is dual to the
following Goldstone action \be {\cal L}_{\rm GS}=
\frac{1}{2\kappa}\sum_\mu(\p_\mu\chi-\hbar A^S_\mu)^2.\label{eft3}
\ee Here the phase angle $\chi$ transforms under a angle
$\beta$-$S_z$ rotation as $\chi\ra\chi+\beta$. It can also be shown
that computing the spin-spin correlation function $\langle
e^{i\chi(\v x_1,t)} e^{-i\chi(\v x_2,t)}\rangle$ using action
\Eq{eft3} is equivalent to computing the monopole-anti monopole
correlation function using \Eq{eft2}, i.e., \be \langle e^{i\chi(\v
x_1,t)} e^{-i\chi(\v x_2,t)}\rangle_{{\cal L}_{\rm GS}}=\langle
V^+(\v x_1,t)V^-(\v x_2,t)\rangle_{{\cal L}_{\rm
gauge}}.\label{mono}\ee Here $V^+$ and $V^-$ are the creation
operators of the monopole and antimonopole in $a$-gauge field.
Since a monopole (anti-monopole) is the event at which ${1\over
2\pi}(\gr\times a)_0=S_z/\hbar$ changes by $1$ ($-1$), we conclude
that \be\langle V^+(\v x_1,t) V^-(\v x_2,t)\rangle\sim \langle
S^+(\v x_1,t) S^-(\v x_2,t)\rangle.\ee Thus the $S_z$ spin
superfluid described by \Eq{eft2} and \Eq{eft3} is an $XY$ ordered
magnet (from the previous discussion we know this order is AF) and
$z^\dagger_1$,$z^\dagger_2$ create charged XY vortices
(anti-vortices). Finally $z_1^{\dagger}z_2$ inserts charge $2e$
hence creates a monopole in $\alpha$-gauge field, and
$\Psi_{\uparrow}^{\dagger}\Psi_{\downarrow}$ raise $S_z$ by $1$
hence is a monopole operator $V^{\dagger}$ in $a$-gauge field. (We
will come back to this point later in the slave-rotor description
Sec.\ref{slave rotor} and Sec.\ref{sec:monopole}, where we
explicitly compute the monopole quantum number for $V^\dagger$ and
find it is AF.)

Since $z_1$ and $z_2$ carry the vorticity of magnetic order parameter their
condensation will destroy the AF order. However since $z_1$ and $z_2$ carry
charge, the condensate is a superconductor. Now we have gone
through a full circle. We start from the d-wave superconductor, the
spin-carrying vortex condensation brings the system to a XY ordered
antiferromagnet. In the reverse direction, start from the
antiferromagnet, the condensation of the charge-carrying vortices
brings the system back to the superconductor.

In Lapp.\ref{sec:XY-MF-BdG},\ref{sec:vortex-tunneling} we perform the
parallel study of the AF vortex core states as what we did for SC
vortex in Sec.\ref{sec:SC},\ref{sec:sc-eff},\ref{sec:psi-cond-af}.
We indeed find that there are two mid-gap levels in the AF vortex
core. The filling/unfilling of these levels lead to charge-$\pm1$ AF vortices, whose
condensation triggers the SC order. Analogous to the discussion in section III we can determine the SC pairing symmetry by studying the tunneling property of the magnetic vortices.

\subsection{Spin-Charge Duality}
\label{spincharge}  As mentioned previously, the self duality of the
easy plane NCCP$_1$ model at criticality, leads to spin-charge
duality here. The effective theories near the transition point,
\Eq{eff} and \Eq{eft}, map onto one another with the substitution
$\Psi_\s\leftrightarrow z_\alpha, 2eA \leftrightarrow \hbar A^S$
which implies that the roles of spin and charge are identical at
criticality. We point out some consequences of this duality in the
following.

At criticality we expect both the electric and spin conductivity to
be finite and nonzero. The self duality ensures the electric
conductivity ($\s_c$) and spin conductivity ($\s_s$) to be related
by ${\s_c\over e^2}={\s_s\over (\hbar/2)^2}.$ (In general, self
duality allows another term in the response action :
$\frac{i\theta}{\pi}A\cdot\nabla\times A^S$, i.e., a spin Hall term.
However this is forbidden by particle hole symmetry $PH$:
$A\rightarrow -A$.) Therefore we expect the critical gauge action to
have the following form:
\begin{widetext}
\be
 {\cal L}_{eff}=\frac{\sigma_0}{2}
\left[(\nabla\times eA)_\mu {1\over\sqrt{\nabla^2}}(\nabla\times
eA)_\mu+(\nabla\times {\hbar\over 2}A^S)_\mu
{1\over\sqrt{\nabla^2}}(\nabla\times {\hbar\over 2}A^S)_\mu\right]
.\ee
\end{widetext}
Note, in a regular Fermi liquid, a Wiedemann-Franz like relation
connects the metallic spin and charge conductivity in precisely the
same way, since at low temperatures both quantities are transported
by electronic quasiparticles with a fixed charge to spin ratio. At
the critical point described above, although we have spin-charge
separated excitations, the self duality of the theory restores this
Wiedemann-Franz like relation between the metallic spin and charge
conductivities.

At criticality the correlations of the magnetic order parameter
$<S^+(r,t)S^-(0,0)> \sim \frac{(-1)^r}{r^2-c^2t^2}^{1/2+\eta/2}$
and the Cooper pair order parameter $<\Delta^* (r,t)\Delta(0,0)>
\sim \frac{1}{r^2-c^2t^2}^{1/2+\eta/2}$ are related by the self
duality, and constrained to fall off with the same power law. Since
these are both expressed as bilinears of vortex operators, the
$\eta$ exponent is expected to be large. Thus, the critical point
unites the antiferromagnetism and superconductivity in a remarkable
symmetric fashion.

\section{Slave Boson Formulation}\label{slave rotor}

In Sec.\ref{sec:SC} we started from a lattice Hamiltonian for a
superconductor and showed that the vortices carry spin. We then
condensed these vortices and showed that, as  a result, the system
becomes an AF ordered Mott insulator. In this section we begin with
a lattice theory for this Mott insulator, and show that we can
arrive at the same conclusions from this different angle. The slave
boson gauge theory approach used below, captures both the
superconducting and antiferromagnetic insulator phases within a
single formulation. While the former is obtained by the Higgs
mechanism, the latter, surprisingly, is the Coulomb phase of the
gauge theory. This is an interesting counterexample to the standard
folklore that slave boson theory can readily describe
superconducting states but not magnetically ordered phases. Here we
establish that magnetic order of the easy plane variety can indeed
be captured.
\\

\subsection{Insulating State}
Consider a half filled strongly correlated electronic model on
square lattice.
\begin{equation} H_{\rm Hubbard} = H_{\rm hop} +U\sum_i (n_i-1)^2
\label{Hubbard}
\end{equation}
where the first term is a hopping Hamiltonian, while the interaction
term tends to prefer single occupancy per site ($n_i$ is the
electron number on site $i$). Unlike most other such models which
exhibit spin SU(2) symmetry we assume that there is only $S_z$ spin
rotation symmetry;this is incorporated in $H_{\rm hop}$ via e.g.
spin dependent hopping terms.

Deep in the Mott insulating state, when we turn up the value of $U$,
the physics is described by a spin 1/2 Hamiltonian. For concreteness
we consider obtaining the following simple Hamiltonian:
\begin{eqnarray}
\label{HSpin}
H &=& J\sum_{\langle rr'\rangle}\vec{S}_r\cdot \vec{S}_{r'} + H_{\rm anis}\\
H_{\rm anis}&=& -\frac{J_2}{2}\sum_{\langle\langle
rr''\rangle\rangle} (S_r^{+}S_{r''}^{-} +S_{r}^-S_{r''}^+),
\label{Hanis}
\end{eqnarray}
where the SU(2) breaking term comes in as a ferromagnetic second
neighbor interaction between planar components of the spin. The
ground state of this Hamiltonian is expected to be an easy-plane
Neel ordered state with spins in the XY plane.

Below we will show that the same conclusion can be reached via the
slave boson formulation in a controllable fashion. This new
mechanism for obtaining magnetically ordered states in the
slave boson approach is not just of academic interest. It allows us to
simultaneously capture the magnetically ordered states and
superconductivity (the latter is readily realized using slave
bosons as the Higgs particle) within the same formulation. This is
achieved without having to resort to 'confinement' which in the
slave boson context does not allow for a unique identification of
the ordered state. In contrast, the approach presented here will
allow us to identify both the XY antiferromagnet and the
superconducting phases uniquely. Moreover unifying these two phases within a single
formulation leads to the possibility of a direct
continuous transition between these two phases. The slave boson
approach also allow us to identify the nature of the superconducting
state and the symmetries required to accomplish such a transition.

Consider treating the Hamiltonian (\ref{HSpin}) using a 'Schwinger
fermion' representation of spins, i.e. introduce two component
fermions $(f_{r\uparrow},\,f_{r\downarrow})$ at each site with the
constraint:
\begin{equation}
\label{constraint}
\sum_\sigma f^\dagger_{r\sigma}f_{r\sigma} = 1
\end{equation}
imposed on every site. Then, the spin 1/2 operators are $\vec{S_r} =
\frac12
f^\dagger_{r\sigma}\vec{\sigma}_{\sigma,\,\sigma'}f_{r\sigma'}$.
Substituting this in (\ref{HSpin}) leads to a Hamiltonian quartic in
the fermionic operators. To make progress, consider the mean field
approximation \cite{2986272,PhysRevB.39.8988,PhysRevB.37.3774},obtained by
replacing $\langle f^\dagger_{r\sigma} f_{r'\sigma'}\rangle =
\chi_{rr',\,\sigma,\sigma'}$ and treating the constraint on
average. It is well known that in the absence of the spin anisotropy
term, the optimal mean field ansatz with uniform amplitude is the
staggered flux state
\begin{equation}
H^{SF}_{MF} =-\sum_{\langle
rr'\rangle}\chi_{rr'}f^\dagger_{r\sigma}f_{r'\sigma}
\end{equation}
with the nearest neighbor hopping amplitudes
$\chi_{rr'}=|\chi|e^{\pm i\Phi/4}$ along (against) the directions
shown in the Figure \ref{hop}(a). The optimal
variational state within this set is reached for $\Phi \approx 0.34 \pi$.\cite{ivanov:132501}

Now let us try to minimize the mean-field energy due to the spin anisotropy term
\begin{equation}
\langle H_{anis}\rangle =\frac{J_2}{2}\sum_{\langle\langle
rr''\rangle\rangle}\langle
f^\dagger_{r\uparrow}f_{r''\uparrow}\rangle \langle
f^\dagger_{r''\downarrow}f_{r\downarrow}\rangle. 
\end{equation}
One way in which
this can be achieved is if the two expectation values are real but
with opposite signs, i.e.
\begin{equation}
H_{MF} = -\sum_{\langle
rr'\rangle}\chi_{rr'}f^\dagger_{r\sigma}f_{r'\sigma}
-\sum_{\langle\langle
 rr''\rangle\rangle} \s t^s_{rr''}f^\dagger_{r\sigma}f_{r''\sigma}
\label{MF}
\end{equation}
the spatial structure of the spin dependent next-neighbor hopping
$t^s_{rr''}$ is shown in the figure \ref{hop}(a). This choice of $t^s_{rr''}$
 opens a gap at the Dirac points and hence leads to a
better mean field energy.

The mean field Hamiltonian above has a familiar form. In the absence
of spin dependence of the next neighbor hopping, this is the
``chiral spin liquid'' Hamiltonian\cite{PhysRevB.39.11413}. The spin dependence
leads to opposite Chern numbers \cite{PhysRevLett.49.405} ($\pm 1$) for the spin
up and spin down fermions, and hence describes a spin Hall insulator. However, unlike the honeycomb lattice spin
hall insulator introduced in Ref.\cite{kane:226801},  this band structure
breaks time reversal symmetry, although time reversal invariance
combined with a translation is still a symmetry. Note also, that the
unit cell is doubled once these spin dependent hoppings are
introduced (the staggered flux mean field theory in contrast, is
effectively a translationally symmetric ansatz).\\

{\bf Gauge fluctuations and Neel order}
\\

The U(1) symmetry, corresponding to the conservation of the $f$ fermion number, of \Eq{MF} is not  a physical symmetry, but rather
corresponds to a U(1) gauge redundancy. Including the corresponding
gauge field $a_{rr'}$ on bonds of the lattice, requires
$\chi_{rr'}\rightarrow \chi_{rr'}e^{ia_{rr'}}$ and
$t^s_{rr''}\rightarrow t^s_{rr''}e^{ia_{rr''}}$. The temporal
component of the gauge field $a_0$ arises as the Lagrange multiplier
imposing the constraint (\ref{constraint}).

Since the fermions are gapped, they can be safely integrated out to
obtain an effective action for the gauge degrees of freedom. In
order to interpret this action it is useful to introduce an external
'spin' gauge field $A^S_{rr'}$. This is possible since $S_z$ is a
conserved spin density. Replacing $a_{rr'}\rightarrow
a_{rr'}+\frac{\hbar}2A^S_{rr'}{\bf \sigma^z}$, and integrating out
the fermions, one obtains the Euclidean Lagrangian:
\begin{equation}
{\mathcal L}_E[a,A^S] = i\frac{\hbar}{2\pi} A^S\cdot \nabla \times a
+ \frac K2[ (\nabla \times a)^2+ (\nabla \times \frac{\hbar}{2}A^S)^2]
\end{equation}
where we have used a three vector notation
$\nabla=(\partial_0,\Delta_x,\Delta_y)$ etc., with lattice
derivatives in the space direction. %{\bf Check}.
Note, there is no charge Chern Simons term since the net
Chern number vanishes. The first term
arises because of the opposite Chern numbers for the two spin bands.
It associates a spin of $\hbar/2$ with the $2\pi$ flux of $a$.
%However, there is no Chern Simons term, since the net
%Chern number vanishes.
Finally, the dynamical gauge field ($a$) is
non-compact, i.e. there are no monopole configurations created by
the dynamics, since a monopole event (which inserts $2\pi$ flux) is
associated with a spin flip, which is forbidden by $S_z$
conservation. We now integrate out the dynamical gauge field $a$ to
obtain:
\begin{equation}
{\mathcal L}_E[A^S_\perp] = \frac{\hbar^2}{32\pi^2K} [A^S_\perp]^2
+\frac K2[(\nabla \times \frac{\hbar}{2}A^S)^2]
\end{equation}
the first term is a London term that indicates the phase is a 'spin
superfluid'. This implies XY magnetic order, although it does not
specify the precise pattern of ordering (i.e. ferromagnetic, vs
various Neel ordered states). However, given that the mean field
theory is favored by the microscopic Hamiltonian \Eq{HSpin}, it is
natural to expect this to be the usual Neel ordered state with
opposite moments on the two sublattices. Note, since we have already
doubled the unit cell, the Neel order is at $\v q=(0,0)$, but
transforms nontrivially under the point group.

The nature of the order may be established more rigorously as
follows. Note, due to the absence of monopole events in the
dynamics, one is left with a Maxwell action for the dynamical gauge
field $a$. The gapless photon excitations that this implies,
corresponds to oscillations of the dynamical magnetic flux. However,
due to the binding of flux to spin density, this implies a
fluctuating $S_z$ density. In fact, this gapless photon is simply
the Goldstone mode associated with XY spin symmetry breaking (see the discussions in section IV). In
order to establish the spin order one needs to evaluate $\langle
S^+_rS^-_0\rangle$. This corresponds to evaluating the correlators
of the monopole insertion operator $\langle V_r^\dagger V_0\rangle$.
It is well known\cite{Polyakov}, that in magnetically Coulomb phase
in D=2, the monopole operators have long range order. The precise
ordering pattern corresponds to the transformation properties of the
monopole insertion operator under the lattice symmetries. This is
explicitly evaluated in Appendix \ref{sec:monopole} where it is
confirmed that we indeed obtain the Neel state. Note, this phase is
entirely identical to the regular XY Neel ordered state, although it
is obtained from such an 'exotic' starting point. The magnetic order
induced by gauge fluctuations ensures, for example, that the
counter-propagating edge states implied by the mean field theory are
gapped.

\subsection{Charge Fluctuations}\label{subsec:charge-fluctuation}
Consider lowering the charge gap to approach the Mott transition so
that the electron occupation at each site can now fluctuate into the
$n=0$ and $n=2$ state as well. This is achieved by lowering the
value of $U$ in equation \ref{Hubbard}. These fluctuations are
typically incorporated via the slave rotor
formulation\cite{lee:036403,PhysRevB.70.035114}, by introducing a
rotor variable $z_r=e^{i\phi_r}$ and its conjugate number variable
$N_r$ at each site. The electron operator is then written as
$c^\dagger_{r\sigma} = z_r f^\dagger_{r\sigma}$, and the constraint
now reads $n^f_r+N_r=1$. This however does not capture all Mott
transitions, for example it was pointed out that Hubbard models with
psuedospin symmetry requires introducing a SU(2) slave rotor
\cite{Hermele} which is  a simple extension of the SU(2) slave boson
theory\cite{PhysRevLett.76.503} to include both electron and hole
fluctuations. A pair of complex fields $z_{1r},\,z_{2r}$ are
introduced, which are SU(2) 'rotor' variables i.e.
$|z_{1r}|^2+|z_{2r}|^2=1$. The electron operator is written as:
\begin{align}
 c_{r\uparrow}&=f_{r\uparrow}z_{r1}-\eta_{r}f_{r\downarrow}^{\dagger}z_{r2}^{\dagger}\nonumber\\
c_{r\downarrow}&=f_{r\downarrow}z_{r1}+\eta_rf_{r\uparrow}^{\dagger}z_{r2}^{\dagger}.
\label{eq:rotor-U(1)}
\end{align}
if we define:
\begin{eqnarray}
\Psi_r &=& \left [ \begin{array}{c}
           c_{r\uparrow} \\
           \eta_r c^\dagger_{r_\downarrow}
         \end{array} \right ]
         \,\, F_r = \left [ \begin{array}{c}
           f_{r\uparrow} \\
           \eta_r f^\dagger_{r_\downarrow}
         \end{array} \right ]\\
         Z_r &=& \begin{pmatrix}z_{i1}&
         -z^\dagger_{i2}\\z_{i2}&z_{i1}^{\dagger}\end{pmatrix}\\
\Psi_r &=& Z_r F_r \label{eqn:slaverotor}
\end{eqnarray}
the electron operator may be written compactly as in the last line;
where $Z_r$ is an SU(2) matrix. The mean field Hamiltonian
(\ref{MF}) takes on a particularly simple form in this notation:
\begin{equation}
H_{MF} =\sum_{rr'}
F^\dagger_r(\chi^R_{rr'}+t^s_{rr'}+i\chi^I_{rr'}\mu_z)F_{r'}
\label{HMFcompact}
\end{equation}
with $\chi_{rr'}=\chi^R_{rr'}+i\chi^I_{rr'}$ and with diagonal spin
dependent hoppings $t^s_{rr'}$ as shown.

This definition has an SU(2) redundancy, clearly any set of SU(2)
matrices $U_r$ can generate the transformation $Z_r\rightarrow Z_r
U_r^\dagger$, $F_r\rightarrow U_r F_r$ leaving the form above
invariant. Naturally, there is a close connection between this
redundancy and the constraint that needs to be implemented to obtain
the physical Hilbert space. If the operators ${\bf T}_r$ generates
the above transformations, then the constraint on physical states is
that ${\bf T}_r=0$.

Luckily for our purposes we will not need the full SU(2)
formulation. Instead, we only retain the fact that there are a
doublet of charged bosons, but assume that they are rotor variables
with equal amplitude $z_{1r}=\frac1{\sqrt2} e^{i\phi_{1r}},\,
z_{2r}=\frac1{\sqrt2} e^{i\phi_{2r}}$. This form is necessary, and
sufficient, to incorporate the gauge fluctuations and symmetries of
the mean field theory \Eq{MF}. The SU(2) matrices $U$ that preserve
this structure are generated by $U_\epsilon=e^{i\epsilon
\hat{\mu}_z}$, and $U' = i\mu_x$ where $\bf \mu$ are 2x2 Pauli
matrices in the usual representation. While the former is simply the
$U(1)$ gauge transformation that leaves the mean field Hamiltonian
invariant, the latter appears as the gauge transformation required
to implement reflection symmetry (Appendix
\ref{subsec:slave-rotor-sym}). Hence, in the absence of a two
component slave rotor theory, we would be unable to implement this
physical symmetry starting with the mean field theory \Eq{MF}. If we
now define $N_{1r}$ and $N_{2r}$ as the operators conjugate to the
two phase variables, the physical states in the basis $|N_1, N_2;
n_f^{\uparrow} , n_f^{\downarrow}> $ satisfy the constraints
$(N_1+N_2)+(n_f^\uparrow+n_f^\downarrow) = 1$ and invariance under
the combination of $|N_1,N_2>\rightarrow (-1)^{N_1}|-N_2,-N_1>$ and
$|n_f> \rightarrow
\eta_r(f_{r\uparrow}f_{r\downarrow}+f^\dagger_{r\downarrow}f^\dagger_{r\uparrow})|n_f>$.
This constrains the physical states on a site to be:
\begin{eqnarray}
|\uparrow>&=&|0,0;\uparrow>\\ |\downarrow>&=&|0,0;\downarrow>\\
|0> &=& (|0,1; \,\,>+\eta_r |-1, 0;\uparrow\downarrow>) \\
|\uparrow\downarrow>&=& (-\eta_r|1,0;\,\,>+ |0,
-1;\uparrow\downarrow> )
\end{eqnarray}
A physical interpretation of the charge boson doublet is provided at
the end of appendix \ref{subsec:doublet-chargeon}.\\

{\bf The superconducting phase}\\

 An advantage of the two component slave
rotor formulation above is that the superconducting state is very
naturally described. As the Mott transition is approached, the gap
to the charge excitations closes, at which point $z_1,z_2$ condense. If both bosons condense individually,
$<z_1>=\Phi_1,\, <z_2>=\Phi_2$ a superconductor is obtained. This
can be verified by noting that the gauge invariant quantity
$<z^\dagger_1z_2>=\Phi_1^*\Phi_2$ which is the superconducting order
parameter, has long range order.

Once the charged bosons are condensed, the electron $\Psi_r$ has a finite
overlap with the `spinon' $F_{r\s}$, and inherits its dispersion.
Since $\Psi_r\sim \langle Z \rangle F_r$, and $\langle Z \rangle
=\sqrt{|\Phi_1|^2+|\Phi_2|^2}U_\Phi$, where $U_\Phi$ is a unitary
matrix, an effective Hamiltonian that describes the electron
excitation can be readily obtained from the mean field Hamiltonian
(\ref{HMFcompact}):
\begin{equation}
H_{e}=\sum_{rr'}\Psi^\dagger_r
U_\Phi(\chi^R_{rr'}+t^s_{rr'}+i\chi^I_{rr'}\mu_z)
U^\dagger_\Phi\Psi_{r'}
\end{equation}
when both bosons are condensed with equal magnitude
$\Phi_1=e^{i\theta}\Phi_2$, the electronic Hamiltonian is just:
\begin{equation}
H_{dSc} =
\sum_{rr'}c^\dagger_{r\sigma}(\chi^R_{rr'}+t^s_{rr'})c_{r'\sigma'} +
\Delta_{rr'}\epsilon_{\sigma\sigma'}c^\dagger_{r\sigma}c^\dagger_{r'\sigma'}
+ {\rm h.c.}
\end{equation}
where the pairing function
$\Delta_{rr'}=ie^{i\theta}\eta_{r'}\chi^I_{rr'}$ is symmetric upon
$r\leftrightarrow r'$ and changes sign under $\pi/2$ rotation. Hence
it is a spin singlet d-wave pairing function. If $t^s=0$ this is the
regular $d_{x^2-y^2}$ superconductor with translation symmetry and
nodes at $(\pm \pi/2,\pm \pi/2)$. The addition of the spin dependent
hopping opens a gap at these nodes. It doubles the unit cell and
breaks time reversal symmetry (although the combination of a unit
translation and time reversal remains a symmetry). Note however,
since the above SC Hamiltonian  is a unitary transformation of the
hopping Hamiltonian $H_{MF}$, it shares the same spectrum. In
particular, since the hopping Hamiltonian had a nontrivial band
structure and was argued to have counter-propagating spin filtered
edge modes. The same is true of the superconducting Hamiltonian
above. However, once the charge bosons condense, additional terms
are allowed in the fermion Hamiltonian, which are forbidden in the
insulating state. For example, one can add terms corresponding to
`spinon pairing', which are forbidden by the gauge U(1) symmetry on
the insulating state. Such terms can arise simply from the
electron's next neighbor hopping, which in the two component
notation corresponds to $\psi^\dagger_r \mu^z\psi_{r'} =
F^\dagger_{r'}Z^\dagger \mu^zZ F_r=|\Phi|^2F^\dagger_{r'} \mu^x F_r
$.   Note however, the strength of such a term is proportional to
the condensate density, and becomes very small near the transition.
Including these terms removes the nontrivial topological properties
(such as gapless edge states) of the superconductor. Therefore, the
superconducting state obtained on boson condensation is a generic
d-wave superconductor with doubled unit cell. If, however, we demand
particle hole symmetry, then terms like the second neighbor hopping
term above are forbidden, and the superconductor indeed has
nontrivial band topology (see appendix \ref{subsec:pseudospin}). In
our previous analysis it was convenient to consider a particle hole
symmetric superconductor, although it is difficult to demand that as
the microscopic symmetry of a physical system. Luckily, we show in
Sec.\ref{sec:conclusion} that demanding complete particle hole
symmetry is unnecessary - instead we need only tune one chemical
potential type term to zero, to obtain a direct relativistic
transition.

\begin{figure}
 \includegraphics[width=0.3\textwidth]{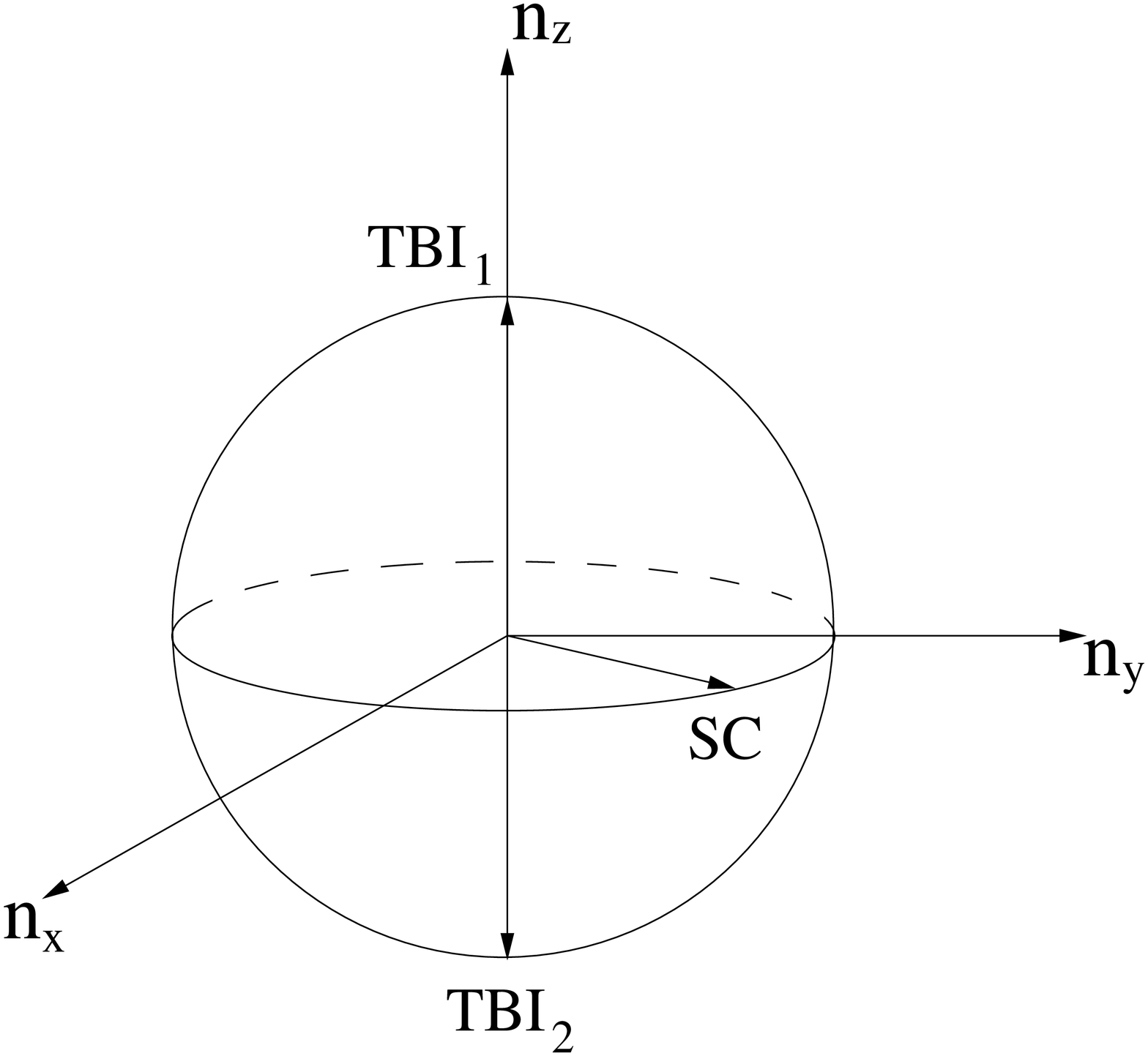}
\caption{The $\vec{n}$ sphere and the corresponding electronic
phases.} \label{fg:pseudospin-rotation}
\end{figure}

We note in passing that condensing just one of the two $(z_1,\,z_2)$
fields leads to a topological band insulator (TBI).  (At the end of
Sec.\ref{sec:qft-transition} we show that these states are realized
in the SC vortex core.) In choosing one of these fields to condense,
one spontaneously breaks the reflection symmetry. If $z_1$
condenses,  the appropriate mean-field electronic Hamiltonian can
 be obtained from \Eq{MF}
by replacing $f_\sigma \rightarrow c_\sigma$. Similarly if $z_2$
condenses we simply make a different replacement: $f_\sigma
\rightarrow \eta_r\epsilon_{\sigma\sigma'}c^\dagger_{\sigma'}$. The
above different condensation scenario of $z_1$ and $z_2$ can be
summarized by the vector $\vec n=z^{\dagger}\vec\sigma z$ where
$z=(\langle` z_1\rangle,\langle z_2\rangle)$. Using $\vec{n}$ we
represent the topological insulator (TBI) associated with $\langle
z_1\rangle\ne 0$ and $\langle z_2\rangle=0$ (TBI$_1$) by $\vec{n}$
pointing to the north pole. Similarly the TBI associated with
$\langle z_1\rangle=0$ and $\langle z_2\rangle\ne 0$ (TBI$_2$) by
$\vec{n}$ pointing to the south pole.  The particle hole
transformation transform TBI$_1$ into TBI$_2$ and vice versa. The
d-wave SC discussed throughout the early part of the paper is
represented by $\vec{n}$ lying in the xy plane. This is illustrated
in Fig.(\ref{fg:pseudospin-rotation}).

 \subsection{The field theory of transition}\label{sec:qft-transition}
Consider approaching the transition from the insulating state. The
charged bosons are gapped but acquire a dispersion from the electron
hopping Hamiltonian \ref{Hubbard}. This includes the regular nearest
neighbor hopping $H^{\rm NN}_{\rm hop} = -t \sum_{\langle
rr'\rangle} c^\dagger_{r\sigma}c_{r'\sigma'}$, which we treat in
detail below (other hopping terms lead to similar results).
Expanding the electron operator as in \Eq{eq:rotor-U(1)}, and
assigning the mean field expectation values to the spinon bilinears
$J\langle \sum_\sigma f^\dagger_{r\sigma}f_{r'\sigma}\rangle =
\chi_{rr'}$, one also obtains a staggered flux dispersion for the
$z$-bosons: $H_{\rm hop} = -\frac{t}{J}
\sum\chi_{r'r}(z^\dagger_{1r}z_{1r'}+z^{\dagger}_{2r}z_{2r'})$. This
dispersion has a unique minimum (for flux $\Phi \neq \pi$) which is
relevant to the low energy theory.

The low energy field theory of the transition is argued below to
take the following form:
\begin{eqnarray}
\label{Leff}
{\mathcal L}_E &=& {\mathcal L}^z + {\mathcal L}^f \\
{\mathcal L}_f &=& i\frac{\hbar}{2\pi} A^S\cdot \nabla \times a
\nonumber\\ \nonumber {\mathcal L}^z &=&
\sum_{\mu,\,\alpha}|(\partial_\mu -ia_\mu)z_\alpha|^2 +\frac K2
(\nabla \times a)^2 + V
+ \Delta {\mathcal L} \\
V&=& r(|z_1|^2+|z_2|^2) + U(|z_1|^2+|z_2|^2)^2 - \lambda |z_1|^2|z_2|^2\nonumber\\
\Delta {\mathcal L} &=& \mu[z_1^* (\partial_\tau -ia_0)z_1 -
z_2^*(\partial_\tau -ia_0)z_2]\nonumber
\end{eqnarray}
The Euclidean Lagrangian consists of two pieces, one arising from
integrating out the fermionic spinons ${\mathcal L}_f$ and one for
the $z$-bosons ${\mathcal L}^z$, coupled to an low-energy U(1)
fluctuating gauge field $a$. Note, the representation of the
electron operator in Eq. \ref{eqn:slaverotor} implies that the
$z$-bosons are also minimally coupled to the U(1) gauge field of the
spinons. The remaining terms appearing in ${\mathcal L}^z$ are the
ones allowed by the symmetries of the problem, which act on these
bosons in a nontrivial manner as described in Appendix
\ref{subsec:slave-rotor-sym}. Since the fermionic spinons are gapped throughout the
transition, they can be safely integrated out. This produces the
spin Hall effective action ${\mathcal L}_f$ since the mean-field
state of the spinon is a spin hall insulator. (Note that to the
spinons
 $a$ plays the role of the charge gauge field.) 
Symmetries prohibit other terms in this low energy
action. For example, (see appendix \ref{subsec:slave-rotor-sym} )
under reflections about the x or y axis passing through a plaquette center, the fields transform according to
$z_1 \rightarrow z^*_2, \, z_2 \rightarrow -z^*_1$, in addition to
the reflection of coordinates. This symmetry rules out terms like
$|z_1|^2-|z_2|^2$ and $z_1^*\partial_\tau z_1 + z_2^*\partial_\tau
z_2$. It does however allow the linear time derivative term $\Delta
{\mathcal L}$. While point group symmetries do not forbid  a
background flux term like $h(\partial_xa_y-\partial_ya_x)$, taking into account symmetry under
time reversal (followed by translation) rules out such a term. (Only
a staggered background flux is allowed, but this has already been
incorporated while deriving the $z$-boson dispersion.)

In the absence of $\Delta {\mathcal L}$, the effective action for the transition is in the easy-plane NCCP$_1$ universality class\cite{PhysRevB.70.075104}. As we discussed
at the beginning of the paper, the nature of direct transition (whether it is generically fluctuation driven first order) in the relativistic easy-plane NCCP$_1$ model
is controversial.  In contrast, in the presence of  $\Delta {\mathcal L}$  the non-relativistic NCCP$_1$ transition in $d+z=4$ has been studied using perturbative RG \cite{PhysRevLett.32.292}. The conclusion is that fluctuation generically drives the transition to first order.

In order to access the relativistic transition, one needs to tune
both $r$ and $\mu$ to zero in equation (\ref{Leff}). The latter
corresponds to a chemical potential since $z_1$ and $z_2$
carry opposite electrical charge. 
\emph{It is important that one only need to tune two parameters to reach the relativistic
critical point. Because it implies this critical point is realizable in a two dimensional phase diagram}.
Physically $\mu$ measures the deviation from
half-filling. In the AF insulator phase the charge density remains fixed for a range of $\mu$, due to the
charge incompressibility. In the SC phase the charge density does
change with chemical potential and one would need to fine tune $\mu$ to reach the relativistic NCCP$_1$ quantum critical point.

The above situation is analogous to the simpler case of accessing the relativistic XY
transition in the superfluid-Mott insulator transition of the Bose
Hubbard model \cite{SachdevBook}. There, the relativistic transition
occurs at the tip of the superfluid lobe, while moving away from
this point leads to a non-relativistic Bose Einstein condensation
transition with dynamical critical exponent $z=2$. In Fig.(\ref{fig:phase-diagram}) we
present an analogous (schematic) phase diagram in the plane spanned
by the chemical potential $\mu$ and interaction strength $U$.

As being discussed in Sec.V A, the electromagnetic field associated with the gauge field $a$
corresponds to the 3-current of $S_z$ arising from the gaussian fluctuation in the magnetic XY order.
Since $z_1$ and $z_2$ couple to $a$ as charges, it is natural to associated them with the magnetic vortices.
From the earlier discussions we see that they
carry electrical
charge. A related
observation is that in the superconducting state near the critical
point, the superconducting vortices carry spin. This can be seen as
follows. Consider an equal amplitude condensate $\langle z_1\rangle
=\sqrt{\rho}e^{i\phi_1},\, \langle z_2 \rangle =\sqrt{\rho}
e^{i\phi_2}$. The energy cost of a static vortex is then obtained
from the lagrangian above as $E = \rho (\nabla \phi_1 -a
-eA/\hbar)^2+\rho(\nabla \phi_2 -a +eA/\hbar)^2$, where $A$ is the
external vector potential that induces vorticity. A unit vortex with
$\oint A \cdot dl = h/2e$ can be created either with (1) $\oint
\nabla \phi_1 \cdot dl = 2\pi,\, \oint \nabla \phi_2 \cdot dl = 0 $
and $\oint a \cdot dl = \pi$ or with (2) $\oint \nabla \phi_1 \cdot
dl = 0,\, \oint \nabla \phi_2 \cdot dl = 2\pi $ and $\oint a \cdot
dl = -\pi$. Due to the presence of ${\mathcal L}_f$ which attaches
spin to emergent flux, these vortices carry spin $\pm \hbar/2$.
Thus, near the transition, the low energy vortices in the
antiferromagnet carry charge $\pm e$ while vortices in the
superconductor carry spin $\pm \hbar/2$. Thus, a direct transition
between these phases can be explained on the basis of the
condensation of these defects.

Finally before closing this section we note that the mechanism that
generates magnetic order studied here does not require ``confinement''
which is invoked in other slave boson theories. The advantage of our mechanism
is that it allows us to identify both the nature of the magnetic order, which has not been possible in
confinement type approach, and
the superconducting pairing symmetry unambiguously.

\section{Road map to the appendices}
Before concluding, we outline the content of a number of appendices.
They are presented either to elaborate points made in the text, or
because they are interesting digression which contribute to the
general understanding of subjects covered by this paper. Appendix
\ref{duality} contains the technical details of the duality
transformation. Such transformation allows one to go back and forth
between \Eq{eff} and \Eq{eft}. Appendix
\ref{sec:SC-BdG} contains the Bogoliubov deGennes treatment of the
core state of superconducting vortices. It compliments our numerical
results presented earlier.
Appendix \ref{sec:XY-MF-BdG} present a mean-field theory
for the XY ordered state and study the core states of the AF
vortices. Appendix \ref{sec:vortex-tunneling} provides details of the vortex tunneling amplitude calculation in both the SC and the AF phases, where we explicitly compute the XY order pattern (AF) from the SC vortex tunneling, and the SC pairing symmetry (d-wave) from the AF vortex tunneling. Appendix \ref{subsec:slave-rotor-sym} is
devoted to the symmetry analysis of the slave boson theory in the AF
phase and the transformation properties of the antiferromagnetic
vortices. Appendix \ref{sec:monopole} presents the quantum number of
the monopole operator of \Eq{mono} within the slave rotor theory. Appendix
\ref{sec:sc-more} gathers some miscellaneous  discussions mentioned in the main text. Finally Appendix \ref{sec:honeycomb} discusses the direct transition between an f-wave SC and AF magnetic phase on the honeycomb lattice, which is a direct application of the techniques that we learned in the current square lattice case.

\section{Conclusion}\label{sec:conclusion}

In this paper we present a theory exhibiting a d-wave superconductor
and an easy-plane antiferromagnetic insulator as its two phases.
Unlike ordinary d-wave superconductor, our superconductor has a full
gap due to nodes mixing caused by a translation symmetry breaking.
In addition, the topological defects, namely vortices, of each of
these phases carry the quantum number of the order parameter for the
other phase. We derived the effective field theory describing the
superconductor to AF insulator phase transition. It is the
easy-plane non-compact CP$_1$ model. The possible phase transition
scenarios of this model, include:  (1) a direct continuous SC-AF
transition, (2) a direct first order SC-AF phase transition, and (3)
two consecutive continuous transitions linking SC to a SC+AF
coexistence phase and finally to AF. In particular, if scenario (1)
is realized, it would constitutes an realization of Landau forbidden
transition. While the first two scenarios correspond to the
condensation of the fundamental vortices in theory, the third
requires the condensation of a composite of the fundamental
vortices.

\begin{figure}
 \includegraphics[width=0.32\textwidth]{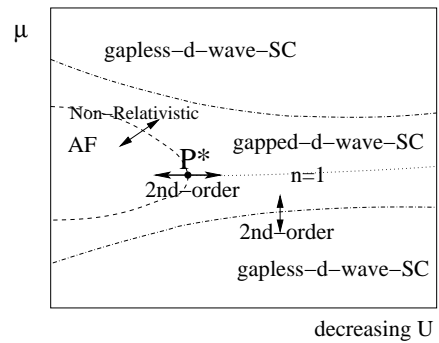}
\caption{The schematic global phase diagram. The horizontal axis is
the interaction strength, $U$, and the vertical axis is the chemical
potential $\mu$. The dashed lines mark the non-relativistic phase transition at mean-field level (which is likely to be first-order beyond mean-field theory).  This transition line terminates at the $PH$ symmetric
critical point $P^*$ discussed in the present work. Along the dotted
line the average charge density is $\langle n\rangle=1$, i.e.,
half-filling. The SC phase surround the the AF phase is a fully
gapped d-wave superconductor. Upon changing of chemical potential
the quasiparticle gap closes and the SC undergoes a second-order
Lifshitz transition into the nodal gapless d-wave SC at the
dash-dotted lines. The quasiparticle spectrum of the gapless d-wave
superconductor is identical to that seen in high $T_c$
superconductors.} \label{fig:phase-diagram}
\end{figure}

\begin{figure}
 \includegraphics[scale=0.6]{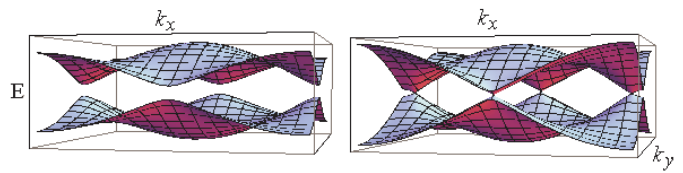}
\caption{In the presence of a sufficiently large chemical potential
$\mu$ the nodes of a conventional d-wave superconductor is restored.
The parameters used in this figure are $\phi=\pi/5, t=0.2$ and left
panel $\mu=0$, right panel $\mu= 0.5$. } \label{doped}
\end{figure}

Physically this phase transition requires tuning the interaction
strength (or the bandwidth) at fixed, half, filling factor. We
specified the symmetry requirements for its existence. When the
filling factor deviates from $1/2$ the transition generically
becomes first order. In Fig.\ref{fig:phase-diagram} we present a
schematic phase diagram in the interaction strength - chemical
potential plane. For a range of chemical potential where the
superconductor remains fully gapped with the minimum gap (left panel
of Fig.\ref{doped}) occurring at $(\pm \pi/2,\pm\pi/2)$ in the
square lattice Brillouin zone. For sufficiently large chemical
potential deviation (from that of half filling)  the superconducting
gap closes. When that happens the quasiparticle spectrum becomes
that of a conventional $d_{x^2-y^2}$ superconductor (right panel of
Fig.\ref{doped}), and the positions of the node change with chemical
potential.

In the above discussion we have assumed that upon doping the AF-SC
transition is direct. It is also possible that such transition
proceeds in an indirect manner. For example, imagine a situation the
doped charge in the AF insulator are accommodated as charged
vortices $z_1$ and $z_2$. To minimize the cost in kinetic energy, it
is favorable to keep the total vorticity zero, so that there are
equal number of vortex and antivortex. Thus, if the doping is
n-type, there will be an equal number of $z_1$ and $z^\dagger_2$
vortices (recall that $z_1$ and $z_2$ vortex carry opposite charge),
while for p-type doping there will be the same number of $z_2$ and
$z^\dagger_1$ vortices. Due to the logarithmic attraction, it is
energetically favorable for $z_1,z^\dagger_2$ or $z^\dagger_1,z_2$
to form charge $\pm 2e$ bound pairs. Because these pairs carry no
net vorticity, their condensation does not destroy the AF order.
Hence a SC and AF coexistence phase emerge. Such a transition is of
the usual non-relativistic XY universality. In the coexistence
phase, due to the non-zero $\langle z^\dagger_2 z_1\rangle$ or
$\langle z^\dagger_1 z_2\rangle$, the $z_1$ and $z_2$ AF vortices
mixes, and as the result only one AF vortex $\tilde z=\alpha
z_1+\beta z_2$ (where $\alpha,\beta$ are complex mixing coefficient)
is left as low energy excitations. When $\tilde z$ condenses the AF
order is finally destroyed and the system becomes a pure
superconductor. The coexistence $\rightarrow$ AF transition is also
of the usual XY universality class. 

It is also interesting to look at the indirect transition from the
superconducting side. As discussed at the beginning, such transition
is triggered by the condensation of the composite vortex
$\Psi^\dagger_\ua\Psi_\da$. Such composite vortex carries spin
quantum number $1$ but no net vorticity (of the superconducting
order parameter). As the SC to SC+AF coexistence transition is
approached, this excitation will become the lowest-energy magnetic
excitation. Like the roton in a superfluid, such vortex-antivortex
pair will carry a finite momentum. Such momentum will shift the
momentum of the minimum energy $\Psi^\dagger_\ua\Psi_\da$ away from
$(\pi,\pi)$, similar to the incommensurate magnetic excitation of
the high $T_c$ compounds.

We note that the similar ideas that the defects in the AF phase can carry charge and the defects in the SC phase can carry spin, are introduced in the phase string theory\cite{PhysRevLett.80.5401,kou:235102,weng-2007-21} in the context of High-Tc superconductor. Although there are some significant differences between the current work and the phase string theory: for example, we require the doubling of unit cell, easy-plane spin anisotropy and full energy gap on both AF and SC phases, it is still possible that there are underlying relations which are left for future investigations.

The above discussions are intended  to point out some similarity of
the phenomenology in our model and that of the cuprate
superconductors. However, due to the symmetry difference in the
superconducting state, the connection is by no means clear.
Nonetheless it is interesting to explore the possible relation with
the cuprates in future studies.

This work is supported by NSF-DMR 0645691, LBNL DOE-504108 and DOE
DE-AC02-05CH11231.

\appendix
\section{Duality and Symmetries of the Superconductor Vortex Theory}\label{duality}

We first fix the relation between the relativistic vortex fields
$\Psi_\da,\, \Psi_\ua$ and the canonical vortex creation and
destruction operators.  In \Eq{eff} \be &&\Psi_\ua=(\psi_{{\rm
AV}\ua}+e^{-i\beta}\bar{\psi}_{{\rm B}\bar{{\rm
V}}\da})/\sqrt{2}\nn&&\Psi_\da=(\psi_{{\rm
BV}\da}+e^{-i\beta}\bar{\psi}_{{\rm A}\bar{{\rm
V}}\ua})/\sqrt{2},\label{rel}\ee where $\psi_{{\rm AV_\ua}}$ bosonic
vortex field for the ${\rm AV_\ua}$ vortices, and etc (For more
discussions see the end of this paragraph). The phase $\beta$ is the
phase of the annihilation matrix elements of (AV$\ua$, B$\bar{{\rm
V}}\da$) and (BV$\da$, A$\bar{{\rm V}}\ua$), respectively (the
reason that these two phase factor are the same is due to the
$T\circ\ua\leftrightarrow\da$ symmetry). We note that in terms of
$\Psi_{\ua,\da}$ \Eq{eff} has a relativistic form. It can be derived
from a theory where $\psi_{{\rm AV}_\ua}, \psi_{{\rm
A\bar{V}}_\ua},\psi_{{\rm BV}_\da}$ and $\psi_{{\rm B\bar{V}}_\da}$
are treated as non-relativistic fields; in the presence of pair
creation and pair annihilation terms $-|J|e^{i\beta}\psi_{{\rm
AV}_\ua}\psi_{{\rm B\bar{V}}_\da}+c.c$ the combination $(\psi_{{\rm
AV}\ua}-e^{-i\beta}\bar{\psi}_{{\rm B}\bar{{\rm V}}\da})/\sqrt{2}$
and $(\psi_{{\rm BV}\da}-e^{-i\beta}\bar{\psi}_{{\rm A}\bar{{\rm
V}}\ua})/\sqrt{2}$ become the more massive fields and hence can be
dropped from the low energy theory. The left low energy combinations
is given in \Eq{rel}. The important thing is that in the derivation
sketched above it is assumed that $\psi_{{\rm AV}_\ua}, \psi_{{\rm
A\bar{V}}_\ua},\psi_{{\rm BV}_\da}$ and $\psi_{{\rm B\bar{V}}_\da}$
are all degenerate with respect to one another (in the sense that
permuting any two fields will leave the (non-relativistic) action
invariant. This important degeneracy is guaranteed by the symmetry
of operations of \Eq{transv}.

Note that according to Eq.(\ref{transv}) $PH$ transforms a vortex
into an anti-vortex, while preserving spin, while mirror reflection
$\mathcal{P}$ also transforms a vortex into an anti-vortex, while
preserving the spin. Finally, $T\circ \mathcal{TR}$ transforms a
vortex into an anti-vortex and flips spin. Thus we have:
\begin{align}
 \mbox{$PH$: }&\mbox{A}\mbox{V}_{\uparrow}\rightarrow\mbox{A} \bar{\mbox{V}}_{\uparrow}&&\mbox{BV}_{\downarrow}\rightarrow  \mbox{B}\bar{\mbox{V}}_{\downarrow},\notag\\
\Rightarrow& \Psi_{\uparrow}\rightarrow
\Psi_{\downarrow}^{\dagger}&&\Psi_{\downarrow}\rightarrow
\Psi_{\uparrow}^{\dagger}\\
 \mbox{$\mathcal{P}$: }&\mbox{A}\mbox{V}_{\uparrow}\rightarrow\mbox{A} \bar{\mbox{V}}_{\uparrow}&&\mbox{BV}_{\downarrow}\rightarrow  \mbox{B}\bar{\mbox{V}}_{\downarrow},\notag\\
\Rightarrow& \Psi_{i \uparrow}\rightarrow
\Psi_{\mathcal{P}(i)\downarrow}^{\dagger}&&\Psi_{\downarrow}\rightarrow
\Psi_{\mathcal{P}(i)\uparrow}^{\dagger}\\
  \mbox{$T\circ \mathcal{TR}$: }&\mbox{A}\mbox{V}_{\uparrow}\rightarrow\mbox{B} \bar{\mbox{V}}_{\downarrow}&&\mbox{BV}_{\downarrow}\rightarrow  \mbox{A}\bar{\mbox{V}}_{\uparrow},\notag\\
\Rightarrow&  \Psi_{\uparrow}\rightarrow
\Psi_{\uparrow}^{\dagger}&&\Psi_{\downarrow}\rightarrow
\Psi_{\downarrow}^{\dagger}\mbox{ (\textbf{anti-unitary})}
\end{align}
this allows us to restrict our field theory to the form in equation
\Eq{eff}.

{\bf Duality:}

We now carry through the duality starting with \Eq{eff}. We first
set $\Psi_\s=\sqrt{\bar{\rho}}\phi_\s$ at long wavelength/time,
where $\bar{\rho}$ is the average condensate density of the up and
down vortices, and $\phi_\s$ is a U(1) {\it phase factor}.
Subsequently we can introduce auxiliary fields $J^\s_\mu$ so that
\begin{widetext}
\begin{large}
\be e^{\{-\int d^2xdt
\sum_{\s}\sum_\mu|(\partial_{\mu}+i\alpha_{\mu}-i\frac{\hbar}{2}A^S_{\mu}\sigma^3)\Psi_{\s}|^2\}}
=\int D[J^\s_\mu]e^{\{-\int d^2dt\left[\sum_{\mu\s}
{4\over\bar{\rho}}
(J^\s_\mu)^2-J^\s_\mu(\bar{\phi}_\s\p_\mu\phi_\s+i\alpha_{\mu}
-i\frac{\hbar\s}{2}A^S_{\mu})\right]\}}\ee \label{dec}
\end{large}
\end{widetext}
Next,we separate $\bar{\phi}_\s\p_\mu\phi_\s$ into
$\bar{\phi}_\s\p_\mu\phi_\s=i\p_\mu\theta_\s+\bar{\phi}^v_\s\p_\mu\phi^v_\s$,
where $\phi^v_\s$ denotes the topological non-trivial part (i.e.,
the vortex containing part) of $\phi_\s$. Integrating out
$\theta_\s$ generates the constraint $\p_\mu J^\s_\mu=0$. This
constraint is solved via the introduction of two new ``gauge
fields''$J^\s_\mu={1\over 2\pi} (\gr\times a^\s)_\mu.$
Now we collect all the $\alpha_\mu$ and $a_\mu$ dependent terms in the
action to obtain
\begin{widetext}
\be \int
d^2xdt\sum_{\s\mu}\left\{{1\over\pi^2\bar{\rho}}(\gr\times
a^\s)_\mu^2-ia^\s_\mu \left[K^\s_\mu+{1\over 2\pi}(\gr\times
\alpha)_\mu-{\hbar\s\over 4\pi}(\gr\times A^S)_\mu\right]+{\kappa\over
2}(\gr\times a)^2+i{e\over \pi}A\cdot(\gr\times \alpha)
\right\}.\label{det}\ee
\end{widetext}
In \Eq{det} $K^\s_\mu=-{i\over 2\pi}\phi^v_\s\p_\mu\phi^v_\s$ is
the 3-current of the vortex of superconducting vortices (i.e., the
vortex of $\psi_\s$ bosons). Next, we integrate out $\alpha_\mu$ which
leads to \be&&\sum_\s (\gr\times a^\s)_\mu=2e(\gr\times
A)_\mu+{\rm~less~relevant~terms}\nn&&{\rm~at~long~wavelength}.\label{qq}
\ee \Eq{qq} fixes the flux in $a^\ua+a^\da$ and leaves the flux of
$a^\ua-a^\da$ free to fluctuate. Let us define
$a_\mu=(a^\da_\mu-a^\ua_\mu)/2,$ then the remaining part of
\Eq{det} reduces to
\begin{widetext}
\be S_{\rm{dual}}=\int d^2x
dt\sum_{\s\mu}\left\{{1\over\pi^2\bar{\rho}}[e^2(\gr\times
A)_\mu^2+(\gr\times a)_\mu^2]-ie
A_\mu(K^\ua+K^\da)_\mu+ia_\mu(K^\ua-K^\da)_\mu-{i\hbar\over
2\pi}(\gr\times a)\cdot A^S\right\}.\label{det2}\ee
\end{widetext}
If we identify \be &&~~K^\ua_\mu\leftrightarrow {\rm ~
the~3-current~of~} z_1{\rm ~boson}\nn&&-K^\da_\mu\leftrightarrow{\rm
~ the~3-current~of~} z_2{\rm ~boson},\ee \Eq{det2} can be recognized
as the Feynman path integral representation of the second quantized
action in \Eq{eft}.

\section{Bogoliubov deGennes analysis of the superconducting vortices}\label{sec:SC-BdG}
In section \ref{sec:SC} we have determined the core structure of the
superconducting vortices numerically. To make sure that our
numerical results are universal, in this section we present an
analytical study in the long wavelength limit. Interestingly, as a
by-product, we obtain a new situation where zero modes exist in a
non-Dirac like Hamiltonian.

\begin{figure}
\includegraphics[width=0.25\textwidth]{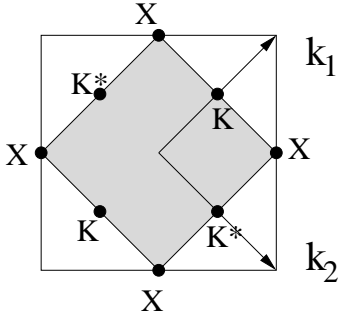}
\caption{The reduced Brillouin Zone (shaded area) of the model
Eq.(\ref{sc}).}\label{fig:bz}
\end{figure}

We first write down the d-wave pairing Hamiltonian Eq.(\ref{sc}) in
the momentum space: \be H=\sum_k
\Psi_k^{\dagger}H(k)\Psi_k,\label{uy}\ee where
\begin{widetext}
\begin{tiny}
\begin{align}
&H(k)=\notag\\
&\begin{pmatrix}
                      2t(\cos k_2-\cos k_1)&\chi_1(1+e^{-i(k_1+k_2)}+e^{-ik_1}+e^{-ik_2})&0&\chi_2^*(-1-e^{-i(k_1+k_2)}+e^{-i k1}+e^{-i k_2})\\
\chi_1(1+e^{i(k_1+k_2)}+e^{ik_1}+e^{ik_2})&-2t(\cos k_2-\cos k_1)&\chi_2^*(-1-e^{i(k_1+k_2)}+e^{i k1}+e^{i k_2})&0\\
0&\chi_2(-1-e^{-i(k_1+k_2)}+e^{-i k1}+e^{-ik_2})&2t(\cos k_2-\cos k_1)&-\chi_1(1+e^{-i(k_1+k_2)}+e^{-ik_1}+e^{-ik_2})\\
\chi_2(-1-e^{i(k_1+k_2)}+e^{i
k1}+e^{ik_2})&0&-\chi_1(1+e^{i(k_1+k_2)}+e^{ik_1}+e^{ik_2})&-2t(\cos
k_2-\cos k_1)
                     \end{pmatrix}
\label{eq:sc-momentum-ham}
\end{align}
\end{tiny}
\end{widetext}
In \Eq{uy} and \Eq{eq:sc-momentum-ham}
\begin{align}
 \Psi_k=\big(c_{A\uparrow k},c_{B\uparrow k},c_{A\downarrow -k}^{\dagger},c_{B\downarrow -k}^{\dagger}\big),
\end{align}
and we denote $Re(\chi)=\chi_1$ and $Im(\chi)=\chi_2$. We allow
$\chi_2$ to be a complex number since the pairing amplitude gain a
phase under the charge-$U(1)$ rotation. $k_1$ and $k_2$ are the
momentum components along the reduced reciprocal lattice vectors in
Fig.(\ref{fig:bz}).

Let us first study the parameter regime: $|\chi_1|\sim 1$,
$|\chi_2|\sim1$ and $|t|\ll 1$. If $t=0$ we have a gapless d-wave
superconductor with two inequivalent Dirac nodes at $K$ and $K^*$
(see Fig.\ref{fig:bz}). The $t$ term mixes  $K$ and $K^*$ and opens
up a small gap. In this limit we can linearize the Hamiltonian
matrix Eq.(\ref{eq:sc-momentum-ham}) around $K$ as:
\begin{align}
H_K=\begin{pmatrix}
-4t&2\chi_1ik_2&0&-2\chi_2^*ik_1\\
-2\chi_1ik_2&4t&2\chi_2^*ik_1&0\\
0&-2\chi_2ik_1&-4t&-2\chi_1ik_2\\
2\chi_2ik_1&0&2\chi_1ik_2&4t \label{hk}\end{pmatrix}
\end{align}
where $(k_1,k_2)=\vec k-K$.

In the presence of a vortex $\chi_2\ra \chi_2(\vec
x)=|\chi_2|e^{-i\theta(\vec x)}$. \Eq{hk}, in real space, becomes
\begin{align}
 H_K=-4t\mu_3+2i\chi_1\mu_2\tau_3\p_2-2i|\chi_2|e^{i{\theta\over
 2}\tau_3}\mu_2\tau_1e^{-i{\theta\over 2}\tau_3}\p_1.\label{br}
\end{align}
Here the Pauli matrices $\vec \mu$ mix the sublattice labels and
$\vec \tau$ mixes particle and hole. One can perform a series of
rotations to transform $H_K$ into a more familiar form (a form that
has been studied before\cite{hou:186809}):
\begin{align}
 \tilde H_K
&=4t(\cos\theta\tau_1+\sin\theta\tau_2)-2i(|\chi_2|\mu_1\tau_3\p_1+\chi_1
\mu_2\tau_3\p_2)\nn\equiv&4t(\cos\theta\tau_1+\sin\theta\tau_2)-2i(\mu_1\tau_3\tilde\p_1
+\mu_2\tau_3\tilde\p_2)\label{ar}
\end{align}
where we define $\tilde\p_1=2|\chi_2|\p_1$ and $\tilde
\p_2=2\chi_1\p_1$.

The transformations that lead from \Eq{br} to \Eq{ar} are the
following. % we present the rotation from $H_K$ to $\tilde H_K$.
Firstly we perform a rotation along $\tau_1$ such that
$\tau_3\rightarrow -\tau_2$. Afterwards we perform  another rotation
along $\mu_1$ such that $\mu_3\rightarrow -\mu_2$, $\mu_2\rightarrow
\mu_3$. After these transformations
\begin{align}
 H_K\rightarrow 4t\mu_2-2i\chi_1\tau_2\mu_3\p_1-2i|\chi_2|e^{-i\theta/2\tau_2}\tau_1\mu_3e^{i\theta/2\tau_2}\p_2
\end{align}
Second, we perform the rotation
$e^{-i\theta/2\tau_2}e^{i\theta/2\mu_3}$. Although this rotation is
spacial dependent, it commutes with $\vec\p$ in the long wavelength
limit. Another worth noting fact is that
$e^{-i\theta/2\tau_2}e^{i\theta/2\mu_3}$ has no branch cut since for
both $\theta=0$ and $\theta=2\pi$,
$e^{-i\theta/2\tau_2}e^{i\theta/2\mu_3}=1$. After the above
transformation
\begin{align}
 H_K\rightarrow 4t(\cos\theta\mu_2-\sin\theta\mu_1)-
 2i(\chi_1\tau_2\mu_3\p_1+|\chi_2|\tau_1\mu_3\p_2)
\end{align}
After interchanging $\vec \mu\leftrightarrow\vec \tau$ and
redefining $\theta\rightarrow\theta+\pi/2$ we obtain \Eq{ar}.

In Ref.\cite{hou:186809} it is shown that \Eq{ar} has one zero mode.
Similarly one can show there is another zero mode associated with
$K^*$. And same analysis can be applied to anti-vortex. We thus
established the two mid-gap modes in the SC vortex (anti-vortex).
Their existence ensures the existence of the four low energy spinful
vortices discussed in Sec.\ref{sec:SC}.

Next, we analyze a different limit of parameters: $|\chi_1|\sim1$,
$|t|\sim 1$ and $|\chi_2|\ll 1$. Since it is possible to
adiabatically tune the parameter from the previous regime to this
limit, while preserving 90$^{\circ}$ rotation and $PH$ symmetries,
we expect the zero modes to survive. 
In specific, in the previous limit the two zero modes are associated
with two different nodes (for small $t$ and slowly varying
$\theta(\v x)$ momentum is approximately conserved), and transform
into each other under 90$^{\circ}$ rotation. By forming symmetric
and anti-symmetric combination of the two zero modes we obtain two
new ones that transform under 90$^{\circ}$ with s and d symmetries
respectively. On the other hand $PH$ does not change angular
momentum, and transforms each mode into its particle-hole conjugate
hence reverse the sign of energy. Since the entire adiabatic process
preserves the 90$^{\circ}$ rotation and $PH$ symmetries, the in-gap
states can not shift away from zero energy. We have performed
numerical study of the vortices in this parameter limit and find the
mid-gap modes.

The reason we are interested in the second limit is because it gives
rise to an interesting situation where zero mode arise from a
non-Dirac like cone. In the limit of $|\chi_2|\ra 0$ two {\it
quadratically dispersed bands} touch at a single $X$ point (see
Fig.\ref{fig:bz}). Nonzero $|\chi_2|$ opens up a gap. To quadratic
order in momentum departure from $X$ the Hamiltonian looks like:
\begin{align}
 H_X=\begin{pmatrix}
      t(k_2^2-k_1^2)&-\chi_1 k_1k_2&0&-4\chi_2^*\\
      -\chi_1 k_1k_2&-t(k_2^2-k_1^2)&-4\chi_2^*&0\\
      0&-4\chi_2&t(k_2^2-k_1^2)&\chi_1 k_1k_2\\
      -4\chi_2&0&\chi_1 k_1k_2&-t(k_2^2-k_1^2)
     \end{pmatrix}\label{eq:sc-x-bdg}
\end{align}
where $(k_1,k_2)=\vec k-X$. The existence of zero modes when the is
a vortex in $\chi_2$ is very interesting. For up to present all
known zero modes are associated with ``mass vortex'' in Dirac-like
equations. The fact that \Eq{eq:sc-x-bdg} is non-relativistic, yet
in the presence of ``gap vortices'' there are zero modes suggest a
new type of ``index theorem'': For one vortex in $\chi_2$ of Eq.(\ref{eq:sc-x-bdg}) there are two zero modes.

\section{Symmetry analysis for the slave rotor theory }\label{subsec:slave-rotor-sym}

In this appendix
we study the symmetries of the slave rotor theory. As discussed in
appendix \ref{slave rotor}, the relation between the electron
operator and the slave rotor and spinon operator is given
by\cite{hermele:035125}:
\begin{align}
\begin{pmatrix}c_{i\uparrow}&\eta_i c_{i\downarrow}^{\dagger}\\c_{i\downarrow}&-\eta_{i}
c_{i\uparrow}^{\dagger}\end{pmatrix}=\begin{pmatrix}f_{i\uparrow}&\eta_i
f_{i\downarrow}^{\dagger}\\f_{i\downarrow}&-\eta_{i}f_{i\uparrow}^{\dagger}\end{pmatrix}
\begin{pmatrix}z_{i1}& z_{i2}\\-z_{i2}^{\dagger}&z_{i1}^{\dagger}\end{pmatrix},\label{eq:SU(2)-rotor}
\end{align}
or equivalently Eq.(\ref{eq:rotor-U(1)}):
\begin{align}
 c_{i\uparrow}&=f_{i\uparrow}z_{i1}-\eta_{i}f_{i\downarrow}^{\dagger}z_{i2}^{\dagger}\nn
c_{i\downarrow}&=f_{i\downarrow}z_{i1}+\eta_if_{i\uparrow}^{\dagger}z_{i2}^{\dagger}.
\end{align}
According to the above equation one can insert a site-dependent
$SU(2)$ matrix and its inverse between the $f$-spinon and $z$-rotor
matrices while leave the electron operators invariant. In terms of
$f$-spinon and $z$-boson this local $SU(2)$ transformation is:
\begin{align}
 \psi_i&=\begin{pmatrix}f_{i\uparrow}^{\dagger}\\\eta_if_{i\downarrow}\end{pmatrix}
 \rightarrow 
 e^{i\boldsymbol\alpha_i\cdot\frac{ \boldsymbol\sigma}{2}}\cdot
 \psi_i
 \notag\\
Z_i&=\begin{pmatrix}z_{i1}\\-z_{i2}^{\dagger}\end{pmatrix}\rightarrow
e^{i\boldsymbol\alpha_i\cdot\frac{ \boldsymbol\sigma}{2}}\cdot Z_i.
\label{eq:f-SU(2)}
\end{align}
The above $SU(2)$ gauge symmetry is broken by the form of the
mean-field spin-Hall insulator Hamiltonian of the spinon (\Eq{TBI}).
The
remanent gauge symmetry is $U(1)$,
\begin{align}
 f_{i\alpha}&\rightarrow e^{i\theta_i}f_{i\alpha}&z_{ia}&\rightarrow e^{-i\theta_{i}}z_{ia}.\label{eq:U(1)-gauge}
\end{align}
Due to this remanent U(1) symmetry, the fluctuation around the
mean-field theory appear in the form of a U(1) gauge theory.
We have shown in appendix \ref{slave rotor} that if the $U(1)$ gauge
field is in the Coulomb phase, the spin-Hall insulating properties
of the spinon indicates that this is actually the XY ordered phase.
This is because the spin-Hall response implies that the photon of
the gauge field is actually the Goldstone mode of the XY ordered, as
has been shown in text. In the following we study the manifestation
of various physical symmetry of the AF phase in the slave-rotor
gauge theory.

There are two   physical global $U(1)$ symmetries: the global spin
($S_z$) rotation, and the charge-$U(1)$ symmetry. They are
manifested in the slave-rotor theory as:
\begin{align}
\mbox{$S_z$ rotation by $\theta$: }  f_{i\uparrow}&\rightarrow e^{i\theta/2}f_{i\uparrow}&f_{i\downarrow}&\rightarrow e^{-i\theta/2}f_{i\downarrow}\notag\\
z_{ia}&\rightarrow z_{ia}.\notag\\
\Rightarrow c_{i\uparrow}&\rightarrow
e^{i\theta/2}c_{i\uparrow}&c_{i\downarrow}&\rightarrow
e^{-i\theta/2}c_{i\downarrow},
\end{align}
and
\begin{align}
\mbox{charge-$U(1)$: }  f_{i\alpha}&\rightarrow f_{i\alpha}\notag\\
z_{i1}&\rightarrow e^{i\theta }z_{i1}&z_{i2}&\rightarrow e^{-i\theta }z_{i2}.\notag\\
\Rightarrow c_{i\uparrow}&\rightarrow
e^{i\theta}c_{i\uparrow}&c_{i\downarrow}&\rightarrow
e^{i\theta}c_{i\downarrow}.
\end{align}
These equation imply $z_1$ and $z_2$ are spin zero charge-carrying
particles.

Next we study discrete symmetries. Before doing so it is convenient
to write the mean-field spinon Hamiltonian in the $SU(2)$ form.
In this form any transformation which changes the mean-field
Hamiltonian by a $SU(2)$ gauge transformation is regarded as a
symmetry operation due to the gauge redundancy.
Eq.(\ref{TBI}) can be rewritten as:
\begin{widetext}
\be
 H_{TBI}
 =\sum_{\langle ij\rangle}
 \psi_i^{\dagger}\begin{pmatrix}
 -\chi_{ij}^*&0\\0&-\chi_{ij}\end{pmatrix}\psi_j
 +\sum_{\langle\langle ij\rangle\rangle}\psi_i^{\dagger}\begin{pmatrix}
 -t_{ij}&0\\0&-t_{ij}
 \end{pmatrix}\psi_j+h.c.\equiv \psi_i^\dagger U_{ij}\psi_j+h.c.
\ee \end{widetext} Under a $SU(2)$ gauge transformation
,i.e.,$\psi_i\rightarrow W_i\psi_i$, \be U_{ij}\ra\rightarrow
W_i^{\dagger}U_{ij}W_j.\ee The $SU(2)$ gauge invariant quantities
are
\begin{align}
 \Tr\big[U_{\chi_{i,i+\hat x}}U_{t_{i+\hat x,i+\hat y}}U_{\chi_{i+\hat y, i}}\big]=-2t\cos2\phi.
\end{align}
for every triangular loop $i\rightarrow i+\hat x\rightarrow i+\hat
y\rightarrow i$. Two mean-field Hamiltonians with the same loop
trace are related to each other by an $SU(2)$ gauge transformation.

Now we are ready analyze the discrete symmetries. First, it is
obvious that the 90$^{\circ}$ rotation around the center of each
plaquette $\mathcal{R}_{90}$ is a symmetry since it leaves \Eq{TBI}
invariant. Under this operation the $z$ bosons transform as:
\begin{align}
 \mbox{$\mathcal{R}_{90}$: } &z_{i1}\rightarrow z_{\mathcal{R}_{90}(i),1}&&z_{i2}\rightarrow z_{\mathcal{R}_{90}(i),2}.
\end{align}
Next we consider the reflection $\mathcal{P}$ about the
vertical/horizontal lines passing through  the center of each
plaquette. Naively one would claim that this is not a symmetry of
\Eq{TBI} because 
\be
 \mbox{$\mathcal{P}$: }&\chi\rightarrow \chi^*,~t\rightarrow t.\label{eq:tbi-p}
\ee But this is actually a symmetry transformation because it does
not change loop trace; one just needs to perform a global $SU(2)$
gauge transformation $W_i=i\sigma_2$ (see Eq.(\ref{eq:f-SU(2)})) to
restore \Eq{TBI}. Thus reflection, $\mathcal{P}$, is a symmetry.
Under $\mathcal{P}$ the $z$ bosons transform as(see
Eq.(\ref{eq:f-SU(2)})):
\be
 \mbox{$W\circ\mathcal{P}$: } z_{i1}\rightarrow z_{\mathcal{P}(i),2}^{\dagger},~z_{i2}\rightarrow -z_{\mathcal{P}(i),1}^{\dagger}.\label{eq:slave-rotor-p}
\ee This is consistent with the result Eq.(\ref{eq:xy-p}). Next,
come the particle-hole transformation $PH$. According to
Ref.\cite{hermele:035125}, this transformation is implemented on the
$z$-boson via
\begin{align}
\mbox{$PH$: } &f_{i\alpha}\rightarrow f_{i\alpha}& z_{i1}&\rightarrow z_{i2}& z_{i2}\rightarrow -z_{i1}.\notag\\
\Rightarrow& c_{i\uparrow}\rightarrow \eta_i
c_{i\downarrow}^{\dagger}&c_{i\downarrow}&\rightarrow -\eta_i
c_{i\uparrow}^{\dagger}\label{eq:PH}
\end{align}
Again, this transformation law is consistent with result in
Eq.(\ref{eq:xy-ph}).
This transformation also preserves a gapped $z$ boson action and
\Eq{TBI} hence is a symmetry. Now we consider the translation $T$:
\be
 \mbox{$T_{x/y}$: } f_{i}\rightarrow f_{i+x/y},~z_{i}\rightarrow z_{i+x/y}
 ,~c_{i\s}\rightarrow c_{i+x/y\s}.\label{eq:slave-rotor-t} \ee Under this operation
 the parameters in \Eq{TBI} transform as:
\be
 \mbox{$T$: }\chi\rightarrow \chi^*,~t\rightarrow -t.\label{eq:tbi-t}
\ee This causes the loop trace to change sign (i.e., $-2t\cos2\phi
\rightarrow 2t\cos2\phi$) 
hence $T$ is broken by \Eq{TBI}. Next, we proceed to time reversal
$\mathcal{TR}$. Under time reversal \be
&&\mbox{$\mathcal{TR}$(anti-unitary):~}f_{i\alpha}\rightarrow
 (i\sigma^2)_{\alpha\beta}f_{i\beta},~~z_{ia}\rightarrow z_{ia}\nn&&
\Rightarrow~c_{i\uparrow}\rightarrow
c_{i\downarrow},~c_{i\downarrow}\rightarrow
-c_{i\uparrow}\label{eq:slave-rotor-tr} \ee Under the above
transformation the parameters of \Eq{TBI} change according to
\be
 \mbox{$\mathcal{TR}$: }\chi\rightarrow\chi^*,~t\rightarrow -t,\label{eq:tbi-tr}
\ee which also flips sign of the loop trace. Hence $\mathcal{TR}$ is
also broken by \Eq{TBI}.

Next we study the compound operation $T\circ \mathcal{TR}$ .
Combining Eq.(\ref{eq:tbi-t},\ref{eq:tbi-tr}) it is simple to show
that \Eq{TBI} is invariant under this transformation. Thus we
conclude $T\circ \mathcal{TR}$ is a good symmetry. Using
Eq.(\ref{eq:slave-rotor-t},\ref{eq:slave-rotor-tr}) we obtain the
$z$ boson transformation law as \be \mbox{$T\circ \mathcal{TR}~$:
}z_{i1}\rightarrow z_{i+1,1},~z_{i2}\rightarrow
z_{i+1,2}.\label{eq:slave-rotor-t-tr} \ee Once again, this is
consistent with Eq.(\ref{eq:xy-tr}).

Because the slave-rotor theory does not capture the $S_z$ rotation
symmetry breaking at the mean-field level, it appears that
$\mathcal{R}_{90}$, $\mathcal{P}$,$PH$, $T\circ
\uparrow\leftrightarrow\downarrow$ and $T\circ \mathcal{TR}$ are
always good symmetries. After the breaking of the $S_z$ rotation the
above symmetries might be modified; after the operation of the
discrete symmetries it might require an additional $S_z$ rotation to
restore the ground state. For example $\mathcal{R}_{90}$ and
$\mathcal{P}$ are modified to $e^{i\pi S_z}\circ\mathcal{R}_{90}$
and $e^{i\pi S_z}\circ\mathcal{P}$ due to the AF order. For example,
the angle of the $S_z$ rotation can be determined as follows. Assume
\be e^{i\theta S_z}\circ \mathcal{R}_{90}|GS\rangle=|GS\rangle\ee we
obtain
\begin{align}
 &\langle GS|S_i^{\dagger}|GS\rangle=\langle GS|(e^{i\theta S_z}\circ \mathcal{R}_{90})S_i^{\dagger}(e^{i\theta S_z}\circ \mathcal{R}_{90})^{-1}|GS\rangle\notag\\
&=e^{-i\theta}\langle
GS|S_{R_{90}(i)}^{\dagger}|GS\rangle=-e^{-i\theta}\langle
GS|S_i^{\dagger}|GS\rangle,
\end{align}
hence $\theta=\pi$. After some simple calculation it can be shown
that in the antiferromagnetic state $PH$ and $T\circ \mathcal{TR}$
alone remain good symmetries. Finally, since $T\circ
\uparrow\leftrightarrow\downarrow$ transforms XY order parameter $M$
to $-M^*$ (note how the nature of the AF order enters), the
associated $S_z$ rotation must rotate $-M^*$ back to $M$.

In summary, in the AF phase we find while $T$ and $\mathcal{TR}$ are
broken, $e^{i\pi S_z}\circ\mathcal{R}_{90}$,$e^{i\pi
S_z}\circ\mathcal{P}$,$T\circ \mathcal{TR}$ and $PH$ are good
symmetries. This is consistent with a regular easy-plane AF ordered
state with $PH$ symmetry. Thus {\it the AF phase portrayed by the
slave-rotor+ gauge fluctuation theory is the regular AF phase}.

\section{A mean-field theory for the AF phase and the analysis of
the core structure of the AF vortices}\label{sec:XY-MF-BdG}

In the main text we have started with a mean-field superconductor
and show its vortices carries spin. The condensation of these
vortices destroy the superconductor and drive the system into an
insulating easy-plane antiferromagnet. The vortices of this
antiferromagnet carries charge and their condensation destroys the
AF order and drives the system back to the superconducting phase. In
section \ref{subsec:slave-rotor-sym} we have presented a gauged
slave rotor theory for this antiferromagnet. In this appendix we
present a mean-field Hamiltonian for this phase and study its
vortices.

From Sec.\ref{subsec:slave-rotor-sym} we have shown the symmetries
of the AF phase are  charge-$U(1)$, $e^{i\pi S_z}\circ \mathcal{R}_{90}$, $e^{i\pi S_z}\circ \mathcal{P}$, $PH$, $T\circ\mathcal{TR}$. These symmetries
severely limit the possible quadratic mean-field Hamiltonians. The
most general quadratic Hamiltonian can be written as combination of
hopping terms (however in general the hopping does not have to
conserve the spin). Demanding the above symmetries it is possible to
write down all symmetry allowed hopping terms. For example consider
the nearest neighbor hopping. We shall show in the following that
the only symmetry allowed such hopping term is of the form
$\chi_1(c_{i\sigma}^{\dagger}c_{j\sigma}+h.c.)$ where $\chi_1$ us a
real number. Let us first assume the up spin hopping amplitude is
imaginary: $i\lambda c_{i\uparrow}^{\dagger}c_{j\uparrow}+h.c.$.
Under $PH$ transformation it becomes $-i\lambda
c_{i\da}c^\dagger_{j\da}+h.c.=i\lambda
c^\dagger_{j\da}c_{i\da}+h.c.=-i\lambda
c^\dagger_{i\da}c_{j\da}+h.c.$ Hence the spin up and spin down
hopping matrix elements must differ by a sign, i.e.,
$i\sigma(c_{i\sigma}^{\dagger}c_{j\sigma}-c_{j\sigma}^{\dagger}c_{i\sigma})+h.c.$.
Because the sign difference between $i\rightarrow j$ and
$j\rightarrow i$, this hopping has a direction. Let us represent the
direction in which the hopping amplitude is $i\lambda$ by an arrow.
$T\circ\mathcal{TR}$ requires the arrows to be translation
invariant, but this breaks the $e^{i\pi S_z}\circ R_{90}$. So the
imaginary spin preserving nearest neighbor hopping is forbidden.
Next let us consider the spin flipping nearest neighbor hopping
$M^*c_{i\uparrow}^{\dagger}c_{j\downarrow}+h.c.$. $PH$ transforms it
into
$M^*c_{i\da}c^\dagger_{j\ua}+h.c.=-M^*c^\dagger_{j\ua}c_{i\da}+h.c.$
 Thus the hopping also has a direction, i.e., the hopping amplitude
for $c_{i\uparrow}^{\dagger}c_{j\downarrow}$ is the negative of that
of $c_{j\uparrow}^{\dagger}c_{i\downarrow}$. The
$T\circ\mathcal{TR}$ requires the hopping to be translation
invariant, but this again must break the $e^{i\pi S_z}\circ
\mathcal{R}_{90}$. So the spin flipping nearest neighbor hopping is
also forbidden. This leaves the regular real-amplitude
spin-preserving hopping as the only possibility.

One can perform similar analysis for all possible quadratic terms:
on-site terms, nearest neighbor terms, next nearest neighbor terms,
etc. The following Hamiltonian has the most general form up to
second neighbor:
\begin{widetext}
\begin{align}
 H_{AF}^{MF}=\sum_i \eta_i(M_1^* c_{i\uparrow}^{\dagger}c_{i\downarrow}+h.c.)
 +\chi_1\sum_{\langle ij\rangle}(c_{i\sigma}^{\dagger}c_{j\sigma}+h.c.)
+\sum_{\langle\langle i k\rangle\rangle}t_{ik}(\sigma
c_{i\sigma}^{\dagger} c_{k\sigma}+h.c.)+\sum_{\langle\langle i
k\rangle\rangle}\eta_i
M_2^*(c_{i\uparrow}^{\dagger}c_{k\downarrow}+c_{k\uparrow}^{\dagger}c_{i\downarrow})+h.c.\label{eq:xy-mf}
\end{align}
\end{widetext}
where $t_{ij}$ has the staggered pattern as shown in Fig.\ref{hop}(a).

Interestingly it is not possible to choose parameters so that
Eq.(\ref{eq:xy-mf}) will have conic (i.e., Dirac-like) intersecting
bands.
%to have two Dirac nodes. A simple gapless limit is to set
If one set $M_1=M_2=0$, the two bands intersect quadratically at $X$
point. Small $M_1$ and/or $M_2$ open up a gap. In this limit the
momentum space Hamiltonian around $X$ is
\begin{widetext}
\begin{align}
H_{AF,X}^{MF}=\Psi_k^{\dagger}\begin{pmatrix}
      t(k_2^2-k_1^2)&-\chi_1 k_1k_2&M^*&0\\
      -\chi_1 k_1k_2&-t(k_2^2-k_1^2)&0&-M^*\\
      M&0&-t(k_2^2-k_1^2)&-\chi_1 k_1k_2\\
      0&-M&-\chi_1 k_1k_2&t(k_2^2-k_1^2)
                               \end{pmatrix}
\Psi_k
\end{align}
\end{widetext}

where $(k_1,k_2)=\vec k-X,~M=M_1-2M_2$ and
\begin{align}
 \Psi^\dagger_k&=(c^\dagger_{A\uparrow k},c^\dagger_{B\uparrow k},c^\dagger_{A\downarrow k},
 c^\dagger_{B\downarrow k}).
\end{align}

After a simple unitary rotation:
\begin{align}
 \Psi_k\ra \begin{pmatrix}
      \mathbf{1}&0\\
      0&i\sigma_2
\end{pmatrix}\Psi_k
\end{align}
The form of the Hamiltonian becomes the same form as
Eq.(\ref{eq:sc-x-bdg}) in appendix B. Based on the index theorem obtained there, we conclude one vortex (antivortex) in $M$ also has two mid-gap energy levels. If both levels are filled the vortex (anti-vortex) carries electric charge $1$, and if both are empty the vortex (anti-vortex) carries charge $-1$. This is because the vortex with both levels filled and the vortex with both levels empty are related by $PH$ and thus their charges differ by sign change. On the other hand by definition these two vortices' charges differ by 2. We thus prove the existence of the charge-$\pm1$ vortex (anti-vortex) in the AF phase.

\section{The matrix element of the order parameter operator (vortex tunneling operator) in SC and AF phases}\label{sec:vortex-tunneling}
The SC$\leftrightarrow$AF transition is realized by condensing
vortices with non-trivial quantum number. The motivation of studying
the vortex tunneling operator is to compute the order pattern in the
vortex condensed phase. The calculation presented here provides a
rigorous and systematical way to determine the order pattern in the
vortex condensed phase. As a first application, let us study the
vortices in the SC phase to show that the vortex condensed phase is
AF.

What is the order parameter in the SC vortex
$|\langle\Psi_{\ua}\rangle|=|\langle\Psi_{\da}\rangle|\neq
0$condensed phase? Apparently both $\langle\Psi_{\ua}\rangle$ and
$\langle\Psi_{\da}\rangle$ are non-zero and one may think that there
are two spin-1/2 order parameters. However both $\Psi_{\ua}$ and
$\Psi_{\da}$ are non-local operators (since they annihilate vortices
of the SC phase) hence can not serve as order parameter.
However the combination 
$\Psi_{\uparrow}^{\dagger}\Psi_{\downarrow}$ (which does not 
change vorticity but flips spin) is local, and acquires none zero
value in the XY ordered state, hence can serve as an order
parameter.
To determine the nature (i.e. FM versus AF) of the $XY$ order, we
compute the quantum number of the
$\Psi_{\uparrow}^{\dagger}\Psi_{\downarrow}$ operator. For example,
if $\Psi_{\uparrow}^{\dagger}\Psi_{\downarrow}$ changes sign under a
90 degree rotation ($\mathcal{R}_{90}$) around the center of a
plaquette , it is consistent with AF order.

In order to determine the quantum number of the operator
$\Psi_{\uparrow}^{\dagger}\Psi_{\downarrow}$ we consider the two
vortex states $|{\rm AV}_{\uparrow}\rangle$ and $|{\rm
BV}_{\downarrow}\rangle$ with the location of the vortices displaced
from one another by one lattice spacing. However we tune the
parameters so that the core size $\xi$ of the vortices is
significantly larger than their
separation. 
As shown in Sec.\ref{sec:SC-BdG}, there are two spin-up in-gap
levels in the vortex core, and AV$_{\uparrow}$ and BV$_{\downarrow}$
differ by whether the two levels are filled or not. If
$\gamma^\dagger_{1\uparrow}$ and $\gamma^\dagger_{2\uparrow}$ create
Bogoliubov quasiparticle in these levels, then in the limit $\xi>>1$
$|$AV$_{\uparrow}\rangle$=$\gamma_{1\uparrow}^{\dagger}\gamma_{2\uparrow}^{\dagger}|$BV$_{\downarrow}\rangle$.
On the other hand from \Eq{rel} we know that the operator
$\Psi_{\uparrow}^{\dagger}\Psi_{\downarrow}$ sends BV$_{\downarrow}$
to AV$_{\uparrow}$, i.e., \emph{it is the vortex tunneling
operator}:
\begin{align}
 |\mbox{AV}_{\uparrow}\rangle=\Psi_{\uparrow}^{\dagger}
 \Psi_{\downarrow}|\mbox{BV}_{\downarrow}\rangle.
\end{align}
Therefore the quantum number of
$\Psi_{\uparrow}^{\dagger}\Psi_{\downarrow}$ is simply the quantum
number difference of $|\mbox{AV}_{\uparrow}\rangle$ and
$|\mbox{BV}_{\downarrow}\rangle$. For example let us consider the
rotation $\mathcal{R}_{90}$. If the vortex configuration is
symmetric under $\mathcal{R}_{90}$ we expect
$|\mbox{AV}_{\uparrow}\rangle$ and $|\mbox{BV}_{\downarrow}\rangle$
to be eigenstates of $\mathcal{R}_{90}$:
\begin{align}
 \mathcal{R}_{90}|\mbox{AV}_{\uparrow}\rangle&=e^{i\theta_{\uparrow}}|\mbox{AV}_{\uparrow}\rangle\notag\\
 \mathcal{R}_{90}|\mbox{BV}_{\downarrow}\rangle&=e^{i\theta_{\downarrow}}|\mbox{BV}_{\downarrow}\rangle
\end{align}
To determine the quantum number difference,
$e^{i(\theta_{\uparrow}-\theta_{\downarrow})}$, we can calculate the
following matrix element ratio
\begin{align}
\frac{\langle \mbox{AV}_{\uparrow}|S_{i}^{+}|\mbox{BV}_{\downarrow}
\rangle}{\langle
\mbox{AV}_{\uparrow}|S_{j}^{+}|\mbox{BV}_{\downarrow}\rangle}.
\label{eq:matrix-element-ratio}
\end{align}
To see that let us consider $j=\mathcal{R}_{90}(i)$ and thus
$S_{j}^{+}=\mathcal{R}_{90}^{-1}S_{i}^{+}\mathcal{R}_{90}$. In this
case  Eq.(\ref{eq:matrix-element-ratio}) becomes
\begin{align}
 \frac{\langle \mbox{AV}_{\uparrow}|S_{i}^{+}|\mbox{BV}_{\downarrow}\rangle}
 {\langle \mbox{AV}_{\uparrow}|\mathcal{R}_{90}^{-1}S_{i}^{+}
 \mathcal{R}_{90}|\mbox{BV}_{\downarrow}\rangle}
 =e^{i(\theta_{\uparrow}-\theta_{\downarrow})}.\label{eq:ratio-sym}
\end{align}
From the above it is clear that 
one can replace the $S_i^\dagger$ in \Eq{eq:ratio-sym} by any spin-1
operator $\hat O_i$ and get the same result so long as $\langle
\mbox{AV}_{\uparrow}|\hat{O}_i|\mbox{BV}_{\downarrow}\rangle\ne 0$.

In Eq.(\ref{eq:ratio-sym}) we have assumed that the vortex
configuration is $\mathcal{R}_{90}$-symmetric so that
$|\mbox{AV}_{\uparrow}\rangle$ and $|\mbox{BV}_{\downarrow}\rangle$
are eigenstates of $\mathcal{R}_{90}$. The easiest way to implement
the 90-degree rotation symmetry is to put in a vortex under open
boundary condition. 
However this brings in edge states in the gap since the
superconductor in question is a topological one. To avoid the
complication of edge states it is better to use periodic boundary
condition. However in that case one has to put in a vortex and an
antivortex hence necessarily breaks the 90 degree rotation symmetry.
Fortunately, it turns out that the ratio in
\Eq{eq:matrix-element-ratio} is almost completely determined by the
the wavefunctions of the mid-gap levels. The latter is localized in
the core region and can only sense a region $D$ around the center of
the vortex. Thus if we separate the vortex and the antivortex by
sufficient distance and make sure that the vortex configuration
within region $D$ is rotation symmetric, we expect the matrix
element ratio would converge to the desired result
$e^{i(\theta_{\uparrow}-\theta_{\downarrow})}$ in the following
limit: (1) large vortex core size (2) large system size ($D$ size).
In practice we input the vortices by Jacobi theta function as
discussed in detail in Sec.\ref{sec:vortex-tunneling} and the vortex
configuration around the vortex center tends to rotational symmetric
in the afore mentioned limit.

\begin{figure}
 \includegraphics[width=0.233\textwidth]{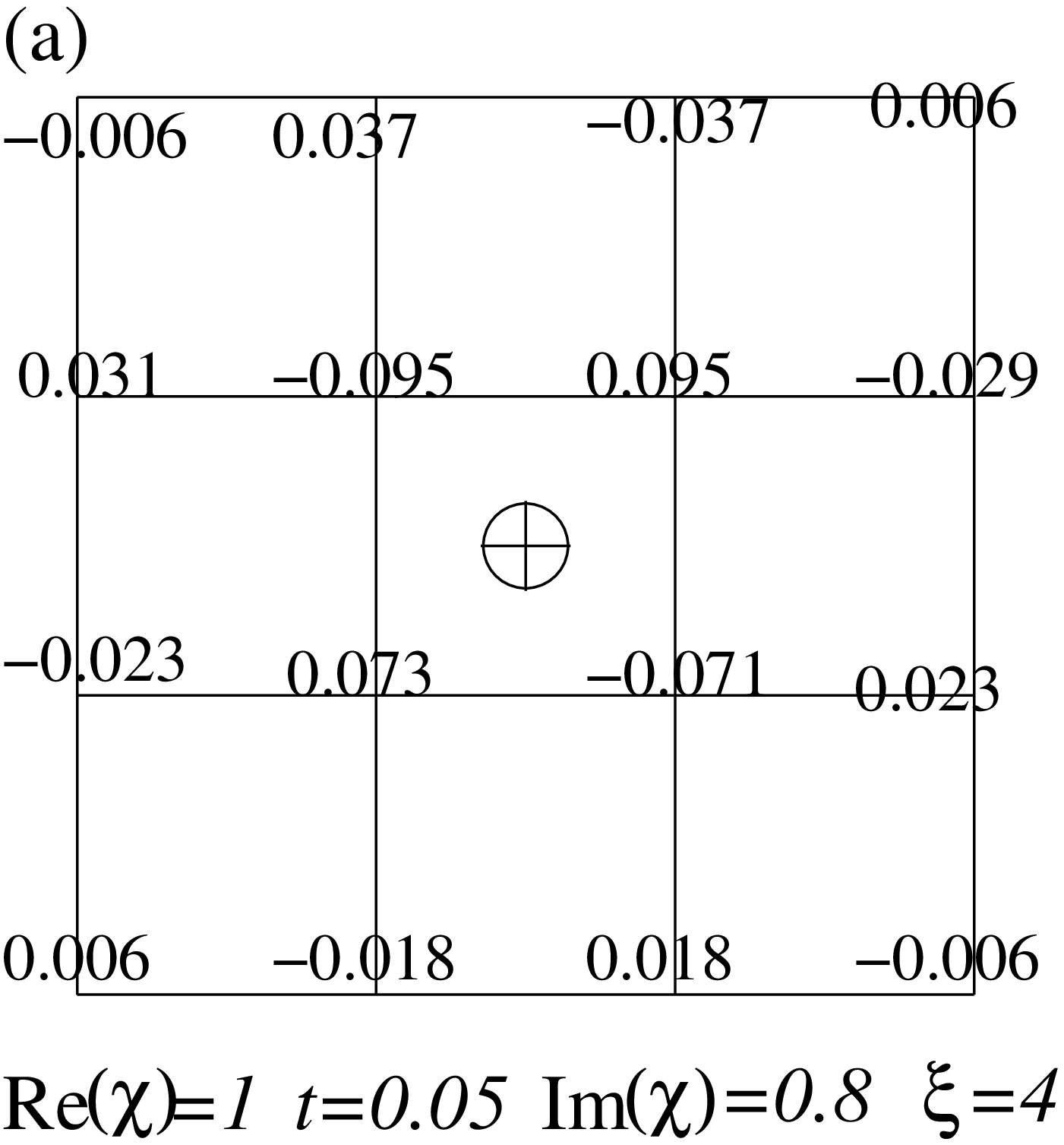}\;\;\;\includegraphics[width=0.24\textwidth]{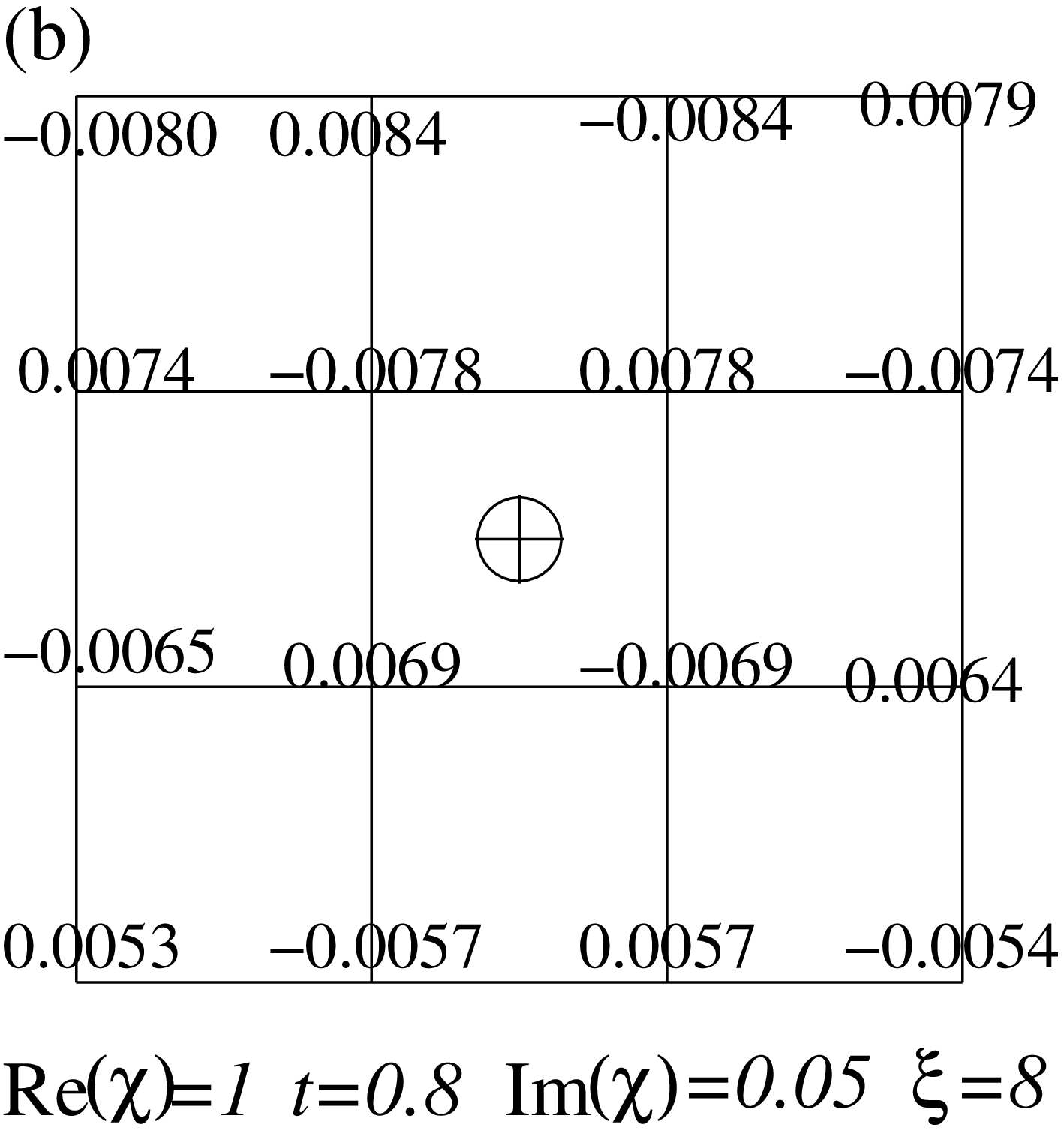}
\caption{ $\langle
\mbox{AV}_{\uparrow}|S_{i}^{+}|\mbox{BV}_{\downarrow}\rangle$ as a
function of $i$ for sites close to the center of the vortex (labeled
by $\oplus$). We input a vortex and an anti-vortex separated by half
system size on a 24x24x2 ('2' means 2 sites per unit cell) system on
torus (details of the vortex configurations are given in
Sec.\ref{sec:vortex-tunneling}). Re$(\chi)$, $t$ and Im$(\chi)$ are
the parameters in Eq.(\ref{sc}). To make the core size $\xi$ big, we
input the pairing order parameter in the vicinity of the vortex core
as $\mbox{Im}(\chi)(1-e^{-r/\xi})e^{i\theta}$, where $r,\theta$ are
the polar coordinates with respect to the the center of the vortex.
We used two sets of parameters: (a) large pairing and small $t$ and
(b) small pairing and large $t$. In limit (a) the vortex core is
relatively small, while (b) the vortex core is large (for details
see Sec.\ref{sec:SC-BdG} and \ref{sec:vortex-tunneling}). In both
limit we see staggered signs, implying AF order.
}
\label{fig:sc-vortex-core-af}
\end{figure}

In Fig.\ref{fig:sc-vortex-core-af} we present the result for
$\langle
\mbox{AV}_{\uparrow}|S_{i}^{+}|\mbox{BV}_{\downarrow}\rangle$ as a
function of $i$ for a fixed vortex configuration. We find in the
vortex core the matrix elements have staggered signs, which means
the XY order is AF.

\begin{figure}
 \includegraphics[width=0.15\textwidth]{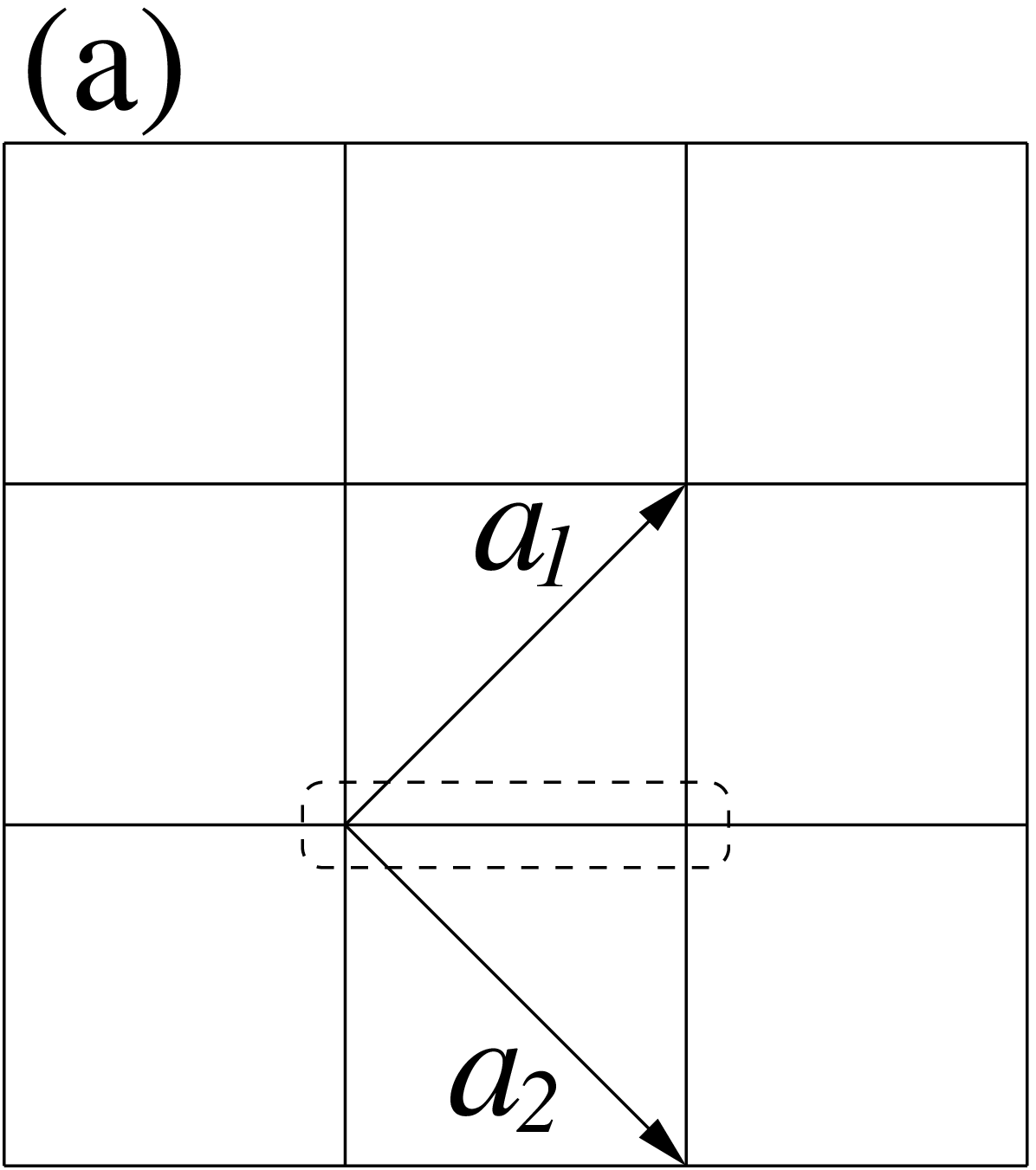} \;\;\;\;\;\includegraphics[width=0.25\textwidth]{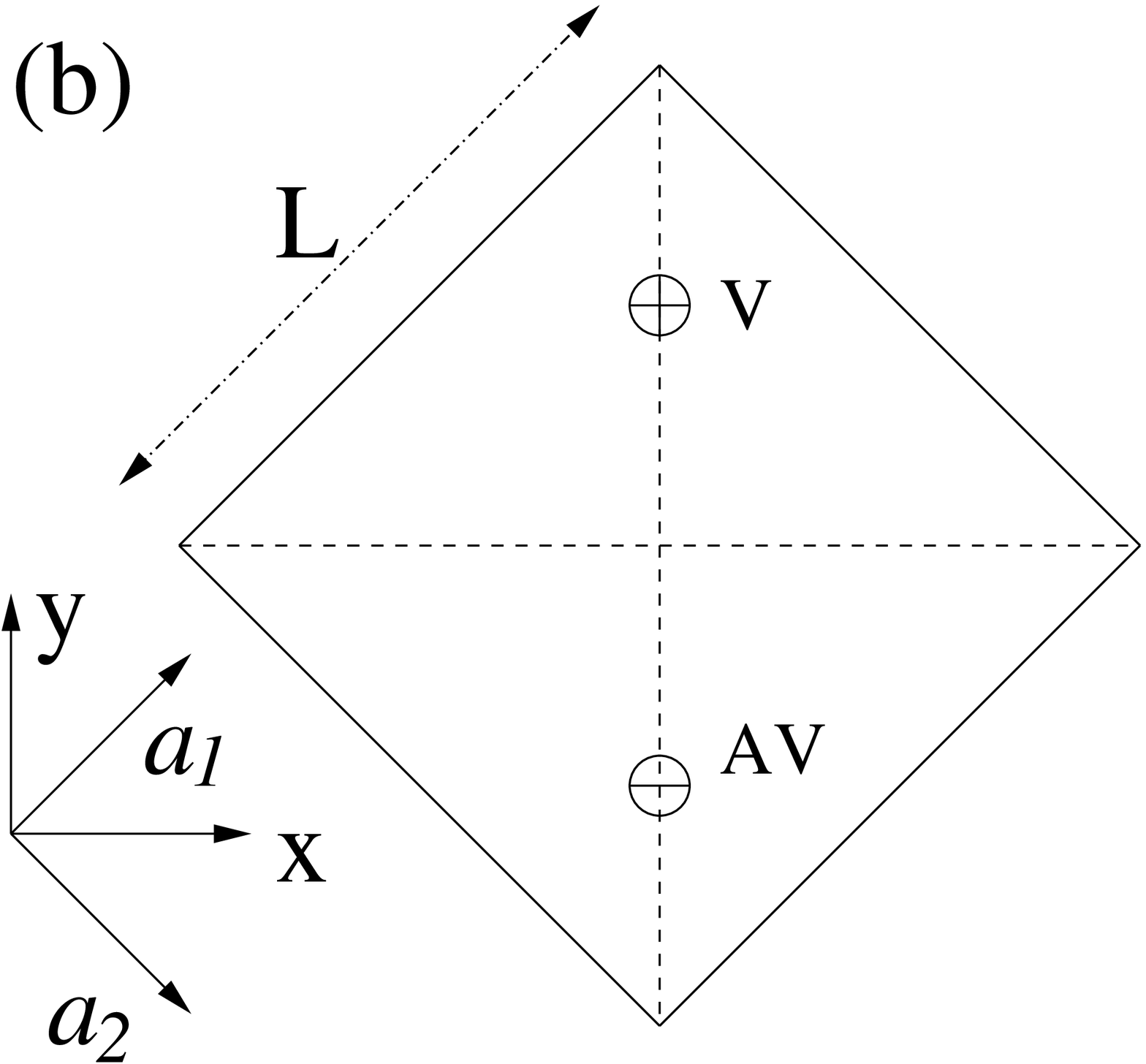}
\caption{(a) The real space unit cell of the square lattice model Eq.(\ref{sc}) and the basis vectors $\boldsymbol a_1$,$\boldsymbol a_2$. (b) In our numerical study, we put one vortex $\oplus$ and one anti-vortex $\ominus$ separated by half system size on a sample with periodic boundary conditions.}\label{fig:vortex-position}
\end{figure}

We showed $\Psi_{\uparrow}^{\dagger}\Psi_{\downarrow}$ is an AF order parameter. Here we present some details of the numerics that we performed. In Fig.\ref{fig:vortex-position} we show the positions of the vortex and the anti-vortex centers on the torus. We use complex number $w_1,w_2$ to represent the position of the vortex and anti-vortex. Given these positions, and using complex number $z=x+iy$ to represent the position of the pairing bond center, the pairing amplitude is given by:
\begin{align}
 \Delta(z)=|\Delta_0|e^{i\phi(z)}(1-e^{-|z-w1|/\xi})(1-e^{-|z-w2|/\xi}),
\end{align}
where $\xi\ll L$ is the size of the vortex core, and the phase $\phi(z)$ is given by:
\begin{align}
 \phi(z)
=&\arg\Big(\theta_1\big(\frac{\pi(z-w_1)}{\sqrt{2}L},e^{i\pi\frac{1+i}{2}}\big)\bar\theta_1\big(\frac{\pi(z-w_2)}{\sqrt{2}L},e^{i\pi\frac{1+i}{2}}\big)\Big)\notag\\
&-\frac{2\pi\mbox{Im}(z)}{L^2}\mbox{Re}(w_1-w_2).
\end{align}
And the Jacobi theta function is defined as:
\begin{align}
 \theta_1(z,q)=\sum_{n=-\infty}^{\infty}(-1)^{n-\frac{1}{2}}q^{\big(n+\frac{1}{2}\big)^2}e^{(2n+1)iz}.
\end{align}
One can show that $e^{i\phi(z)}$ is a  periodic function on the LxL
sample with a vortex and an anti-vortex at $w_1,w_2$.

We mention that $\xi$ determines the input vortex size, but it does
not determine the size of the zero mode wavefunction (which actually
weakly dependent on $\xi$). We find that size of the zero mode
wavefunction is mainly determined by the pairing $|\Delta_0|$ by
numerical study. This behavior at least can be understood in the
Dirac limit where the analytical solution of the zero mode
wavefunction is known\cite{hou:186809} to be $\sim e^{-|\Delta_0|
r}$. To provide a quantitative understanding of the size of the zero
mode wavefunction $\psi(\vec r)$ we numerically fit the density
$|\psi(\vec r)|^2$ by $A e^{-\frac{(\vec r-\vec r_0)^2}{l^2}}$. We
find for the vortex configuration in
Fig.\ref{fig:sc-vortex-core-af}(a) the zero mode size $l=1.5$, and
for Fig.\ref{fig:sc-vortex-core-af}(b) $l=6.0$. This finishes the
discussion of the SC vortex tunneling amplitude.

From now on we study the AF vortex tunneling amplitude. We adopt the mean-field Hamilton Eq.(\ref{eq:xy-mf}) from Sec.\ref{sec:XY-MF-BdG} and only consider the on-site magnetization:
\begin{align}
 H_{AF}^{MF}&=\sum_i \eta_i(M^* c_{i\uparrow}^{\dagger}c_{i\downarrow}+h.c.)
 +\chi\sum_{\langle ij\rangle}(c_{i\sigma}^{\dagger}c_{j\sigma}+h.c.)\notag\\
&+\sum_{\langle\langle i k\rangle\rangle}t_{ik}(\sigma
c_{i\sigma}^{\dagger} c_{k\sigma}+h.c.)
\end{align}
From Sec.\ref{sec:XY-MF-BdG} we know that there are two zero modes in the vortex core $\eta_1,\eta_2$ in the long wavelength limit. Therefore there are two charged vortex states $|V_{Q=1}\rangle$ and $|V_{Q=-1}\rangle$ are related by $|V_{Q=1}\rangle=\eta_1^{\dagger}\eta_2^{\dagger}|V_{Q=-1}\rangle$. On the other hand because the $z$-bosons in Eq.(\ref{eft}) is the relativistic boson fields of these vortices and anti-vortices, we conclude that:
\begin{align}
 |V_{Q=1}\rangle=z_1^{\dagger}z_2|V_{Q=-1}\rangle
\end{align}
This is because $z_1^{\dagger}z_2$ does not change vorticity but create a pair of electrons. $z_1^{\dagger}z_2$ is nothing but a cooper pair creation operator. In the $\langle z_1\rangle\neq 0,\langle z_2\rangle\neq 0$ phase $z_1^{\dagger}z_2$ acquires non-zero expectation value and it is the SC order parameter. What is the SC order pattern?

\begin{figure}
 \includegraphics[width=0.23\textwidth]{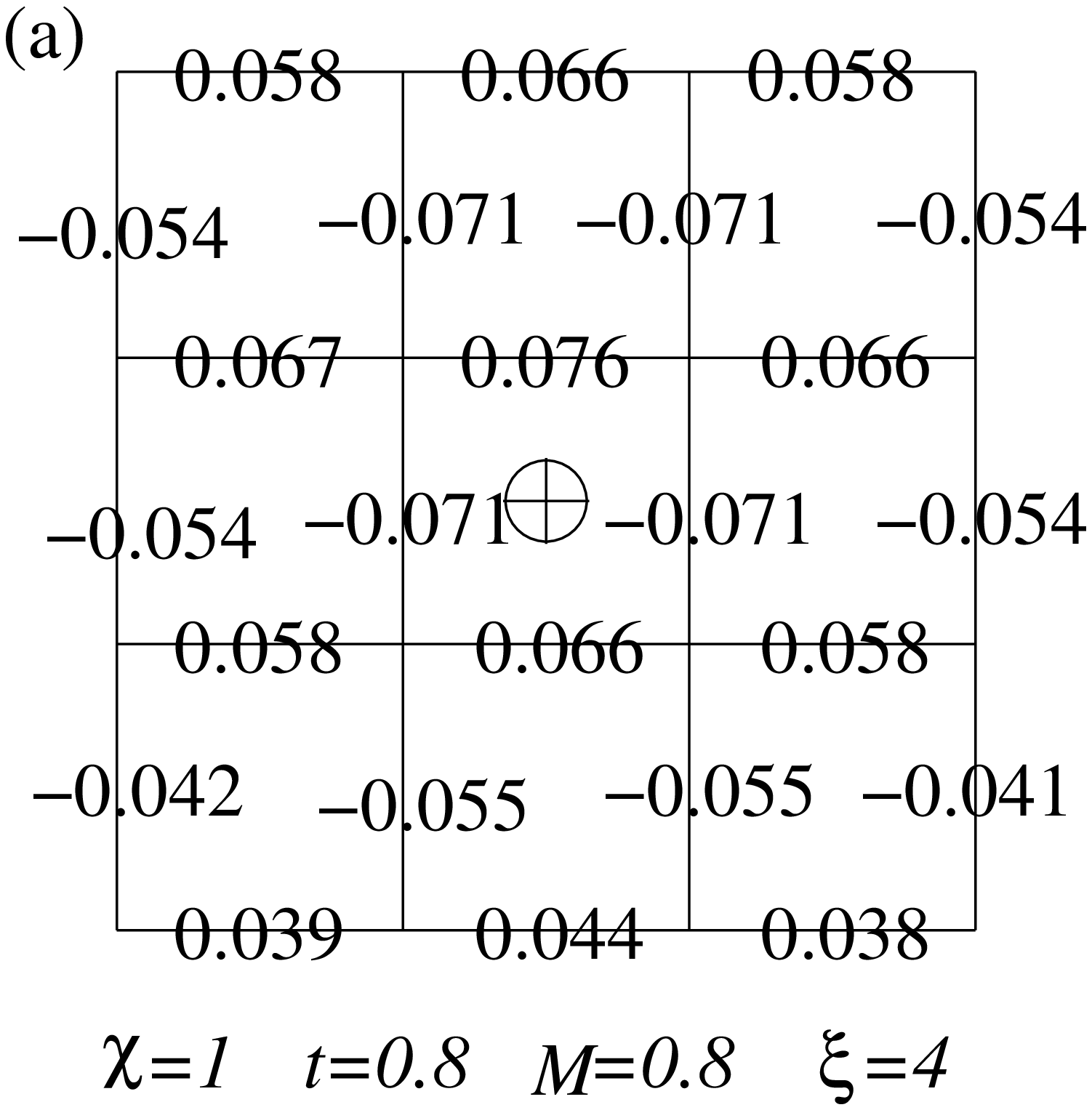}\;\;\; \includegraphics[width=0.23\textwidth]{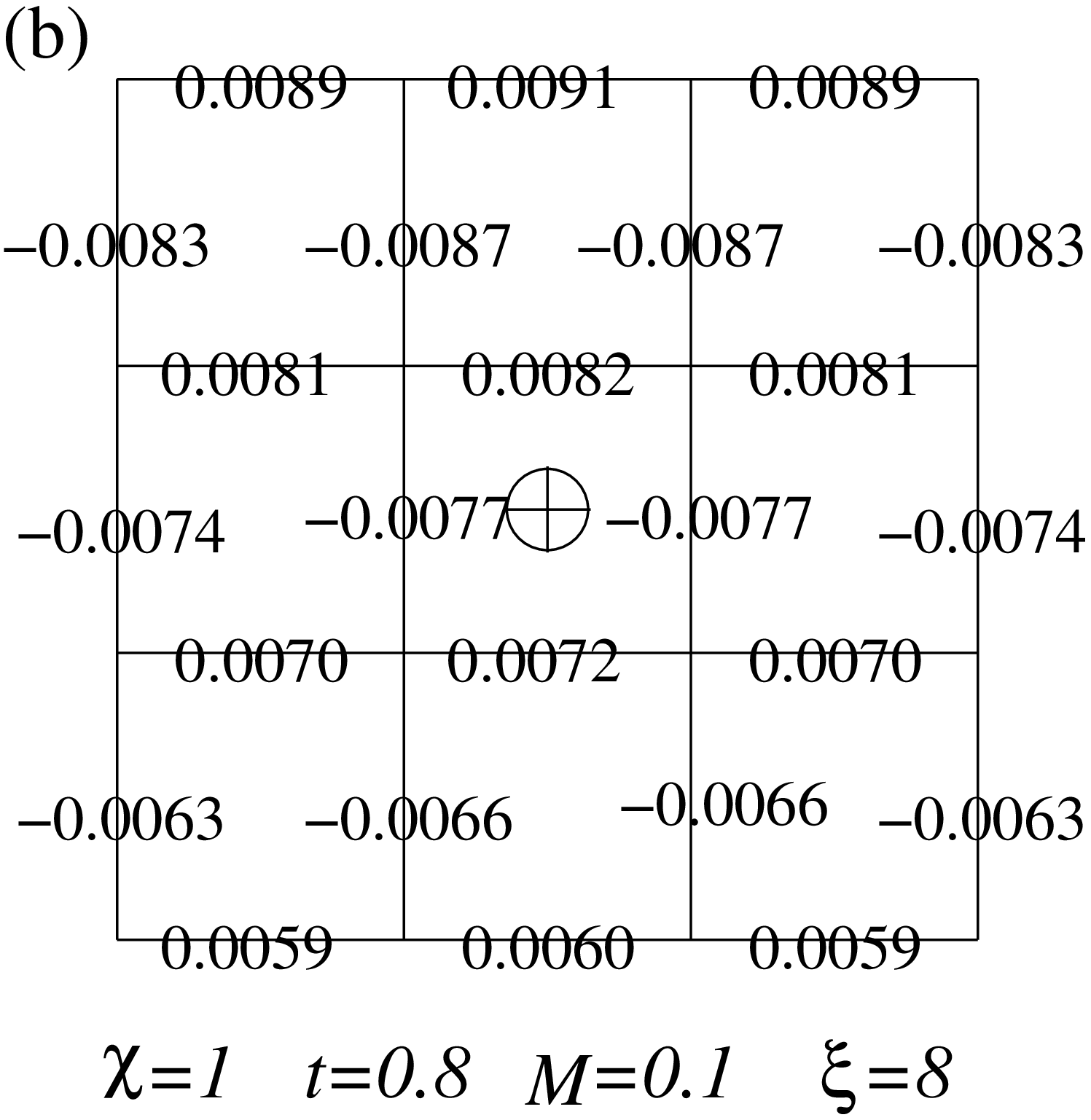}
\caption{The AF vortex matrix elements $\langle V_{Q=1}|(c_{i\uparrow}^{\dagger}c_{i+\{x,y\}\downarrow}^{\dagger}-c_{i\downarrow}^{\dagger}c_{i+\{x,y\}\uparrow}^{\dagger})|V_{Q=-1}\rangle$. We input the vortex and anti-vortex in the same fashion as we did for SC vortex, and choose two sets of parameters: (a) large magnetization with small vortex size, and (b) small magnetization with large vortex size. The staggered sign indicates that the pairing is d-wave in the AF vortex condensed phase.}
\label{fig:af-vortex-core-sc}
\end{figure}

To answer this question we perform the following matrix element ratio computation.
\begin{align}
 \frac{\langle V_{Q=1}|(c_{i\uparrow}^{\dagger}c_{i+\{x,y\}\downarrow}^{\dagger}-c_{i\downarrow}^{\dagger}c_{i+\{x,y\}\uparrow}^{\dagger})|V_{Q=-1}\rangle}{\langle V_{Q=1}|(c_{i\uparrow}^{\dagger}c_{j+\{x,y\}\downarrow}^{\dagger}-c_{j\downarrow}^{\dagger}c_{i+\{x,y\}\uparrow}^{\dagger})|V_{Q=-1}\rangle}
\end{align}
where site $i$ and site $i+\{x,y\}$ form a n.n. bond. Similar to the discussion in Sec.\ref{sec:psi-cond-af}, one can convince oneself that if the bond in the numerator and the bond in the denominator are related by $\mathcal{R}_{90}$, the 90 degree rotation around a plaquette center, then this ratio computes the $\mathcal{R}_{90}$ quantum number of $z_1^{\dagger}z_2$ operator. In Fig.\ref{fig:af-vortex-core-sc} we see the staggered sign of the bond pairing matrix elements, which indicates the d-wave pairing symmetry in the AF vortex condensed phase.

\section{The quantum number of the monopole operator and the magnetic order pattern in the slave-rotor theory}\label{sec:monopole}
In this section we determine the magnetic order pattern in the slave
rotor theory. We first show the quantum number of the gauge monopole
determines the magnetic order pattern,
then we find a way to 
compute this quantum number. We find the magnetic order pattern is
antiferromagnetic consistent with the conclusion in the main text.

In the magnetic ordered state, the $z_\alpha$'s are absent at long
distance/time. Dropping them from the slave-rotor gauge theory we
arrive at a spinon theory interacting with a $U(1)$ gauge field:
\begin{widetext}
\be L_{\rm latt}=\sum_{i,\s}
f^\dagger_i(\p_t+ib_{0,i}-{\hbar}A^s_{0,i}\s^3)f_i+\Big[\sum_{\langle
i,j\rangle,\s}e^{i(b_{ij}-A^S_{ij}\s^3)}\chi_{ij}f_{i\sigma}^{\dagger}f_{j\sigma}
+\sum_{\langle\langle i,j\rangle\rangle,\s} \s
t_{ij}e^{i(b_{ij}-A^S_{ij}\s^3)}f_{i\s}^{\dagger}f_{j\s}+h.c\Big].\label{lt}\ee
\end{widetext}
Physically \Eq{lt} describes a half-filled (i.e., one particle per
site) system of fermions, or more precisely, a spin state.

We now show that this theory describes an XY ordered phase. As
discussed in the paragraph containing \Eq{mono} the monopole
operator serves as the order parameter of the magnetic order, i.e.,
we expect
\begin{align}
 V_{D}^{\dagger}\sim\sum_{i\in D} e^{i\theta_i} S_i^{+},
\end{align}
where $D$ is the spatial extent of the monopole (i.e., the region
spanned by the inserted gauge flux). We expect $V_D^\dagger$ to be
the antiferromagnetic order parameter hence \be
e^{i\theta_i}=\eta_i\ee where $\eta_i$ is the staggered sign. In
general there is an arbitrary global phase factor relating
$V_{D}^{\dagger}$ to $\sum_{i\in D} \eta_i S_i^{+}$, however this
phase does not affect
our determination of the the magnetic order.

It is convenient to choose $D$ and the inserted flux distribution so
that they do not break lattice symmetry of \Eq{lt}. The
transformation
properties of 
$V_{D}^{\dagger}$ under the symmetry operation is what we
referred to as the monopole quantum number. 
As discussed in appendix \ref{subsec:slave-rotor-sym} the symmetry
of \Eq{lt} include rotation $\mathcal{R}_{90}$ and/or
$T\circ\mathcal{TR}$. If $e^{i\theta_i}=1\forall i$ (ferromagnetic)
$V_{D}^{\dagger}$ would be invariant under $\mathcal{R}_{90}$. On
the other hand if $e^{i\theta_i}=\eta_i$ (antiferromagnetic)
$V_{D}^{\dagger}$ would change sign under rotation. Thus the
transformation property (or the quantum number) of $V_D^\dagger$
crucially depends on the phase $e^{i\theta_i}$ (or more precisely on
the relative phase $e^{i(\theta_i-\theta_j)}$ for $i\in$ A or B
sublattices. 

How to compute the relative phases? In the following we will show
that 
\begin{align}
 e^{i(\theta_j-\theta_i)}=\Big(\frac{\langle GS,
 \mbox{1-flux}|PS_{i}^+P|GS,\mbox{0-flux}\rangle}{\langle GS,
 \mbox{1-flux}|PS_j^+P|GS,\mbox{0-flux}\rangle}\Big),\label{eq:relative-phase}
\end{align}
where $|GS,\mbox{0-flux}\rangle$ is the spinon mean-field
wavefunction of \Eq{TBI} in zero gauge flux and
$|GS,\mbox{1-flux}\rangle$ is the mean-field wavefunction in the
background of a uniform flux  integrated to one flux quantum. $P$ is
the operator that executes the projection onto the Hilbert space of
one-fermion-per-site. This formula of computing the monopole quantum numbers is firstly proposed by one of the authors in Ref.\cite{ran-2007-kagome-monopole} to address the similar issue of the monopole quantum number in a U(1) Dirac spin liquid on a Kagome lattice. One way to understand this formula, is to
consider the wavefunction of a monopole condensed phase. Clearly, it
is a superposition of different fluxes $|M\rangle \cong
P|\mbox{0-flux}\rangle + \alpha P|\mbox{1-flux}\rangle +\alpha^*
P|\mbox{-1 flux}\rangle$. Hence the expectation value of the spin
operator is given by $\langle M|S^+_r |M\rangle \sim \langle
\mbox{0-flux}|P S^+_r P |\mbox{1-flux}\rangle$.
\begin{figure}
 \includegraphics[scale=0.25]{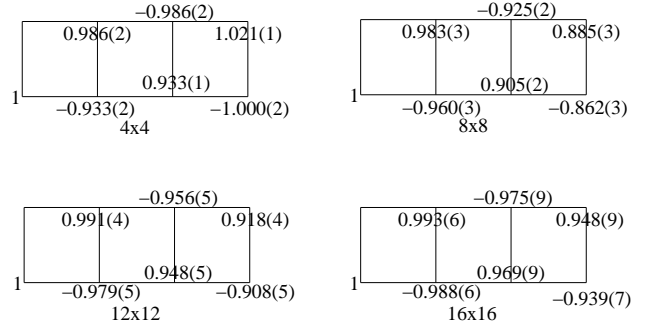}
\caption{The numerical result for \Eq{eq:relative-phase} for
$4\times 4, 8\times 8, 12\times 12$ and $16\times 16$ lattices under
periodic boundary condition. We fix $i$ to be the bottom left site
and show the result as $j$ goes through six different sites as
shown. In the limit of large system size the magnitude of the ratio
approaches one and the phases factor is $1$ for i,j on the same
sublattice and $-1$ for the opposite sublattices. In constructing
this fiture we choose parameters $\Phi=0.6\pi$, $|t/\chi|=0.2$ in
\Eq{TBI}. The relative phase factor are found to be independent of
the parameter values.} \label{fig:mono}
\end{figure}

We have performed numerical calculation of the relative phases in
\Eq{eq:relative-phase}. Specifically, we fix $i$-site to be an
arbitrary site on the square lattice and compute $\theta_j-\theta_i$
as $j$ varies through the sites of a finite lattice. (We employ the
periodic boundary condition in this calculation.) 
The result for different lattice sizes are illustrated in
Fig.(\ref{fig:mono}).  We find $e^{i(\theta_j-\theta_i)}$ is $1$ if
$i$ and $j$ are on the same sublattice and $-1$ if $i$ and $j$ are
on different sublattices. Moreover the result is independent of the
value of $\chi$ and $t$. This proves that the magnetic order
described by \Eq{lt} is {\em anti-ferromagnetic}.

In the following we present a discussion of why
Eq.(\ref{eq:relative-phase}) is the right quantity to calculate. As
discussed in appendix \ref{subsec:slave-rotor-sym} symmetry
operation is manifested in gauge theory modulo gauge transformation.
In other words, any change of the $f$-Hamiltonian in \Eq{lt} induced
by a transformation that can be undone by a gauge transformation
will preserve the gauge theory defined in \Eq{lt}. The collection of
such compounded transformation form the so-call projective symmetry
group (PSG)\cite{PhysRevB.65.165113}. 
However, unlike the physical symmetry operations, the elements of
the PSG depends on which gauge the mean-field $f$-Hamiltonian is
written. Certain gauge are particularly convenient (in the sense
that the accompanied gauge transformation is trivial) for some
physical symmetry operations (but not for others). However, so long
as two gauges are related by a gauge transformation \be f_{i\s}\ra
e^{i\phi_i}f_{i\s},~~{\rm where~~}\prod_i e^{i\phi_i}=1\ee the
corresponding PSG operation will yield the same result when acting
upon a spin state (i.e., a spinon state with one particle per site).
This degrees of freedom works in our favor when we try to determine
the symmetry properties of spin state. For example, it allows us to
choose the ``best gauge'' for each different symmetry operation.
Thus in order to determine the symmetry property of a spin state
(written in spinon variables) one can adapt the ``best gauge'' for
each symmetry operations.

Now we are ready to define the monopole quantum number. Let 
$|\Psi,\mbox{0-flux}\rangle$ and $|\Psi,\mbox{1-flux}\rangle$ be the
ground spin state (written in
terms of the spinon variables) 
with zero and one quantum of background flux, respectively. 
(By definition
$|\Psi,\mbox{1-flux}\rangle=V^{\dagger}|\Psi,\mbox{0-flux}\rangle$.)
The quantum number of  $V^{\dagger}$ is determined by the relative
transformation properties of $|\Psi,\mbox{1-flux}\rangle$ and
$|\Psi,\mbox{0-flux}\rangle$ under the elements of PSG (hopefully
under the best gauge). Let us again use rotation $\mathcal{R}_{90}$
as an example. If $|\Psi,\mbox{0-flux}\rangle$ is invariant but
$|\Psi,\mbox{1-flux}\rangle$ acquires a phase factor $e^{i\phi}$
under the appropriate PSG operation
\begin{align}
 \mathcal{R}_{90}|\Psi,\mbox{0-flux}\rangle&=|\Psi,\mbox{0-flux}\rangle\notag\\
\mathcal{R}_{90}|\Psi,\mbox{1-flux}\rangle&=e^{i\phi}|\Psi,\mbox{1-flux}\rangle,\label{eq:mono-def}
\end{align}
then by definition $V^{\dagger}$ has a rotational quantum number
$e^{i\phi}$. Because $P|GS,\mbox{0-flux}\rangle$ and
$P|GS,\mbox{1-flux}\rangle$ also describes state with one spinon per
site, they also enjoy the best gauge freedom discussed earlier. It
is also important to point out that although
$P|GS,\mbox{0-flux}\rangle$ and $P|GS,\mbox{1-flux}\rangle$ are not
the exact states $|\Psi,\mbox{0-flux}\rangle$ and
$|\Psi,\mbox{1-flux}\rangle$. So long as they have the same PSG
transformation property as them, Eq.(\ref{eq:relative-phase}) (in
the thermodynamic limit) will yield the exact monopole quantum
number. Again using $\mathcal{R}_{90}$ as an example we should have
\begin{align}
&\frac{\langle GS,\mbox{1-flux}|PS_{i}^+P|GS,\mbox{0-flux}\rangle}{\langle GS,\mbox{1-flux}|PS_{\mathcal{R}_{90}(i)}^+P|GS,\mbox{0-flux}\rangle}\notag\\
=&\frac{\langle GS,\mbox{1-flux}|PS_{i}^+P|GS,\mbox{0-flux}\rangle}
{\langle GS,\mbox{1-flux}|P\mathcal{R}_{90}^{-1}S_{i}^+
\mathcal{R}_{90}P|GS,\mbox{0-flux}\rangle}.
\end{align}
The advantage of $P|GS,\mbox{0-flux}\rangle$ and
$P|GS,\mbox{1-flux}\rangle$ are that their PSG transformation
properties can be be read off from those of
$|GS,\mbox{0-flux}\rangle$ and  $|GS,\mbox{1-flux}\rangle$.

Even so determine the transformation properties of
$|GS,\mbox{0-flux}\rangle$ and  $|GS,\mbox{1-flux}\rangle$ can be
quite involved.
For example if we want to obtain the transformation property of
$|GS,\mbox{0-flux}\rangle$ under $\mathcal{R}_{90}$, one can choose
a gauge (the best gauge) in which the the spinon hopping matrix
elements are fully translation invariant. In this gauge we expect
$|GS,\mbox{0-flux}\rangle$ to be rotation symmetric.
\begin{figure}
 \includegraphics[width=0.18\textwidth]{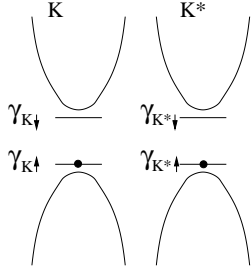}
\caption{The in-gap landau levels when a single quantum of magnetic
flux is added to \Eq{TBI}. $K$ and $K^*$ mark the momentum space
location of the Dirac points when the second neighbor hopping is
switched off. For up-spinons the Landau level are close to the top
of the valence band, and for down spinons they are close to the
bottom of the conduction band. In the depicted filling of Landau
levels, the total $S_z=1$.} \label{fig:landau-level}
\end{figure}
However to figure out the symmetry properties of
$|GS,\mbox{1-flux}\rangle$, we face a more challenging situation.
For example it is impossible to choose a gauge which preserves the
rotation symmetry on torus (because the fluxes through the torus
holes are gauge invariant and breaks rotation). Under open boundary
condition one can choose the symmetric gauge $\vec
A=\frac{1}{2}B\hat z\times \vec r$ which is rotation invariant.
Under this (best) gauge it is possible to analytically study the
transformation law of $|GS,\mbox{1-flux}\rangle$ under rotation.
More specifically, due to the topological nature of the spinon
mean-field Hamiltonian, one can show that in the presence of one
quantum of uniform flux, there are two in-gap Landau levels near the
top of the valence band for the up spinons and another two near the
bottom of the conduction band for the down spinons. This is
illustrated in Fig.\ref{fig:landau-level}. Because these in-gap
levels have definite angular momentum in the symmetric gauge an
analytical study of the rotation quantum number is possible. In
numerical study of Eq.(\ref{eq:relative-phase}) one does not need to
worry about any of the above; one simply computed in the ratio in
the any gauge. We have done just that for the results presented in
Fig.\ref{fig:mono}.

\section{Some small sections}\label{sec:sc-more}

\subsection{Microscopic symmetries and action Eq.(\ref{eft})}\label{subsec:sym-eft}
As a double check we show in the following that
the validity of action Eq.(\ref{eft}) is ensured by microscopic
symmetries. We should check that under
symmetry transformations (1) $z_1\rightarrow z_2$ (2) $z\rightarrow
z^{\dagger}$ (which ensures theory to be relativistic.), and (3)
$z_1$ and $z_2$ particles do not experience any background magnetic
field. In the AF phase the spin $S_z$ rotation is broken and
charge-$U(1)$, $PH$, $T\circ\mathcal{TR}$ remains good symmetries.
Due to the antiferromagnetic order, reflection $\mathcal{P}$ has to
be followed by a $\pi$ spin $S_z$ rotation, i.e., $\mathcal{P}\ra
e^{i\pi S_z}\circ\mathcal{P}$ so that the ground state will be
preserved.

$PH$ transforms a spin vortex into a vortex and flip the charge,
therefore
\be
 \mbox{$PH$:~}
z_1\rightarrow z_2,~z_2\rightarrow z_1\label{eq:xy-ph} \ee Under
$e^{i\pi S_z}\circ\mathcal{P}$ a vortex is transformed into an
anti-vortex while preserving the charge, thus we have \be
 \mbox{$e^{i\pi S_z}\circ\mathcal{P}$:~}
z_1\rightarrow z_2^{\dagger},~z_2\rightarrow
z_1^{\dagger}\label{eq:xy-p}\ee Eq.(\ref{eq:xy-ph},\ref{eq:xy-p})
ensures condition (1) and (2). $T\circ \mathcal{TR}$ transforms a
vortex into a vortex while preserving the charge:
\be
\mbox{$T\circ \mathcal{TR}$: }
z_1\rightarrow z_1&& z_2\rightarrow z_2\label{eq:xy-tr} \ee Although
$z_\alpha$ transform trivially under $T\circ \mathcal{TR}$, it is
important to bare in mind that because it is an anti-unitary
transformation it changes sign of $a_\mu$ in \Eq{eft}. This ensures condition (3).

\subsection{The edge modes of the
topological superconductor}\label{subsec:pseudospin}

Since the rotation Eq.(\ref{trans}) preserves the spectrum of the
Hamiltonian, we can use the existence of gapless edge states in TBI
to imply the same for the SC. Moreover since 
the rotation preserves the spin, we conclude that, as in TBI, the
edge modes in the superconductor must have opposite spin propagate
in opposite direction. However unlike the edge mode in TBI, that of
the superconductor do not carry definite charge.

It is important to ask under what condition are these edge modes
stable. In the TBI phase it is the charge and spin conservation law
that guarantee the stability. 
Translated into the SC, spin conservation law remains the same but
the charge conservation law becomes the conservation of
$\sum_i\eta_i(c_{i\uparrow}c_{i\downarrow}+h.c.)$,
does not seem to be a physical conservation law for a realistic
system. Fortunately it turns out that the particle-hole symmetry plus the 
$S_z$ conservation are sufficient to guarantee the stability of the
gapless edge modes in the SC phase.

This is easily
seen as follows. Particle hole symmetry is implemented by
$\psi_r\rightarrow i\mu_x \psi_r$. For convenience, let us translate
this into the spinon variables where it implies $F_r \rightarrow Z_r
i\mu_xZ^\dagger_rF_r$, which is $F_r \rightarrow i\mu_z F_r$ in a
particular gauge. If the quadratic Hamiltonian of the $F_r$ is
invariant under this transformation, then it should only depend on
the identity and $\mu_z$. This implies it has a U(1) rotation
symmetry corresponding to spinon number conservation. Combined with
U(1) spin rotation invariance, this is sufficient to protect the
edge modes.

\subsection{ Physical Interpretation of the doublet of Chargeons}\label{subsec:doublet-chargeon}
 Since the
preceeding justification for introducing a charged boson doublet is
rather formal, phrased in terms of gauge theories and symmetries, we
provide here a more physical motivation for the appearance of the
pair of fields $z_1,\,z_2$. Consider an insulating state composed of
local moments described by a fermionic mean field theory as in
\ref{MF}. Consider introducing charge fluctuations by including
electron excitations $c^\dagger_r$. In a single band model, these
electrons or holes will form a singlet with the spin on that site.
Hence, if we consider low energy charge excitations which are spin
singlets, then one can construct two different operators $b_1 =
c^\dagger_\sigma f_\sigma$ and $b^\dagger_2= {\bf \epsilon}
_{\sigma\sigma'}c^\dagger_\sigma f^\dagger_{\sigma'}$, where ${\bf
\epsilon}_{\sigma\sigma'}$ is the antisymmetric symbol and a sum on
spin indices is assumed. Keeping in mind the constraint
\ref{constraint} one can readily show that these bosonic operators
satisfy:
\begin{align}
c_{\uparrow}&=b_{1}f_{\uparrow}-f_{\downarrow}^{\dagger}b_{2}^{\dagger}\nonumber
\\
c_{r\downarrow}&=b_{1}f_{r\downarrow}+f_{\uparrow}^{\dagger}b_{2}^{\dagger}
\end{align}
clearly these are very similar to the $z_1,\,z_2$ fields. Moreover,
one can check that the on site Cooper pair operator: ${\bf
\epsilon}_{\sigma\sigma'} c^\dagger_\sigma c^\dagger_{\sigma'} = b_1
b^\dagger_2$. Hence, one can consider these operators as a
'Schwinger boson' representation of the Cooper pair.

\section{XY$\leftrightarrow$SC direct transition on the honeycomb lattice}\label{sec:honeycomb}
From the vortex core zero mode study (see Sec.\ref{sec:SC}$\sim$\ref{sec:transition-from-af}) we found there is a direct transition from AF phase to d-wave SC phase on square lattice. However that approach does not seem to have a predicting power. For example it is unclear how to generalize the square lattice results to a different lattice. On the other hand the slave-rotor theory (see Sec.\ref{slave rotor}) can reproduce all those results in a systematical way, and it is obviously can be applied to other lattices. In this section we will apply the slave-rotor theory on honeycomb lattice, following exactly the same route as in Sec.\ref{slave rotor}, and discuss its XY$\leftrightarrow$SC direct transition.

We again firstly present the minimal requirement of symmetries for microscopic
models containing the XY$\leftrightarrow$SC direct transition. In
summary the symmetries are: charge-$U(1)$, $S_z-U(1)$, $PH$ (particle-hole), $\mathcal{P}$ (reflection of the mirror along the diagonal direction of the hexagonal plaquette) and $\mathcal{TR}$ ($\mathcal{TR}$ is time-reversal).
These transformations affect the electron operators as follows:
\begin{align}
 \mbox{charge-$U(1)$: }& c_{i\uparrow}\rightarrow e^{i\theta}c_{i\uparrow}&&c_{i\downarrow}\rightarrow e^{i\theta}c_{i\downarrow}\notag\\
 \mbox{$S_z-U(1)$: }& c_{i\uparrow}\rightarrow e^{i\theta/2}c_{i\uparrow}&&c_{i\downarrow}\rightarrow e^{-i\theta/2}c_{i\downarrow}\notag\\
\mbox{$PH$: }& c_{i\uparrow}\rightarrow \eta_i c_{i\downarrow}^{\dagger}&&c_{i\downarrow}\rightarrow -\eta_i c_{i\uparrow}^{\dagger}\notag\\
\mbox{$\mathcal{TR}$: }& c_{i\uparrow}\rightarrow c_{i\downarrow}&&c_{i\downarrow}\rightarrow -c_{i\uparrow}.\notag\\
\mbox{$\mathcal{P}$: }& c_{i\uparrow}\rightarrow  c_{\mathcal{P}(i)\uparrow}&&c_{i\downarrow}\rightarrow  c_{\mathcal{P}(i)\downarrow}\label{eq:honeycomb-sym-def}
\end{align}

To describe the XY magnetic ordered phase on the honeycomb lattice by slave-rotor theory, we again assume that the bosonic chargon $z$ are fully gapped, and the fermionic spinon $f$ band structure to be TBI\cite{kane:226801}:
\begin{align}
 H^{TBI}_f&=-t\sum_{\langle i,j\rangle}f_{i\sigma}^{\dagger}f_{j\sigma}+i\sum_{\langle\langle i,j\rangle\rangle}\lambda_{i,j}\left(f_{i\uparrow}^{\dagger}f_{j\uparrow}-f_{i\downarrow}^{\dagger}f_{j\downarrow}\right)\notag\\
&+h.c.,\label{eq:TBI-bs}
\end{align}
where $\lambda_{ij}=\lambda\frac{\hat d_{jl}\times \hat d_{li}\cdot \hat z}{|\hat d_{jl}\times \hat d_{li}|}\equiv \nu_{ij}\lambda$, and $\hat d_{jl}$,$\hat d_{li}$ are the two bonds connecting site $j$ and $i$, $\nu_{ij}=\pm1$. This band structure is the same as the Kane-Mele spin-hall insulator\cite{kane:226801} if one replaces the spinon operator $f$ by electron operator $c$. But the difference is that here the spinon $f$ automatically couples with a dynamical 2+1D $U(1)$ gauge field. Due to the ``spin-hall" effect of the TBI band structure, we learn that the photon mode of the $U(1)$ gauge field is actually the Goldstone mode of the XY magnet. Similar to the square lattice case, the low energy effective theory of the XY magnet has the form of Eq.(\ref{eft}). One can show that the microscopic symmetries Eq.(\ref{eq:honeycomb-sym-def}) ensure the correctness of this low energy effective theory.

\begin{figure}
\bigskip
 \includegraphics[width=0.4\textwidth]{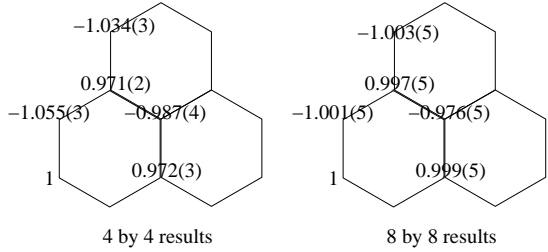}
\caption{We present the numerical result of the ratio of matrix elements $\frac{\langle GS,\mbox{1-flux}|PS_{j}^+P|GS,\mbox{0-flux}\rangle}{\langle GS,\mbox{1-flux}|PS_i^+P|GS,\mbox{0-flux}\rangle}$ for the projected TBI wavefunction on 4 by 4 and 8 by 8 unit cell lattices on torus. We always fix site $i$ to be the left-bottom one and show the result of site $j$ on figure. The $P|GS,\mbox{1-flux}\rangle$ breaks physical translations and rotations since there are fluxes in the torus holes, and thus the magnitude of this ratio is not exactly one. But in the thermodynamic limit where the boundary condition is irrelevant, the magnitude of the ratio should approaches one, which seems to happen if one compares the 4 by 4 result and 8 by 8 result. The relative phases of the matrix element is exactly $0$ for same sublattice and $\pi$ for opposite sublattices. We choose hopping parameters $\lambda/t=0.1$ in these calculations, but the relative phases are found to be independent of choice of the parameters's value.}
\label{fg:honeycomb-monopole-phases}
\end{figure}

The issue is to determine the XY and SC order patterns. We first study the XY order pattern. Is it ferromagnetic or anti-ferromagnetic? To answer this question we again perform two calculations both showing the order is AF. First we computed the matrix element ratio to study the quantum number of monopoles of the $U(1)$ gauge field (see Sec.\ref{sec:monopole}). We find the magnetic order is anti-ferromagnetic and in Fig. \ref{fg:honeycomb-monopole-phases} we presented the matrix element ratio results.

\begin{figure}
 \includegraphics[width=0.4\textwidth]{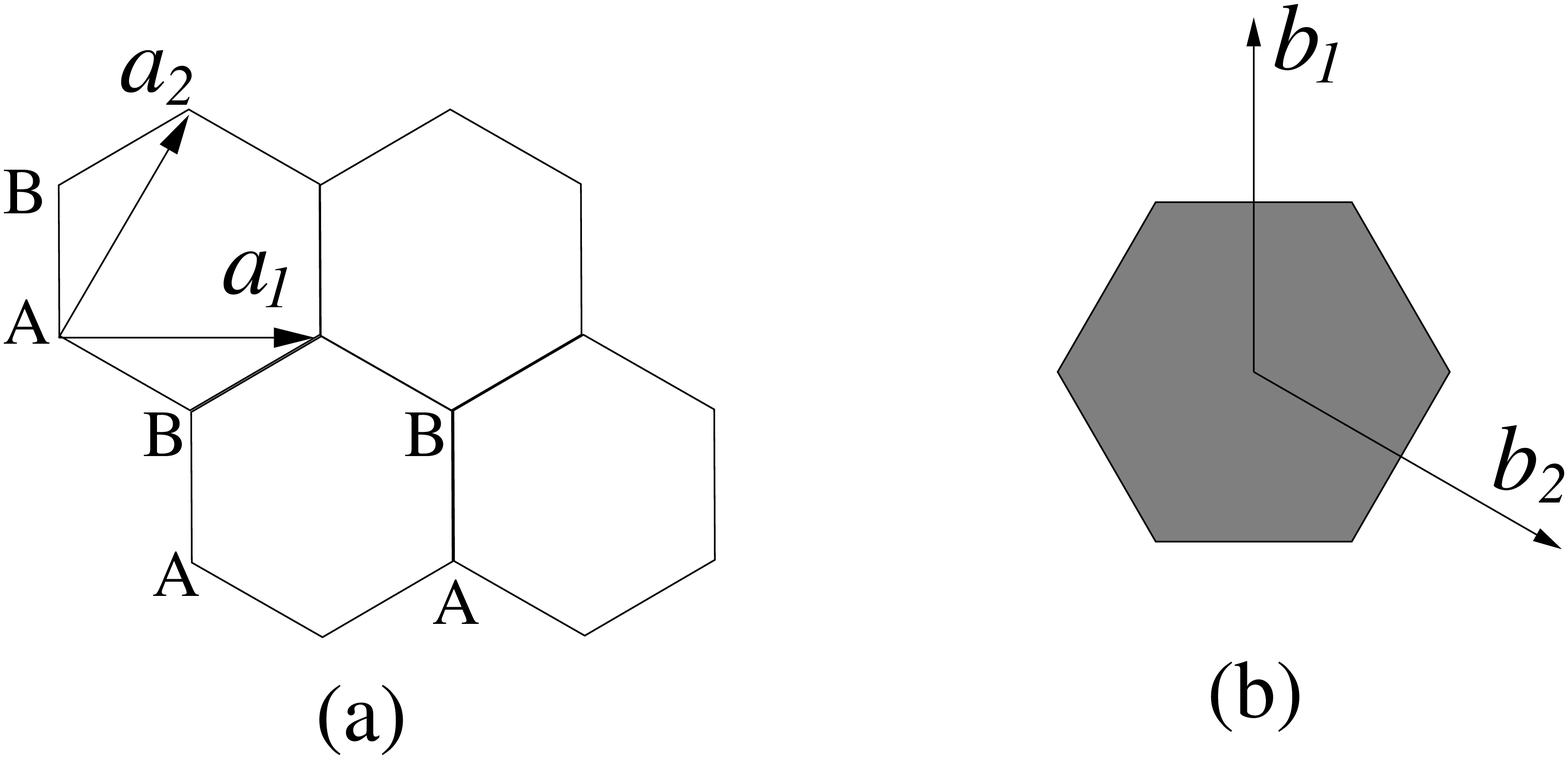}
\caption{(a) The honeycomb lattice in real space. The two sublattices are labeled by $A$ and $B$. $\boldsymbol{a}_1$ and $\boldsymbol a_2$ are the two unit cell basis vectors. (b) The Brillouin zone of the honeycomb lattice. The vectors $\boldsymbol b_1$ and $\boldsymbol b_2$ are the two basis of the reciprocal lattice satisfying $\boldsymbol a_i\cdot \boldsymbol b_j=\delta_{ij}$.}
\label{fg:honeycomb}
\end{figure}

Secondly we directly measure the spin-spin correlation function of the projected TBI wavefunction. We label the two sites in one honeycomb unit cell by $A$ and $B$ as shown in Fig.\ref{fg:honeycomb}, and define $|\Psi_0\rangle=P|GS,\mbox{0-flux}\rangle$. The correlation functions that we measure are:
\begin{align}
C^{AA}(\vec r)&=\langle  S_A^x(\vec r)S_A^x(0)+S_A^y(\vec r)S_A^y(0)\rangle_{\Psi_0}, \notag\\
C^{AB}(\vec r)&=\langle  S_B^x(\vec r)S_A^x(0)+S_B^y(\vec r)S_A^y(0)\rangle_{\Psi_0}.
\end{align}
Because the projected wavefunction is the simplest way to mimic the effect of the gauge fluctuations (i.e., the one-fermion-per-site constraint), we expect that this correlation function to show AF order.

\begin{figure}
 \includegraphics[width=0.23\textwidth]{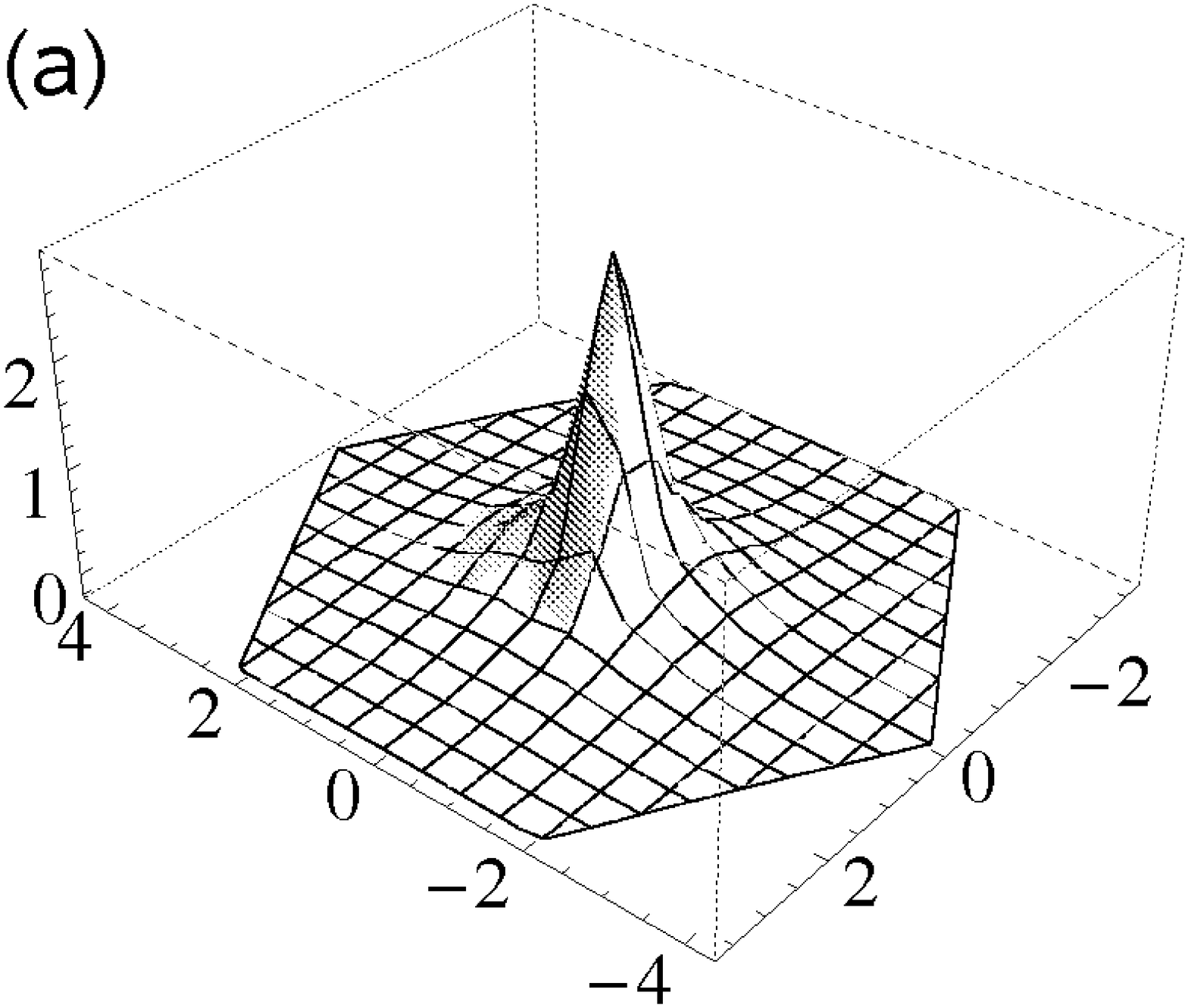}\includegraphics[width=0.23\textwidth]{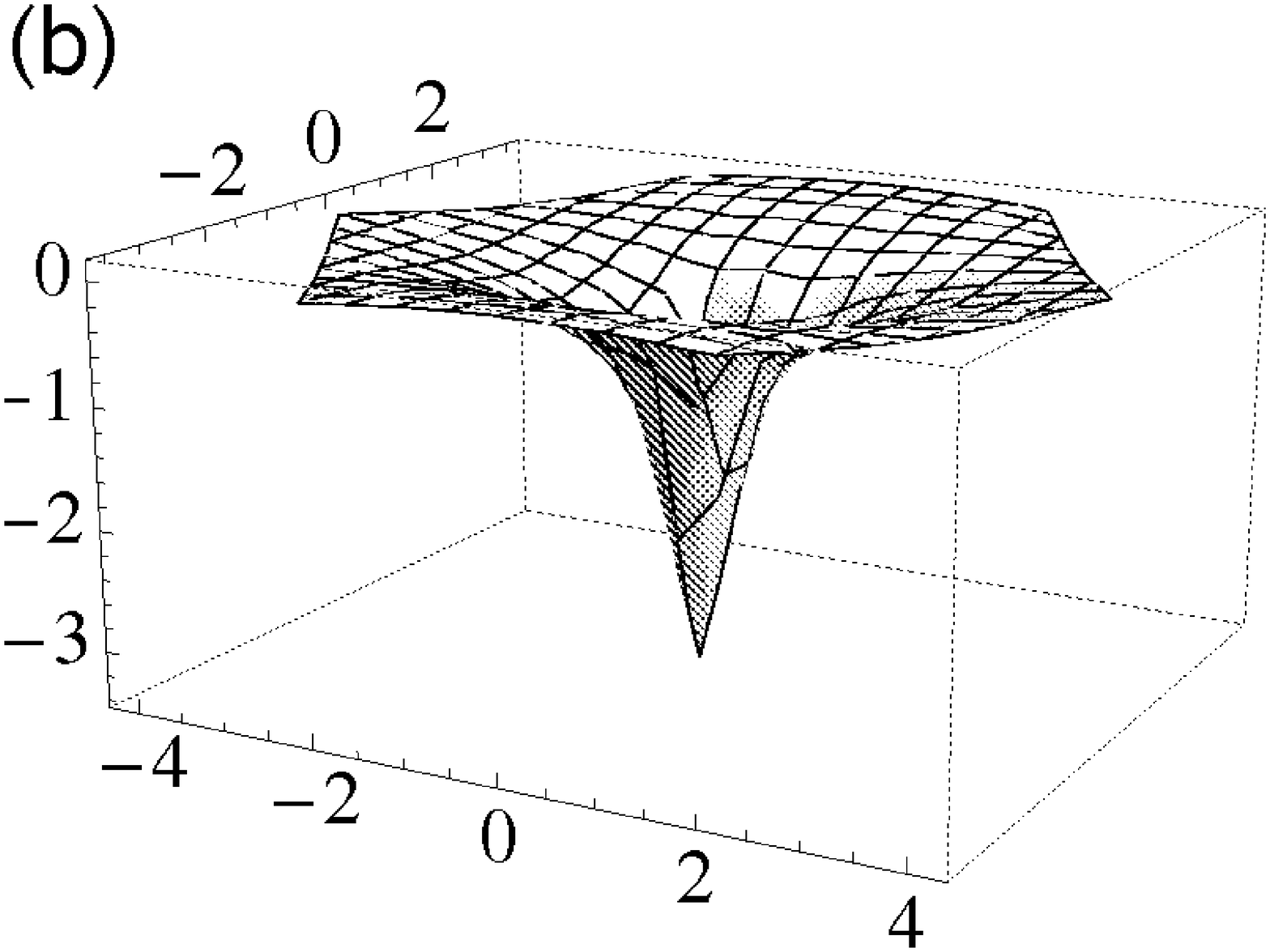}
\caption{For 8 by 8 unit cells, we fix hopping parameters $t=1$, $\lambda=0.1$ and plot Fourier transform of the real space spin spin correlation function $\langle S^x(r)S^x(0)+S^y(r)S^y(0)\rangle$ of the projected wavefunction in the hexagonal Brillouin zone of the honeycomb lattice. (a) is the correlation function of the same sublattice $C^{AA}$. (b) is the correlation function of different sublattices $C^{AB}$. The large $q=0$ positive peak for $C^{AA}$ and negative peak for $C^{AB}$ is a strong evidence that this projected wavefunction is AF ordered.}
\label{fg:corr-fft}
\end{figure}

We firstly present the projected wavefunction result for a fixed lattice size, namely 8 by 8 unit cells (see Fig.\ref{fg:corr-fft}). The momentum space peak at $q=0$ for Fourier transform of the XY spin correlation (positive for $C^{AA}$ and negative for $C^{AB}$) is a strong indication that the magnetic order is AF.

Next we present the finite size scaling study. When $\lambda/t=0$, the projected wavefunction has two Dirac nodes and is expected to show power law scalings of the spin correlation function. We choose periodic boundary condition for difference system sizes $L=8,10,14,16,20,22$ (Note that $6N\times 6N$ lattice has nodal fermions, namely fermion modes right at the nodes. Thus we  avoid those lattice sizes.), and compute the spin correlation functions $C^{AA}$ and $C^{AB}$ at the half way of the system sizes.

\begin{figure}
\begin{center}
 \includegraphics[width=0.4\textwidth]{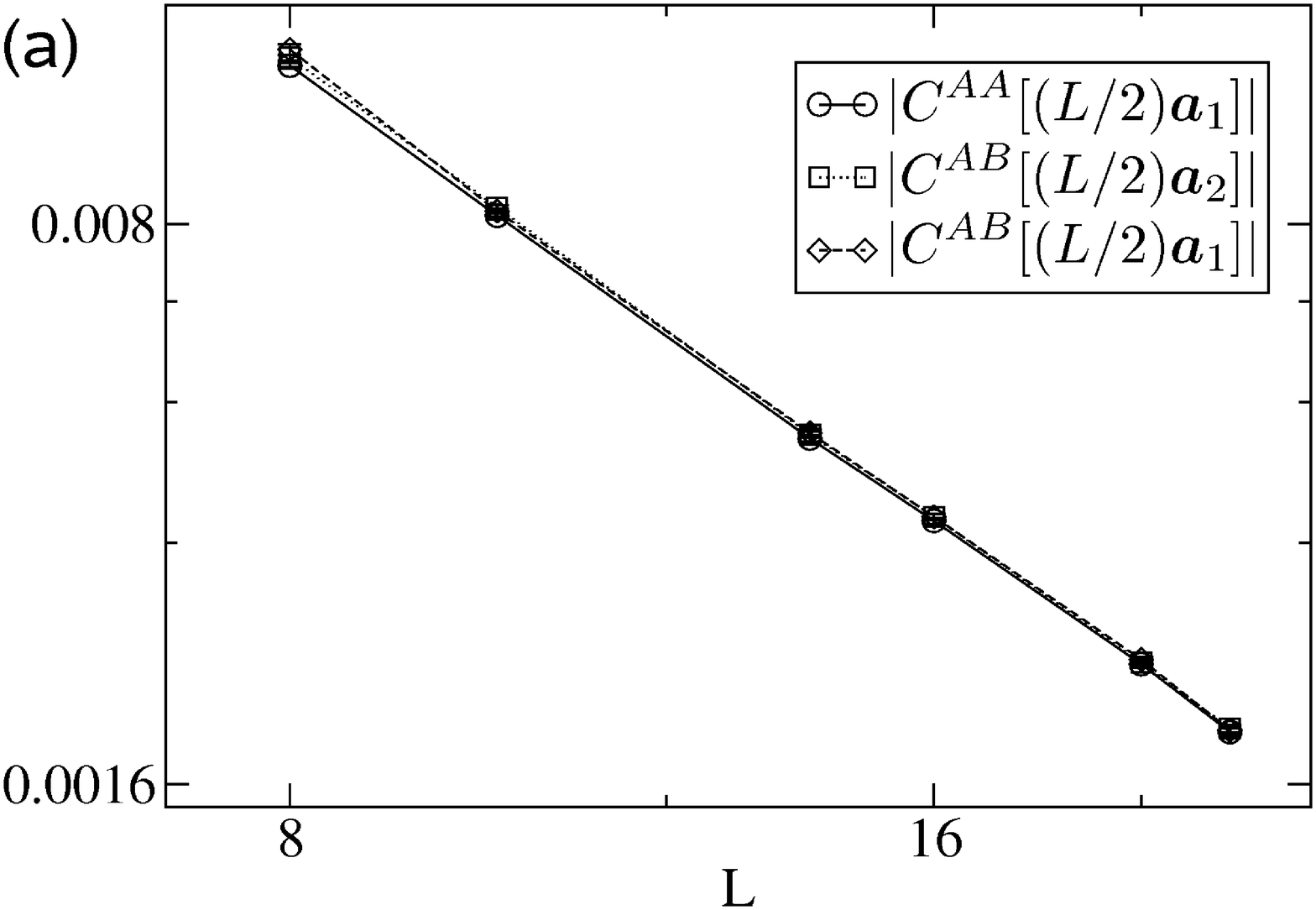}\\
\smallskip
\includegraphics[width=0.45\textwidth]{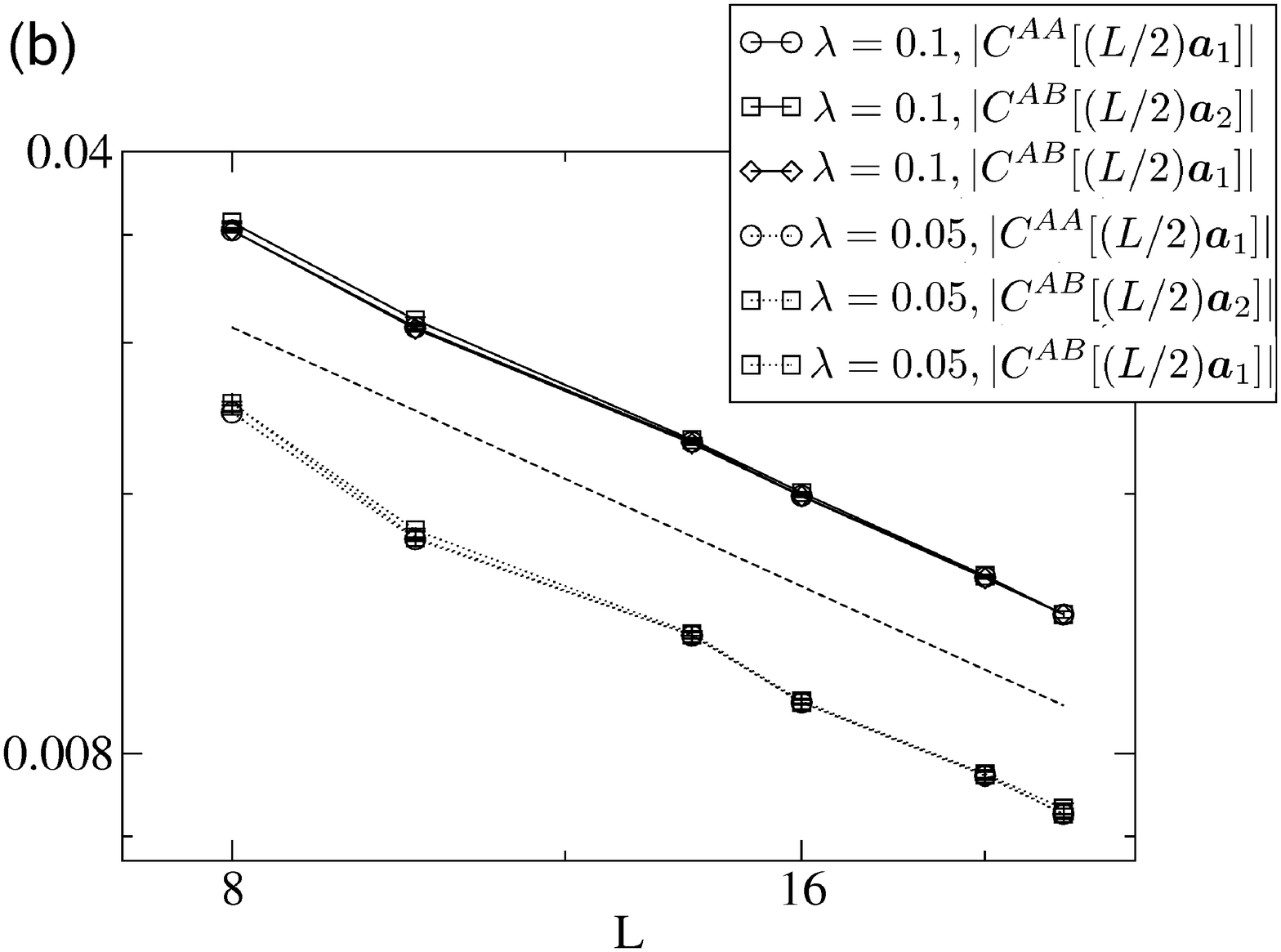}
\end{center}
\caption{We present the log-log plot of the finite size scaling of the spin-spin correlation functions for parameters $t=1$ and (a) $\lambda=0$ and (b) $\lambda=0.05$ and $0.1$. From (a), we find when $\lambda/t=0$ the scaling power law is $|C|\propto \frac{1}{L^{1.89(2)}}$. From (b) we also find power-law scalings for $\lambda/t=0.05$ and $\lambda/t=0.1$, and the exponents for these two cases are the same up to statistical error, which is $C\propto \frac{1}{L^{1.00(2)}}$. The dashed line in (b) is the plot of $C=\frac{0.2}{L}$.}
\label{fg:scaling}
\end{figure}

We checked that the projected wavefunctions that we used respect the full lattice space group symmetry. By the lattice space group symmetry, we find the three spin correlations at half system sizes for $C^{AA}$, namely $C^{AA}(\frac{L\boldsymbol{a}_1}{2})$, $C^{AA}(\frac{L\boldsymbol{a}_2}{2})$ and $C^{AA}(\frac{L(\boldsymbol{a}_1+\boldsymbol{a}_2)}{2})$, are related by symmetry transformation and thus identical. Here $\boldsymbol{a}_1$ and $\boldsymbol{a}_2$ are the real space unit cell basis of the honeycomb lattice as shown in Fig.\ref{fg:honeycomb}. On the other hand for $C^{AB}$ we find $C^{AB}(\frac{L\boldsymbol{a}_2}{2})$ and $C^{AB}(\frac{L(\boldsymbol{a}_1+\boldsymbol{a}_2)}{2})$ are identical, but $C^{AB}(\frac{L\boldsymbol{a}_1}{2})$ are different from them. Therefore we present the results of $C^{AA}(\frac{L\boldsymbol{a}_1}{2})$, $C^{AB}(\frac{L\boldsymbol{a}_2}{2})$ and $C^{AB}(\frac{L\boldsymbol{a}_1}{2})$ in Fig.\ref{fg:scaling}(a). (Note that $C^{AA}$ data are positive and $C^{AB}$ data are negative, and we present the absolute values only.) We find
\begin{align}
 \mbox{$\lambda/t=0$: }C(\vec r)\propto\frac{1}{|\vec r|^{1.89(2)}}.
\end{align}

Now we turn on $\lambda$. Based on the low energy effective theory (see text) we know the ground state should be long-range XY spin ordered. But by studying the projected wavefunction, we find the spin correlation function still have power law scaling. The scaling exponent, however, is strongly modified from the $\lambda=0$ value (see Fig.\ref{fg:scaling}(b)). We find
\begin{align}
 \mbox{$\lambda/t=0.05,0.1$: }C(\vec r)\propto\frac{1}{|\vec r|^{1.00(2)}}.
\end{align}
This means first the projected wavefunction fails to produce the true long-range correlation, and second the AF correlation is strongly enhanced with $\lambda\neq 0$ projected wavefunction compared with the $\lambda=0$ one.

Now we study the SC order pattern. This is straightforward because we should simply condense $z_1$ and $z_2$ together. Deep in the SC phase the easy-plane condensation of $z$-boson can be achieved by doing replacement $z_1=\frac{1}{\sqrt{2}}e^{i\theta_1}$ and  $z_2=\frac{1}{\sqrt{2}}e^{i\theta_2}$ in Eq.~(\ref{eq:rotor-U(1)}):
\begin{align}
  c_{i\uparrow}&=\frac{1}{\sqrt{2}}(f_{i\uparrow}e^{i\theta_1}-\eta_{i}f_{i\downarrow}^{\dagger}e^{-i\theta_2})\\
c_{i\downarrow}&=\frac{1}{\sqrt{2}}(f_{i\downarrow}e^{i\theta_1}+\eta_if_{i\uparrow}^{\dagger}e^{-i\theta_2}).
\end{align}
We can express $f$ fermions in terms of the electron operator $c$:
\begin{align}
f_{i\uparrow}&=\frac{1}{\sqrt{2}}(c_{i\uparrow}e^{-i\theta_1}+\eta_i c_{i\downarrow}^{\dagger}e^{-i\theta_2})\\
f_{i\downarrow}&=\frac{1}{\sqrt{2}}(c_{i\downarrow}e^{-i\theta_1}-\eta_i c_{i\uparrow}^{\dagger}e^{-i\theta_2})
\end{align}
Plugging these into Eq.(\ref{eq:TBI-bs}), the superconductor Hamiltonian is:
\begin{align}
 H^{SC}=&t\sum_{\langle i,j\rangle}c_{i\alpha}^{\dagger}c_{j\alpha}+i\lambda\sum_{\langle\langle i,j\rangle\rangle}\nu_{ij}\eta_ie^{i(\theta_2-\theta_1)}(c_{i\uparrow}c_{j\downarrow}+c_{i\downarrow}c_{j\uparrow})\notag\\
&+h.c.\label{eq:sc-ham}
\end{align}

If we absorb $e^{i(\theta_2-\theta_1)}$ into $\lambda$ so that $\lambda$ is complex, the full n.n.n. pairings are
\begin{align}
(-)^i\nu_{ij}\big[i\lambda(c_{i\uparrow}c_{j\downarrow}+c_{i\downarrow}c_{j\uparrow})+i\lambda^*(c_{i\uparrow}^{\dagger}c_{j\downarrow}^{\dagger}+c_{i\downarrow}^{\dagger}c_{j\uparrow}^{\dagger}),\big]\label{eq:pairing}
\end{align}
these are $f$-wave triplet pairings with $S_z=0$. Note that this $f$-wave SC is fully gapped (its spectrum is the same as TBI). Therefore we show that there is a direct AF$\leftrightarrow$$f$-wave SC transition on honeycomb lattice, whose low energy effective theory is again the easy-plane limit of NCCP$^1$.

Finally we remark that there are zero modes in the vortex cores of the AF and f-wave SC phases by solving the BdG equations. We already presented the mean-field Hamiltonian of the SC phase in Eq.(\ref{eq:sc-ham}). To write down the AF mean-field Hamiltonian one should study which terms are allowed by symmetry (in AF phase, the good symmetries are charge-$U(1)$, $PH$, $\mathcal{P}$ and $e^{i\pi S_z}\circ \mathcal{TR}$). The simplest AF mean-field Hamiltonian consistent with symmetries is:
\begin{align}
H^{XY}=t\sum_{\langle ij\rangle}c_{i\alpha}^{\dagger}c_{j\alpha}+\eta_i\sum_i(M^* c_{i\uparrow}^{\dagger}c_{i\downarrow}+M c_{i\downarrow}^{\dagger}c_{i\uparrow}).\label{eq:honeycomb-xy}
\end{align}
where $M$ is the on-site magnetization.

When the order parameters are small both Eq.(\ref{eq:sc-ham}) and Eq.(\ref{eq:honeycomb-xy}) are in the Dirac limit. The AF side BdG equation has already been solved by Herbut\cite{herbut:206404}. And one can show that the SC side BdG equation turns out to be the same as the AF side BdG equation after appropriate rotations. One can show that the whole analysis on the square lattice, namely zero modes$\rightarrow$vortex carry quantum number$\rightarrow$direct transition, also carries through on the honeycomb lattice and reproduce the same low energy effective theory as the one given by slave-rotor theory.

\bibliographystyle{apsrev}
\bibliography{/home/ranying/downloads/reference/simplifiedying}

\begin{thebibliography}{40}
\expandafter\ifx\csname natexlab\endcsname\relax\def\natexlab#1{#1}\fi
\expandafter\ifx\csname bibnamefont\endcsname\relax
  \def\bibnamefont#1{#1}\fi
\expandafter\ifx\csname bibfnamefont\endcsname\relax
  \def\bibfnamefont#1{#1}\fi
\expandafter\ifx\csname citenamefont\endcsname\relax
  \def\citenamefont#1{#1}\fi
\expandafter\ifx\csname url\endcsname\relax
  \def\url#1{\texttt{#1}}\fi
\expandafter\ifx\csname urlprefix\endcsname\relax\def\urlprefix{URL }\fi
\providecommand{\bibinfo}[2]{#2}
\providecommand{\eprint}[2][]{\url{#2}}

\bibitem[{\citenamefont{Senthil and Lee}(2005)}]{senthil:174515}
\bibinfo{author}{\bibfnamefont{T.}~\bibnamefont{Senthil}} \bibnamefont{and}
  \bibinfo{author}{\bibfnamefont{P.~A.} \bibnamefont{Lee}},
  \bibinfo{journal}{Phys. Rev. B} \textbf{\bibinfo{volume}{71}},
  \bibinfo{pages}{174515} (\bibinfo{year}{2005}).

\bibitem[{\citenamefont{Senthil et~al.}(2004)\citenamefont{Senthil, Vishwanath,
  Balents, Sachdev, and Fisher}}]{senthil-2004-303}
\bibinfo{author}{\bibfnamefont{T.}~\bibnamefont{Senthil}},
  \bibinfo{author}{\bibfnamefont{A.}~\bibnamefont{Vishwanath}},
  \bibinfo{author}{\bibfnamefont{L.}~\bibnamefont{Balents}},
  \bibinfo{author}{\bibfnamefont{S.}~\bibnamefont{Sachdev}}, \bibnamefont{and}
  \bibinfo{author}{\bibfnamefont{M.~P.~A.} \bibnamefont{Fisher}},
  \bibinfo{journal}{Science} \textbf{\bibinfo{volume}{303}},
  \bibinfo{pages}{1490} (\bibinfo{year}{2004}).

\bibitem[{\citenamefont{Read and Sachdev}(1990)}]{PhysRevB.42.4568}
\bibinfo{author}{\bibfnamefont{N.}~\bibnamefont{Read}} \bibnamefont{and}
  \bibinfo{author}{\bibfnamefont{S.}~\bibnamefont{Sachdev}},
  \bibinfo{journal}{Phys. Rev. B} \textbf{\bibinfo{volume}{42}},
  \bibinfo{pages}{4568} (\bibinfo{year}{1990}).

\bibitem[{\citenamefont{Balents et~al.}(2005)\citenamefont{Balents, Bartosch,
  Burkov, Sachdev, and Sengupta}}]{balents:144508}
\bibinfo{author}{\bibfnamefont{L.}~\bibnamefont{Balents}},
  \bibinfo{author}{\bibfnamefont{L.}~\bibnamefont{Bartosch}},
  \bibinfo{author}{\bibfnamefont{A.}~\bibnamefont{Burkov}},
  \bibinfo{author}{\bibfnamefont{S.}~\bibnamefont{Sachdev}}, \bibnamefont{and}
  \bibinfo{author}{\bibfnamefont{K.}~\bibnamefont{Sengupta}},
  \bibinfo{journal}{Phys. Rev. B} \textbf{\bibinfo{volume}{71}},
  \bibinfo{pages}{144508} (\bibinfo{year}{2005}).

\bibitem[{\citenamefont{Grover and Senthil}(2007)}]{grover:247202}
\bibinfo{author}{\bibfnamefont{T.}~\bibnamefont{Grover}} \bibnamefont{and}
  \bibinfo{author}{\bibfnamefont{T.}~\bibnamefont{Senthil}},
  \bibinfo{journal}{Phys. Rev. Lett.} \textbf{\bibinfo{volume}{98}},
  \bibinfo{pages}{247202} (\bibinfo{year}{2007}).

\bibitem[{\citenamefont{Burkov and Balents}(2005)}]{burkov:134502}
\bibinfo{author}{\bibfnamefont{A.~A.} \bibnamefont{Burkov}} \bibnamefont{and}
  \bibinfo{author}{\bibfnamefont{L.}~\bibnamefont{Balents}},
  \bibinfo{journal}{Phys. Rev. B} \textbf{\bibinfo{volume}{72}},
  \bibinfo{pages}{134502} (\bibinfo{year}{2005}).

\bibitem[{\citenamefont{Grover and Senthil}(2008)}]{grover:156804}
\bibinfo{author}{\bibfnamefont{T.}~\bibnamefont{Grover}} \bibnamefont{and}
  \bibinfo{author}{\bibfnamefont{T.}~\bibnamefont{Senthil}},
  \bibinfo{journal}{Phys. Rev. Lett.} \textbf{\bibinfo{volume}{100}},
  \bibinfo{pages}{156804} (\bibinfo{year}{2008}).

\bibitem[{\citenamefont{Assaad et~al.}(1996)\citenamefont{Assaad, Imada, and
  Scalapino}}]{PhysRevLett.77.4592}
\bibinfo{author}{\bibfnamefont{F.~F.} \bibnamefont{Assaad}},
  \bibinfo{author}{\bibfnamefont{M.}~\bibnamefont{Imada}}, \bibnamefont{and}
  \bibinfo{author}{\bibfnamefont{D.~J.} \bibnamefont{Scalapino}},
  \bibinfo{journal}{Phys. Rev. Lett.} \textbf{\bibinfo{volume}{77}},
  \bibinfo{pages}{4592} (\bibinfo{year}{1996}).

\bibitem[{\citenamefont{Motrunich and Vishwanath}(2004)}]{PhysRevB.70.075104}
\bibinfo{author}{\bibfnamefont{O.~I.} \bibnamefont{Motrunich}}
  \bibnamefont{and}
  \bibinfo{author}{\bibfnamefont{A.}~\bibnamefont{Vishwanath}},
  \bibinfo{journal}{Phys. Rev. B} \textbf{\bibinfo{volume}{70}},
  \bibinfo{pages}{075104} (\bibinfo{year}{2004}).

\bibitem[{\citenamefont{Weng et~al.}(1998)\citenamefont{Weng, Sheng, and
  Ting}}]{PhysRevLett.80.5401}
\bibinfo{author}{\bibfnamefont{Z.~Y.} \bibnamefont{Weng}},
  \bibinfo{author}{\bibfnamefont{D.~N.} \bibnamefont{Sheng}}, \bibnamefont{and}
  \bibinfo{author}{\bibfnamefont{C.~S.} \bibnamefont{Ting}},
  \bibinfo{journal}{Phys. Rev. Lett.} \textbf{\bibinfo{volume}{80}},
  \bibinfo{pages}{5401} (\bibinfo{year}{1998}).

\bibitem[{\citenamefont{Kou et~al.}(2005)\citenamefont{Kou, Qi, and
  Weng}}]{kou:235102}
\bibinfo{author}{\bibfnamefont{S.-P.} \bibnamefont{Kou}},
  \bibinfo{author}{\bibfnamefont{X.-L.} \bibnamefont{Qi}}, \bibnamefont{and}
  \bibinfo{author}{\bibfnamefont{Z.-Y.} \bibnamefont{Weng}},
  \bibinfo{journal}{Phys. Rev. B} \textbf{\bibinfo{volume}{71}},
  \bibinfo{pages}{235102} (\bibinfo{year}{2005}).

\bibitem[{\citenamefont{Weng}(2007)}]{weng-2007-21}
\bibinfo{author}{\bibfnamefont{Z.-Y.} \bibnamefont{Weng}},
  \bibinfo{journal}{International Journal of Modern Physics B}
  \textbf{\bibinfo{volume}{21}}, \bibinfo{pages}{773} (\bibinfo{year}{2007}).

\bibitem[{\citenamefont{A.B.~Kuklov and Troyer}(2006)}]{Prokofiev}
\bibinfo{author}{\bibfnamefont{B.~S.} \bibnamefont{A.B.~Kuklov},
  \bibfnamefont{N.V.~Prokof’ev}} \bibnamefont{and}
  \bibinfo{author}{\bibfnamefont{M.}~\bibnamefont{Troyer}},
  \bibinfo{journal}{Annals of Physics} \textbf{\bibinfo{volume}{321}},
  \bibinfo{pages}{1602} (\bibinfo{year}{2006}).

\bibitem[{\citenamefont{Sandvik}(2007)}]{sandvik:227202}
\bibinfo{author}{\bibfnamefont{A.~W.} \bibnamefont{Sandvik}},
  \bibinfo{journal}{Phys. Rev. Lett.} \textbf{\bibinfo{volume}{98}},
  \bibinfo{pages}{227202} (\bibinfo{year}{2007}).

\bibitem[{\citenamefont{Melko and Kaul}(2008)}]{melko:017203}
\bibinfo{author}{\bibfnamefont{R.~G.} \bibnamefont{Melko}} \bibnamefont{and}
  \bibinfo{author}{\bibfnamefont{R.~K.} \bibnamefont{Kaul}},
  \bibinfo{journal}{Phys. Rev. Lett.} \textbf{\bibinfo{volume}{100}},
  \bibinfo{pages}{017203} (\bibinfo{year}{2008}).

\bibitem[{\citenamefont{Motrunich and Vishwanath}(2008)}]{motrunich-2008}
\bibinfo{author}{\bibfnamefont{O.~I.} \bibnamefont{Motrunich}}
  \bibnamefont{and}
  \bibinfo{author}{\bibfnamefont{A.}~\bibnamefont{Vishwanath}}
  (\bibinfo{year}{2008}), \urlprefix\url{arXiv.org:0805.1494}.

\bibitem[{\citenamefont{Kuklov et~al.}(2008)\citenamefont{Kuklov, Matsumoto,
  Prokof'ev, Svistunov, and Troyer}}]{kuklov-2008}
\bibinfo{author}{\bibfnamefont{A.~B.} \bibnamefont{Kuklov}},
  \bibinfo{author}{\bibfnamefont{M.}~\bibnamefont{Matsumoto}},
  \bibinfo{author}{\bibfnamefont{N.~V.} \bibnamefont{Prokof'ev}},
  \bibinfo{author}{\bibfnamefont{B.~V.} \bibnamefont{Svistunov}},
  \bibnamefont{and} \bibinfo{author}{\bibfnamefont{M.}~\bibnamefont{Troyer}}
  (\bibinfo{year}{2008}), \urlprefix\url{arXiv.org:0805.4334}.

\bibitem[{\citenamefont{Affleck and Marston}(1988)}]{PhysRevB.37.3774}
\bibinfo{author}{\bibfnamefont{I.}~\bibnamefont{Affleck}} \bibnamefont{and}
  \bibinfo{author}{\bibfnamefont{J.~B.} \bibnamefont{Marston}},
  \bibinfo{journal}{Phys. Rev. B} \textbf{\bibinfo{volume}{37}},
  \bibinfo{pages}{3774} (\bibinfo{year}{1988}).

\bibitem[{\citenamefont{Lee et~al.}(2006)\citenamefont{Lee, Nagaosa, and
  Wen}}]{9015906}
\bibinfo{author}{\bibfnamefont{P.}~\bibnamefont{Lee}},
  \bibinfo{author}{\bibfnamefont{N.}~\bibnamefont{Nagaosa}}, \bibnamefont{and}
  \bibinfo{author}{\bibfnamefont{X.-G.} \bibnamefont{Wen}},
  \bibinfo{journal}{Rev. Mod. Phys.} \textbf{\bibinfo{volume}{78}},
  \bibinfo{pages}{17 } (\bibinfo{year}{2006}).

\bibitem[{\citenamefont{Wen et~al.}(1989)\citenamefont{Wen, Wilczek, and
  Zee}}]{PhysRevB.39.11413}
\bibinfo{author}{\bibfnamefont{X.~G.} \bibnamefont{Wen}},
  \bibinfo{author}{\bibfnamefont{F.}~\bibnamefont{Wilczek}}, \bibnamefont{and}
  \bibinfo{author}{\bibfnamefont{A.}~\bibnamefont{Zee}},
  \bibinfo{journal}{Phys. Rev. B} \textbf{\bibinfo{volume}{39}},
  \bibinfo{pages}{11413} (\bibinfo{year}{1989}).

\bibitem[{\citenamefont{Thouless et~al.}(1982)\citenamefont{Thouless, Kohmoto,
  Nightingale, and den Nijs}}]{PhysRevLett.49.405}
\bibinfo{author}{\bibfnamefont{D.~J.} \bibnamefont{Thouless}},
  \bibinfo{author}{\bibfnamefont{M.}~\bibnamefont{Kohmoto}},
  \bibinfo{author}{\bibfnamefont{M.~P.} \bibnamefont{Nightingale}},
  \bibnamefont{and} \bibinfo{author}{\bibfnamefont{M.}~\bibnamefont{den Nijs}},
  \bibinfo{journal}{Phys. Rev. Lett.} \textbf{\bibinfo{volume}{49}},
  \bibinfo{pages}{405} (\bibinfo{year}{1982}).

\bibitem[{\citenamefont{Kane and Mele}(2005)}]{kane:226801}
\bibinfo{author}{\bibfnamefont{C.~L.} \bibnamefont{Kane}} \bibnamefont{and}
  \bibinfo{author}{\bibfnamefont{E.~J.} \bibnamefont{Mele}},
  \bibinfo{journal}{Phys. Rev. Lett.} \textbf{\bibinfo{volume}{95}},
  \bibinfo{pages}{226801} (\bibinfo{year}{2005}).

\bibitem[{\citenamefont{Ran et~al.}(2008)\citenamefont{Ran, Vishwanath, and
  Lee}}]{ran-2008}
\bibinfo{author}{\bibfnamefont{Y.}~\bibnamefont{Ran}},
  \bibinfo{author}{\bibfnamefont{A.}~\bibnamefont{Vishwanath}},
  \bibnamefont{and} \bibinfo{author}{\bibfnamefont{D.-H.} \bibnamefont{Lee}}
  (\bibinfo{year}{2008}), \urlprefix\url{arXiv.org:0801.0627}.

\bibitem[{\citenamefont{Dasgupta and Halperin}(1981)}]{PhysRevLett.47.1556}
\bibinfo{author}{\bibfnamefont{C.}~\bibnamefont{Dasgupta}} \bibnamefont{and}
  \bibinfo{author}{\bibfnamefont{B.~I.} \bibnamefont{Halperin}},
  \bibinfo{journal}{Phys. Rev. Lett.} \textbf{\bibinfo{volume}{47}},
  \bibinfo{pages}{1556} (\bibinfo{year}{1981}).

\bibitem[{\citenamefont{Fisher and Lee}(1989)}]{PhysRevB.39.2756}
\bibinfo{author}{\bibfnamefont{M.~P.~A.} \bibnamefont{Fisher}}
  \bibnamefont{and} \bibinfo{author}{\bibfnamefont{D.~H.} \bibnamefont{Lee}},
  \bibinfo{journal}{Phys. Rev. B} \textbf{\bibinfo{volume}{39}},
  \bibinfo{pages}{2756} (\bibinfo{year}{1989}).

\bibitem[{\citenamefont{Baskaran et~al.}(1987)\citenamefont{Baskaran, Zou, and
  Anderson}}]{2986272}
\bibinfo{author}{\bibfnamefont{G.}~\bibnamefont{Baskaran}},
  \bibinfo{author}{\bibfnamefont{Z.}~\bibnamefont{Zou}}, \bibnamefont{and}
  \bibinfo{author}{\bibfnamefont{P.}~\bibnamefont{Anderson}},
  \bibinfo{journal}{Solid State Commun.} \textbf{\bibinfo{volume}{63}},
  \bibinfo{pages}{973 } (\bibinfo{year}{1987}).

\bibitem[{\citenamefont{Ioffe and Larkin}(1989)}]{PhysRevB.39.8988}
\bibinfo{author}{\bibfnamefont{L.~B.} \bibnamefont{Ioffe}} \bibnamefont{and}
  \bibinfo{author}{\bibfnamefont{A.~I.} \bibnamefont{Larkin}},
  \bibinfo{journal}{Phys. Rev. B} \textbf{\bibinfo{volume}{39}},
  \bibinfo{pages}{8988} (\bibinfo{year}{1989}).

\bibitem[{\citenamefont{Ivanov and Lee}(2003)}]{ivanov:132501}
\bibinfo{author}{\bibfnamefont{D.~A.} \bibnamefont{Ivanov}} \bibnamefont{and}
  \bibinfo{author}{\bibfnamefont{P.~A.} \bibnamefont{Lee}},
  \bibinfo{journal}{Phys. Rev. B} \textbf{\bibinfo{volume}{68}},
  \bibinfo{pages}{132501} (\bibinfo{year}{2003}).

\bibitem[{\citenamefont{Polyakov}(1977)}]{Polyakov}
\bibinfo{author}{\bibfnamefont{A.~M.} \bibnamefont{Polyakov}},
  \bibinfo{journal}{Nuclear Physics B} \textbf{\bibinfo{volume}{120}},
  \bibinfo{pages}{429} (\bibinfo{year}{1977}).

\bibitem[{\citenamefont{Lee and Lee}(2005)}]{lee:036403}
\bibinfo{author}{\bibfnamefont{S.-S.} \bibnamefont{Lee}} \bibnamefont{and}
  \bibinfo{author}{\bibfnamefont{P.~A.} \bibnamefont{Lee}},
  \bibinfo{journal}{Phys. Rev. Lett.} \textbf{\bibinfo{volume}{95}},
  \bibinfo{pages}{036403} (\bibinfo{year}{2005}).

\bibitem[{\citenamefont{Florens and Georges}(2004)}]{PhysRevB.70.035114}
\bibinfo{author}{\bibfnamefont{S.}~\bibnamefont{Florens}} \bibnamefont{and}
  \bibinfo{author}{\bibfnamefont{A.}~\bibnamefont{Georges}},
  \bibinfo{journal}{Phys. Rev. B} \textbf{\bibinfo{volume}{70}},
  \bibinfo{pages}{035114} (\bibinfo{year}{2004}).

\bibitem[{\citenamefont{Hermele et~al.}(2005)\citenamefont{Hermele, Senthil,
  and Fisher}}]{Hermele}
\bibinfo{author}{\bibfnamefont{M.}~\bibnamefont{Hermele}},
  \bibinfo{author}{\bibfnamefont{T.}~\bibnamefont{Senthil}}, \bibnamefont{and}
  \bibinfo{author}{\bibfnamefont{M.~P.~A.} \bibnamefont{Fisher}},
  \bibinfo{journal}{Phys. Rev. B} \textbf{\bibinfo{volume}{72}},
  \bibinfo{pages}{104404} (\bibinfo{year}{2005}).

\bibitem[{\citenamefont{Wen and Lee}(1996)}]{PhysRevLett.76.503}
\bibinfo{author}{\bibfnamefont{X.-G.} \bibnamefont{Wen}} \bibnamefont{and}
  \bibinfo{author}{\bibfnamefont{P.~A.} \bibnamefont{Lee}},
  \bibinfo{journal}{Phys. Rev. Lett.} \textbf{\bibinfo{volume}{76}},
  \bibinfo{pages}{503} (\bibinfo{year}{1996}).

\bibitem[{\citenamefont{Halperin et~al.}(1974)\citenamefont{Halperin, Lubensky,
  and Ma}}]{PhysRevLett.32.292}
\bibinfo{author}{\bibfnamefont{B.~I.} \bibnamefont{Halperin}},
  \bibinfo{author}{\bibfnamefont{T.~C.} \bibnamefont{Lubensky}},
  \bibnamefont{and} \bibinfo{author}{\bibfnamefont{S.-k.} \bibnamefont{Ma}},
  \bibinfo{journal}{Phys. Rev. Lett.} \textbf{\bibinfo{volume}{32}},
  \bibinfo{pages}{292} (\bibinfo{year}{1974}).

\bibitem[{\citenamefont{Sachdev}(1999)}]{SachdevBook}
\bibinfo{author}{\bibfnamefont{S.}~\bibnamefont{Sachdev}},
  \emph{\bibinfo{title}{Quantum Phase Transitions}}
  (\bibinfo{publisher}{Cambridge University Press, Cambridge U.K.},
  \bibinfo{year}{1999}).

\bibitem[{\citenamefont{Hou et~al.}(2007)\citenamefont{Hou, Chamon, and
  Mudry}}]{hou:186809}
\bibinfo{author}{\bibfnamefont{C.-Y.} \bibnamefont{Hou}},
  \bibinfo{author}{\bibfnamefont{C.}~\bibnamefont{Chamon}}, \bibnamefont{and}
  \bibinfo{author}{\bibfnamefont{C.}~\bibnamefont{Mudry}},
  \bibinfo{journal}{Phys. Rev. Lett.} \textbf{\bibinfo{volume}{98}},
  \bibinfo{pages}{186809} (\bibinfo{year}{2007}).

\bibitem[{\citenamefont{Hermele}(2007)}]{hermele:035125}
\bibinfo{author}{\bibfnamefont{M.}~\bibnamefont{Hermele}},
  \bibinfo{journal}{Phys. Rev. B} \textbf{\bibinfo{volume}{76}},
  \bibinfo{pages}{035125} (\bibinfo{year}{2007}).

\bibitem[{\citenamefont{Ran et~al.}(2007)\citenamefont{Ran, Ko, Lee, and
  Wen}}]{ran-2007-kagome-monopole}
\bibinfo{author}{\bibfnamefont{Y.}~\bibnamefont{Ran}},
  \bibinfo{author}{\bibfnamefont{W.-H.} \bibnamefont{Ko}},
  \bibinfo{author}{\bibfnamefont{P.~A.} \bibnamefont{Lee}}, \bibnamefont{and}
  \bibinfo{author}{\bibfnamefont{X.-G.} \bibnamefont{Wen}}
  (\bibinfo{year}{2007}), \urlprefix\url{arXiv.org:0710.4574}.

\bibitem[{\citenamefont{Wen}(2002)}]{PhysRevB.65.165113}
\bibinfo{author}{\bibfnamefont{X.-G.} \bibnamefont{Wen}},
  \bibinfo{journal}{Phys. Rev. B} \textbf{\bibinfo{volume}{65}},
  \bibinfo{pages}{165113} (\bibinfo{year}{2002}).

\bibitem[{\citenamefont{Herbut}(2007)}]{herbut:206404}
\bibinfo{author}{\bibfnamefont{I.~F.} \bibnamefont{Herbut}},
  \bibinfo{journal}{Phys. Rev. Lett.} \textbf{\bibinfo{volume}{99}},
  \bibinfo{pages}{206404} (\bibinfo{year}{2007}).

\end{thebibliography}

\end{document}